\documentclass[12pt]{article}

\usepackage{a4wide}

\usepackage{url}
\usepackage[hidelinks]{hyperref}
\usepackage{amsfonts, amsmath, amssymb, theorem}
\usepackage{graphics}

\theorembodyfont{\rmfamily}

\newtheorem{theorem}{Theorem}
\newtheorem{example}[theorem]{Example}

\newcommand{\eq}{\leftrightarrow}

\newcommand{\imp}{\rightarrow}
\newcommand{\Imp}{\Rightarrow}

\newcommand{\Pmi}{\Leftarrow}
\newcommand{\et}{\wedge}
\newcommand{\vel}{\vee}
\newcommand{\Et}{\bigwedge}
\newcommand{\Vel}{\bigvee}

\newcommand{\all}{\forall}
\newcommand{\is}{\exists}
\newcommand{\T}{\top}

\newcommand{\Dia}{\Diamond}
\newcommand{\dia}[1]{\langle #1 \rangle}

\renewcommand{\phi}{\varphi}
\newcommand{\union}{\cup}
\newcommand{\Union}{\bigcup}
\newcommand{\inter}{\cap}
\newcommand{\Inter}{\bigcap}

\newcommand{\powerset}{{\mathcal P}}

\newcommand{\simul}{\preceq}
\newcommand{\lumis}{\succeq}
\newcommand{\bisim}{\simeq}

\newcommand{\weg}[1]{}

\newcommand{\pre}{\mathsf{pre}}

\newcommand{\Formulas}{{\mathcal L}}

\newcommand{\langu}{\Formulas}

\newcommand{\arel}{\ensuremath{\mathsf{R}}}
\newcommand{\Actions}{\ensuremath{\mathsf{S}}}

\newcommand{\Domain}{{\mathcal D}}
\newcommand{\domain}{\Domain}

\newcommand{\Nat}{\mathbb N}
\newcommand{\Naturals}{\Nat}

\newcommand{\lang}{\langu}

\newcommand{\F}{\exists}

\newcommand{\var}{\mathit{v}}

\newcommand{\caldia}[1]{\langle \! [ #1 ] \! \rangle}

\newcommand{\apalbox}{[!]}
\newcommand{\apaldia}{\langle{!}\rangle}
\newcommand{\rmlbox}{[\succeq]}
\newcommand{\rmldia}{\langle\succeq\rangle}

\renewcommand{\F}{\apaldia}     
\newcommand{\G}{\apalbox}     

\usepackage{tikz}
\newcommand{\calbox}[1]{[\!\langle #1 \rangle\!]}

\newcommand{\Kw}{\mathit{Kw}}

\newcommand{\aumodel}{U}
\newcommand{\austates}{O}
\newcommand{\austate}{o}

\newcommand{\aufunction}{R\!R}

\newcommand{\AAUL}{\uparrow}

\begin{document}

\title{To Be Announced}
\author{Hans van Ditmarsch\thanks{University of Toulouse, CNRS, IRIT, France.}}
\date{}

\maketitle              

\begin{abstract}
In this survey we review dynamic epistemic logics with modalities for quantification over information change. Of such logics we present complete axiomatizations, focussing on axioms involving the interaction between knowledge and such quantifiers, we report on their relative expressivity, on decidability and on the complexity of model checking and satisfiability, and on applications. We focus on open problems and new directions for research.
\end{abstract}

\section{Introduction} \label{sec.introduction}

\emph{Truthful public announcement} is the misnomer of the year. Which year? Not the year 1989, the publication of Jan Plaza's `Logics of public communications' \cite{plaza:1989}. Plaza does not use `announcement'. Not the year 1998 either, the publication of Alexandru Baltag et al.'s `The logic of public announcements (\dots)' \cite{baltagetal:1998}, as there are prior occurrences such as in Jelle Gerbrandy's \cite{gerbrandyetal:1997} from 1997.  Whereas in 1986, before \cite{plaza:1989}, Yoram Moses et al.'s \cite{mosesetal:1986} uses `announcement' (possibly, because the statements made by the Queen of Mamajorca are announcements by virtue of her being a queen), but not `public' nor `truthful'. So in the end, it is not known which year.

\emph{Truthful} is a misnomer, because a statement may be truthful (what you \emph{believe} to be true, a subjective criterion) but not true (an objective criterion). Only truth is formally required. The twain meet because announcements are assumed to be made by an omniscient and benevolent agent, for whom being truthful is the same as saying the truth, and who will not say what she does not know to be true. More down to earth: the listener assumes that the speaker wants to inform (is truthful) and is also correct (says the truth).

\emph{Announcement} is a misnomer, because the observed information may be heard as well as seen (the light is on, or off). Such forms of observation are also modelled as an announcement in the logics that we deal with in this contribution. So it need not be a real announcement, which is presumably a speech act. There may not be an announcer anyway, in the sense of an agent modelled in a system. The source of the new information is typically assumed to be anonymous, whereas an announcement of a proposition by an agent modelled in the system is formalized as `the agent knows the proposition'. 

\emph{Public} is, after all, not a misnomer. The new information is assumed to be received synchronously. That is a backdoor towards making `announcement' acceptable terminology: `announcement' carries the legal sense that all are supposed to have heard, and liable for the consequences even if they did not. Announcements tend to be public by definition.

We will often omit `truthful' from `truthful public announcement'. Untruthful announcements play no role in this survey. Public announcements have also been called \emph{revelations}, a very suitable name but that did not stick (unlike in game theory and mechanism design). It is on the same delightfully pretentious scale as calling a state a \emph{world}. And public announcements have been called \emph{public events}, another suitable name that might or might not stick but so far survives in the community. Or public actions, which carries a notion of agency. Personally, I like \emph{public observation}, precisely as it lacks agency. Either way, it always concerns a true proposition that is simultaneously or synchronously observed by a set of agents and as a consequence of which they adjust their knowledge bases, they update what they know to be true. 

{\em Public announcement logic} (PAL) \cite{plaza:1989} is an extension of epistemic logic, the logic of knowledge. Let us survey those origins. Jaakko Hintikka proposed the modal logic of knowledge to give a relational semantics for the perceived properties of knowledge, namely that what you know is true, that you know what you know, and that you know what you don't know \cite{hintikka:1962}. He compares these with the properties of other epistemic notions such as belief (you may believe something even when it is false; in this survey we focus on knowledge). Already in  \cite{hintikka:1962} the analysis of Moorean phenomena \cite{moore:1942} played an important role: you cannot be informed that you are ignorant of a (true) fact without `losing' that ignorance: you cannot know that some fact $p$ is true and that you don't know this. This is inconsistent in a single-agent logic of knowledge. In a multi-agent logic of knowledge similar phenomena are not so paradoxical: there is no issue with \emph{me} knowing that $p$ is true and that \emph{you} don't know this. 

Further developments involved on the one hand group epistemic notions such as common knowledge \cite{aumann:1976,mccarthy:1978} and distributed knowledge \cite{hilpinen:1977}, topics that we will largely bypass in this contribution. 

On the other hand, it involved dynamics of knowledge: the multi-agent logic of knowledge led to increased interest in the analysis of agents informing each other of their ignorance and knowledge, often inspired by logic puzzles \cite{mosesetal:1986,mccarthy:1978}. This culminated in the already mentioned public announcement logic, wherein such informative actions (called public communications in \cite{plaza:1989}) became full members of the logical language besides the knowledge modalities.  `Culminated' is  attributed in retrospect, as the importance of this somewhat informal publication only increased over the years. There were also parallel developments of dynamic but not epistemic logics of information change such as \cite{Emde:84,jfak.logcol:1989}. 

{\em Action model logic} (AML) \cite{baltagetal:1998} provided a generalization of knowledge dynamics from public information change to non-public or untruthful information change (also known as {\em epistemic actions}), such as private or secret announcements to some agents while other agents only partially observe that. Parallel, now lesser known, developments were \cite{gerbrandyetal:1997,hvd.thesis:2000}.

A different direction of generalizing PAL is to consider quantification over announcements.  Arbitrary public announcement logic APAL \cite{balbianietal:2008} contains a construct that can be paraphrased as `there is a truthful public announcement after which proposition $\phi$ is true'. If I, a poor student, am uncertain about the exam result (pass $p$ or fail $\neg p$) but you, the examiner, know that I passed, there is a truthful announcement after which I know that $p$, namely you informing me that $p$ is true. You can, so to speak, consider in your mind all possible true announcements and pick the one that ensures that I know $p$ afterwards. That is then to be announced: the announcement `the examiner knows that $p$' results in (realizes) the epistemic goal `the student knows whether $p$'. This is a very simple example. But as one cannot always make a proposition true by announcing it (we recall Moorean phenomena, such as announcing `$p$ is true and you do not know it'), much more complex settings are conceivable. 

{\em To be announced} is what you have to say in order to achieve an epistemic goal. It is the question whether {\em there is} an announcement in order to realize that goal formula. In modal logics, this announcement may be very different from that goal. This is therefore a non-trivial problem, and the quantification involved seems worthy of investigation.

The logic APAL was rooted in various traditions, both in philosophy and in computer science. It has roots in knowability \cite{fitch:1963}, by way of \cite{jfak.fitch:2004}, but also in propositional quantification \cite{Fine:1970}, by way of \cite{hollenberg:1998,french:2006}. All of these involve quantifying over information change, however not in a dynamic epistemic setting of which the public announcement is the typical example, as we will explain later. However, even in that setting there are many other ways to quantify over information change than in APAL, both over public and over non-public information change, and that can still be seen as announcing something, even when not publicly. Let us contrast the APAL quantifier with two other quantifiers among many presented in this survey (where $\phi$ is a formula and $G$ a subset of the set of all agents).
\begin{enumerate}
\item There is a public announcement after which $\phi$ \cite{balbianietal:2008}. \label{i}
\item There is a public announcement by the agents in group $G$ after which $\phi$ \cite{agotnesetal.jal:2010}. \label{ii}
\item There is an epistemic action (such as a private announcement) after which $\phi$ \cite{hales2013arbitrary}; \label{iii}
\end{enumerate}

{\em Item \ref{i} is the APAL quantifier.} 

{\em Item \ref{ii} is a restriction of the APAL quantifier.} Quantifying over announcements involving a group of agents facilitates the formalization of notions of ability and agency in dynamic epistemic logic. For example, the group $G$ may consist of the principals in a security protocol sending and receiving messages to and from each other, in the presence of one or multiple eavesdroppers possibly intercepting all messages, so that such messages should be considered public announcements. Correctly modelling messages {\em known} by agents requires to carefully distinguish true announcements from truthful announcements, to which we will later pay detailed attention.

{\em Item \ref{iii} generalizes the APAL quantifier.} By epistemic or informative action we mean some event that changes the knowledge of beliefs of agents. A public announcement is an epistemic action, but epistemic actions also include private announcements: also then, there is something to be announced, but the announcement is not observed by all agents in the same way. A typical example is that of a player showing a card to another player while the remaining players observe the interaction but cannot see which card has been shown \cite{hvd.jolli:2002}. Which is another example of an `announcement' in the technical sense of this paper that does not involve anything being said.\footnote{If we replace `showing a card' such that the others cannot see, by `whispering the name of the card' such that the others cannot hear, after first having publicly announced to all `I am now going to whisper the name of a card I have to my neighbour', it is again an announcement wherein something is said.} It involves other agents observing the informative event in different ways. Epistemic actions can be represented as the already mentioned action models, but also in other ways presented in this survey: as \emph{arrow updates} \cite{kooietal:2011,kooirenne,hvdetal.aij:2017}, and as the \emph{refinements} of \cite{hales2013arbitrary,bozzellietal.inf:2014}. In all such cases, quantifying over epistemic actions comes with different properties.

\weg{Item \ref{iv} is a different kind of generalization, but much related to item \ref{iii}. The refinement mentioned in this item is the dual of simulation \cite{blackburnetal:2001}: from the bisimulation requirements, a refinement relation must satisfy {\bf atoms} and {\bf back}. There are subtle differences between `there is an informative action' and `there is a refinement', but the two come quite close. This similarity opens perspectives to compare dynamic epistemic logics with so-called relation-changing modal logics. Refinement seems arbitarily iterated sabotage. Refinement modal logic \cite{bozzellietal.inf:2014} has an information quantifier but no modalities for specific, deterministic, information change, unlike most other logics presented in this survey.}

To determine what is to be announced in order to make $\phi$ true, is to synthesize a plan to realize goal $\phi$. There is therefore a strong relation between having quantifiers over information change in the logical language on one side, and performing epistemic planning \cite{bolanderetal:2011} on the other side, which is done in an off-line, meta-level way: there is a plan to realize an epistemic goal $\phi$ if the search tree that connects information states by permitted actions has a leaf satisfying the goal formula. There are also differences. The goal $\phi$ in epistemic planning is typically an epistemic formula without quantifiers. But a formula $\phi$ in a logical language with quantifiers may contain many such quantifiers, each posing a planning problem. We can expect a difference in complexities of decision problems.

For all dynamic epistemic logics in this survey we focus on complete axiomatizations, decidability of satisfiability, expressivity, to a lesser extent also on the complexity of model checking and (decidable) satisfiability, and last but not least we focus on applications. Applications can be found in multi-agent system analysis, in security protocols, in gossip protocols, and in epistemic planning on specific domains. All these and more will be discussed in detail.  We list many open problems but also completely new directions for research.

In this survey we wish to address both the informed reader in modal logics, such as a master or doctoral student in logic, computer science, or artificial intelligence, as well as the expert in dynamic epistemic logic. For the benefit of the former we start the survey with a self-contained introduction in public announcement logic and action model logic. For the benefit of the latter we carefully list the open problems within well-defined subjects and that build on foundations of established results. Either way, we wish to share with the readers of all backgrounds our enthusiasm for the topic and its further development, and thank them for any moment of their time they choose to spend on the survey.

\paragraph*{Outline} We close the introductory section with an outline of the content. Section \ref{sec.prelims} provides the technical background on public announcement logic and on action model logic. Section \ref{sec.apal} is on arbitrary public announcement logic and its variations. Section \ref{sec.galcal} is on group announcement logic and coalition announcement logic. These are variations of arbitrary public announcement logic modelling aspects of agency, ability or strategy. Section \ref{sec.am} is on quantification in action model logic, Section \ref{sec.au} is on quantification in arrow update logics, and Section \ref{sec.rml} is on refinement modal logics. The logics in the Sections \ref{sec.am}, \ref{sec.au}, and \ref{sec.rml} all quantify over arbitrary, not necessarily public, information change. Sections \ref{sec.iterate} and \ref{sec.sabotage} are more exploratory in character than the other sections in this survey: they point towards directions for novel research and are less a survey of work already done. Section \ref{sec.iterate} is on iterated information change, factual change, group epistemic notions, and epistemic planning. These topics are normally considered in dynamic epistemic logics, but play a minor role in this survey on quantification. Section \ref{sec.sabotage} is on so-called relation-changing modal logics. Such logics are not normally considered dynamic epistemic logics, but there are most relevant relations to quantification.

\section{Dynamic epistemic logic} \label{sec.prelims}

In this section we present the technical prerequisites of certain basic dynamic epistemic logics before plunging into extensions of such logics with quantifiers over information change. We will also include intuitive motivations and examples, in order to make the material more accessible and to keep it lightweight. Readers familiar with such basics may wish to skip the section. We present multi-agent epistemic logic, on the general level of a multi-modal logic \cite{hintikka:1962}, as well as public announcement logic \cite{plaza:1989} and action model logic \cite{baltagetal:1998}. Prerequisites for arrow update logic \cite{kooietal:2011} are in Section \ref{sec.au}. 

The language, structures, and semantics are as follows.

\paragraph*{Languages}
Given a non-empty countable (finite or infinite) set of {\em propositional variables} ({\em atoms}) $P$ and a non-empty finite set of {\em agents} $A$, we consider four different basic logical languages that we will later expand with quantifiers. The elements of the language are the {\em formulas}.
\[ \begin{array}{lclll} 
\lang(\emptyset) & \ni & \phi & ::= & p \mid \neg \phi \mid (\phi \et \phi) \\
\lang(\Dia) & \ni & \phi & ::= & p \mid \neg \phi \mid (\phi \et \phi) \mid \Dia_a \phi  \\
\lang(\Dia,!) & \ \ \ni \ \ & \phi & ::= & p \mid \neg \phi \mid (\phi \et \phi) \mid \Dia_a \phi \mid \dia{\phi} \phi \\
\lang(\Dia,\otimes) & \ni & \phi & ::= & p \mid \neg \phi \mid (\phi \et \phi) \mid \Dia_a \phi  \mid \dia{E_e} \phi
\end{array}
\] 
where $p \in P$, $a \in A$, and where $E_e$ is a finite pointed {\em action model} only explained and defined further below. Other propositional connectives are defined by abbreviation and we also define $\Box_a \phi$ by abbreviation as $\neg\Dia_a\neg\phi$ and similarly $[\phi] \psi$ and $[E_e] \psi$. For $\Dia_a \phi$ we read `agent $a$ considers $\phi$ possible' and for $\Box_a \phi$, `agent $a$ knows $\phi$' (in other epistemic settings this may mean `agent $a$ believes $\phi$'). If $|A|=1$ we may write $\Box\phi$ and $\Dia\phi$ instead of $\Box_a\phi$ and $\Dia_a\phi$. For $\dia{\phi} \psi$ we read `($\phi$ is true and) after (truthful) public announcement of $\phi$, $\psi$ (is true)'. In the literature, formulas $[\phi]\psi$ and $\dia{\phi}\psi$ are often written as $[!\phi]\psi$ and $\dia{!\phi}\psi$. This explains the $!$ in the language definition (and later, in the language extended with quantification over announcements).

Given a formula $\phi$, $\var(\phi)$ is the set of propositional variables occurring in $\phi$, and $d(\phi)$ is the {\em modal depth} or $\Box$-depth of $\phi$ (the maximum stack of modalities in a formula). The definition of $\Box$-depth is: $d(p) = 0$, $d(\neg\phi) = d(\phi)$, $d(\phi\et\psi) = \max \{d(\phi),d(\psi)\}$, $d(\Dia_a\phi) = d(\phi)+1$, $d(\dia{\phi}\psi) = d(\phi)+d(\psi)$. The clause for $\dia{E_e}\phi$ will be given later. The possibly surprising difference between the clause for conjunction, which takes the maximum of the modal depth of both arguments, and the clause for announcement, which takes their sum, is that the latter is equivalent to a formula with a stack of modalities of the length of that sum (see e.g.~\cite{hvdetal.papal:2020}).

\paragraph*{Structures}

An {\em epistemic model} (or `Kripke model', or just `model') $M = (S, R, V)$ consists of a non-empty {\em domain} $S$ of {\em states} (or `worlds'; the domain is also denoted $\domain(M)$), an {\em accessibility function} $R: A \imp {\mathcal P}(S \times S)$, where each $R(a)$ is an {\em accessibility relation}, and a {\em valuation} $V: P \imp {\mathcal P}(S)$, where each $V(p)$ represents the set of states where $p$ is true. For $R(a)$ we write $R_a$, and for $(s,t)\in R_a$ we may write $R_a(s,t)$ or $R_ast$. A pair $(S,R)$ is called a \emph{frame}. For $s \in S$, a pair $(M,s)$, for which we write $M_s$, is a {\em pointed (epistemic) model} (often also simply called model --- the context should disambiguate usage). We will occasionally employ {\em multi-pointed models} $M_T$ where $T \subseteq S$. The model class without any restrictions is ${\mathcal K}$. The class of models where all accessibility relations are equivalence relations is ${\mathcal S5}$. If $R_a$ is an equivalence relation we write $s \sim_a t$ instead of $R_ast$, and the relation is then called {\em indistinguishability relation}.

Let models $M= (S, R, V)$ and $M'= (S', R', V')$ be given. A non-empty relation $Z \subseteq S \times S'$ is a \emph{bisimulation} between $M$ and $M'$, notation $Z: M \bisim M'$, if for all pairs $(s,s') \in
Z$ ($Zss'$) and $a\in A$:
\begin{description}
\item[atoms] $s \in V(p)$ iff $s' \in V'(p)$ for all $p \in P$;
\item[forth] if
$R_ast$, then there is a $t'\in
S'$ such that $R'_as't'$ and
$Ztt'$;
\item[back] if $R'_as't'$, then there is a
$t \in S$ such that $R_ast$
and $Ztt'$.
\end{description}
We write $M \bisim M'$ if there is a bisimulation between $M$ and $M'$, and we write $M_s \bisim M'_{s'}$ if there is a bisimulation between $M$ and $M'$ containing pair $(s,s')$. 

Similarly, a {\em simulation} satisfies {\bf atoms} and {\bf forth}, where we write $M \simul M'$ if there is a simulation between $M$ and $M'$, and $M_s \simul M'_{s'}$ if it contains pair $(s,s')$; whereas its dual, the {\em refinement}, satisfies {\bf atoms} and {\bf back}, and where we write $M \lumis M'$ if there exists a refinement between $M$ and $M'$, and $M_s \lumis M'_{s'}$ if it contains pair $(s,s')$. Further, if $M \lumis M'$ the model $M'$ is  called a refinement of $M$ as well. It is sometimes confusing that in the literature the term `refinement' denotes both the refinement relation linking $M$ and $M'$, and the refined model $M'$.

A (to $Q$) {\em restricted bisimulation} $Z^Q$ is a bisimulation that satisfies {\bf atoms} for all variables $Q \subseteq P$, in notation $M \bisim^Q M'$. A (by $n$) {\em bounded bisimulation} $Z^n$ satisfies {\bf forth} and {\bf back} up to depth $n \in \Naturals$. Formally we define this as a set $Z^0 \supseteq Z^1 \dots \supseteq Z^n$ of $i$-bisimulations for $0 \leq i \leq n$ as follows (see also \cite{blackburnetal:2001}).  Relation $Z^0$ merely satisfies {\bf atoms}, and for all $Z^{n+1}ss'$ and $a\in A$: 
\begin{description}
\item[atoms] $s \in V(p)$ iff $s' \in V'(p)$ for all $p \in P$;
\item[forth-$(n+1)$] if
$R_ast$, then there is a $t'\in
S'$ such that $R'_as't'$ and
$Z^ntt'$;
\item[back-$(n+1)$] if $R'_as't'$, then there is a
$t \in S$ such that $R_ast$
and $Z^ntt'$.
\end{description}

We write  $M_s \bisim^n M'_{s'}$ for $n$-bisimilar models $M_s$ and $M'_{s'}$. Bounded bisimulations are used to compare models $M_s$ and $M'_{s'}$ up to a depth $n$ from the respective points $s$ and $s'$.

\paragraph*{Semantics of PAL}

We first give the semantics for truthful public announcement logic (PAL) \cite{plaza:1989}. Assume an epistemic model $M = (S, R, V)$, and let $s \in S$. We define $M_s \models \phi$ (for: $M_s$ \emph{satisfies} $\phi$, or $\phi$ is \emph{true} in $M_s$) by induction.
\[ \begin{array}{lcl}
M_s \models p &\mbox{ \ iff \ } & s \in V(p) \\ 
M_s \models \neg \phi &\mbox{iff} & M_s \not \models \phi \\ 
M_s \models \phi \et \psi &\mbox{iff} & M_s \models \phi  \text{ and } M_s \models \psi \\  
M_s \models \Dia_a \phi &\mbox{iff} & \text{there is a }  t \in S \text{ such that } R_ast \text{ and } M_t  \models \phi \\
M_s \models \dia{\psi}\phi &\mbox{iff} & M_s \models \psi \text{ and } (M|\psi)_s  \models \phi 
\end{array} \] 
where $M|\psi := (S', R', V')$ such that $
S' := \{ s\in S \mid M_s \models \psi\}$, $R'_a := R_a \inter \ (S' \times S')$, and $V'(p) := V(p) \inter S'$. A formula $\phi$ is {\em valid on $M$}, notation $M \models \phi$, if for all states $s$ in the domain of $M$, $M_s \models \phi$, and a formula is \emph{valid},  notation $\models \phi$, if it is valid on all models $M$. Expression $M_T \models \phi$, where $M_T$ is a multi-pointed model, abbreviates that $M_t \models \phi$ for all $t \in T$.

The set $S' := \{ s\in S \mid M_s \models \psi\}$ above is called the \emph{extension} or \emph{denotation} of $\psi$ in model $M$, and the model $M|\psi$ is called the {\em restriction} of $M$ to $\psi$, or, alternatively, we call $M|S'$ the restriction of $M$ to $S'$.

Given $M = (S,R,V)$, formula $\phi$ is {\em distinguishing} between $T \subseteq S$ and $S \setminus T$ (or: $\phi$ is {\em distinguishing for} $T$) if for all $t \in T$, $M_t \models \phi$, and for all $t \notin T$, $M_t \not\models \phi$. A {\em distinguishing formula for $s \in S$} is therefore distinguishing between $\{s\}$ and $S\setminus \{s\}$.

\begin{example} \label{ex.first}

Agent $a$ is uncertain about $p$. After announcement $p$, she knows that $p$. Consider a two-state model $M$ wherein she cannot distinguish a state $1$ where $p$ is true from a state $0$ where $p$ is false. We have that $M_1 \models \dia{p} \Box_a p$ because $M_1 \models p$ and $(M|p)_1 \models  \Box_a p$, where the latter is true because $(M|p)_1 \models p$, and because $1$ is the only accessible state.

We can visualize such models in different ways. As this will happen regularly in this survey we had better be very explicit on our visual conventions. Either we depict it with all pairs in the relation as arrows, as below on the left, or, for $\mathcal S5$ models with equivalence relations, we only link (`relate') indistinguishable states, as below on the right.  When linking states we assume reflexivity, symmetry and transitivity of the relation. The assumption of transitivity is not used in the example below but in other examples in this survey. We underline the actual state, that is, state $1$ in this example. States where $p$ is true are typically named $1$ and states where $p$ is false are typically named $0$. The update relation between the respective models is visualized by $\Imp$. It is labelled with the formula of the announcement inducing the update.
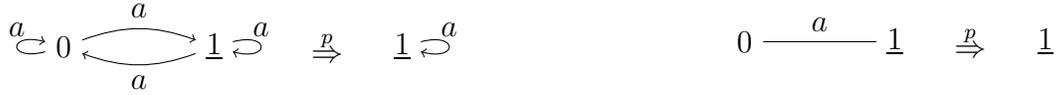
\begin{center}
\begin{tikzpicture}
\node (0) at (0,0) {$0$};
\node (1) at (2,0) {$\underline{1}$};
\draw[->,bend left =20] (0) to node[above] {$a$} (1);
\draw[<-,bend right =20] (0) to node[below] {$a$} (1);
\draw[->] (0) edge[loop left,looseness=10] node[above] {$a$} (0); 
\draw[->] (1) edge[loop right,looseness=10] node[above] {$a$} (1); 
\node (1) at (3.5,0) {$\stackrel{p}{\Imp}$};
\node (1) at (4.5,0) {$\underline{1}$};
\draw[->] (1) edge[loop right,looseness=10] node[above] {$a$} (1); 
\end{tikzpicture}
\quad\quad\quad\quad\quad\quad\quad\quad
\begin{tikzpicture}
\node (corr) at (0,-.5) {$\color{white}0$};
\node (0) at (0,0) {$0$};
\node (1) at (2,0) {$\underline{1}$};
\draw[-] (0) -- node[above] {$a$} (1);
\node (1) at (3,0) {$\stackrel{p}{\Imp}$};
\node (1) at (4,0) {$\underline{1}$};
\end{tikzpicture}
\end{center}
Note that we also have $M_1 \models p\et\neg\Box_ap$. Announcing this, results in the same model restriction to the $1$-state. As $\Box_ap$ implies $\neg p\vel\Box_ap$, which is equivalent to $\neg(p\et\neg\Box_ap)$, we therefore have that $M_1 \models \dia{p\et\neg\Box_ap} \neg(p\et\neg\Box_ap)$: after the announcement $p\et\neg\Box_ap$ this formula has become false. This is an example of a Moorean phenomenon. 
\end{example}

\paragraph*{Axiomatization of PAL}
The axiomatization of {\em minimal modal logic} K contains axiom $\Box_a (\phi \imp \psi) \imp \Box_a \phi \imp \Box_a \psi$ and derivation rule `From $\phi$ infer $\Box_a \phi$' (apart from propositional features). The modal logic S5 additionally contains, for all agents $a$, axioms $\Box_a \phi \imp \phi$, $\Box_a \phi \imp \Box_a \Box_a \phi$, and $\neg\Box_a \phi \imp \Box_a \neg\Box_a \phi$. There are no multi-agent interaction axioms. 

The axiomatization of PAL extends that of K or S5 (or yet other modal logics) with so-called {\em reduction axioms} for the announcement. It follows a pattern that is common in dynamic epistemic logics and that we therefore explain in some detail. The reduction axioms for the announcement are validities that have the shape of equivalences. There is one for each main logical connective of the formula bound by the announcement. 
\[\begin{array}{lll}
{[\psi]}p &\eq &(\psi \imp p) \\
{[\psi]}\neg\phi &\eq & (\psi\imp\neg[\psi]\phi) \\
{[\psi]}(\phi\et\chi) & \eq & ([\psi]\phi\et[\psi]\chi) \\
{[\psi]}\Box_a\phi & \eq & (\psi \imp \Box_a [\psi]\phi) \\
{[\psi]}[\phi]\chi & \eq & [\psi \et [\psi]\phi]\chi
\end{array}\]
Strictly, expressions such as ${[\psi]}(\phi\et\chi)  \eq  ([\psi]\phi\et[\psi]\chi)$ are not axioms but are \emph{axiom schemes}, whereas for example the instantiation ${[p]}(q \et \Box_a p)  \eq ([p]q\et[p]\Box_a p)$ of that schema is an axiom. The difference is relevant because the logic PAL (as well as most other dynamic epistemic logics in this survey) does not have the \emph{substitution property} that we can replace the occurrence of an atom in a validity by any formula. For example, $[p]p$ is valid but $[p \et \neg \Box_a p](p \et \neg \Box_a p)$ is invalid. The shape of the axiom $[\psi]p \eq (\psi \imp p)$ is a giveaway for this lacking feature: it contains an atom as well as an arbitrary formula. That reduction strictly depends on the bound formula being an atom and not just any formula. Whereas in a logic with the substitution property, such as K and S5, atoms can also be any formula.

In other dynamic epistemic logics, many of these reduction axioms are very similar, except the one for the interaction between the update modality (here the announcement) and knowledge. In the continuation we focus on such interaction between update and knowledge.

Except for the first one, the reduction axioms for the announcement all have the shape $[\psi]\mathsf{this} \eq \mathsf{that}$ where {\sf this} is not a subformula of {\sf that}, but where $\mathsf{that}$ contains a subformula $[\psi]\mathsf{other}$ where $\mathsf{other}$ is a subformula of $\mathsf{this}$. Roughly, that feature is what makes it possible to \emph{reduce} (rewrite) a formula with announcements to an equivalent formula without announcements. Let us give an example: \[\begin{array}{ll} {[p \et \neg \Box_a p]} \Box_a p, & \text{iff} \\  (p \et \neg \Box_a p) \imp \Box_a [p \et \neg \Box_a p] p, & \text{iff} \\ (p \et \neg \Box_a p) \imp \Box_a (p \et \neg \Box_a p \imp p). \end{array}\] Note that the last formula is equivalent to $p \et \neg \Box_a p$.  In two steps the reduction terminates. The final step is when an announcement binds an atom, as the announcement modality will then have `disappeared' on the righthand side. In general, to show that such a rewrite procedure terminates one needs to associate a weight or measure (often a natural number) to a formula and prove that the weight of $[\psi]\mathsf{this}$ is larger than the weight of $\mathsf{that}$. Similar reductions exist for many of the dynamic epistemic logics we present in this survey. Each logic may require a different formula weight in order to prove termination of the reduction.

\paragraph*{Expressivity} Given two logics $L_1$ and $L_2$ defined over the same class of models, $L_1$ is {\em at least as expressive as} $L_2$, if for every formula $\phi_2$ in the language of $L_2$ there is a formula $\phi_1$ in the language of $L_1$ such that $\phi_1$ is satisfied by precisely the class of pointed models that satisfy $\phi_2$. 
If $L_1$ is at least as expressive as $L_2$ and $L_2$ is at least as expressive as $L_1$ then $L_1$ and $L_2$ are {\em equally expressive} (or: $L_1$ is as expressive as $L_2$). If $L_1$ is at least as expressive as $L_2$ and $L_2$ is not at least as expressive as $L_1$ then $L_1$ is {\em more expressive than} $L_2$. 

By way of reductions as the above example, it can be shown that every formula with announcements is equivalent to a formula without announcements, and therefore public announcement logic PAL is as expressive as multi-agent epistemic logic K (with respect to model class $\mathcal K$) and as expressive as the logic S5 (with respect to model class $\mathcal S5$). The announcement modality is pushed ever more inward until it gets eliminated by an application of the first axiom $[\psi]p \eq (\psi\imp p)$ above, where on the right-hand side of the equation there is one less announcement modality than on the left-hand side. We now proceed to eliminate the next announcement modality by such reductions, and so on, until there are none left. See \cite{kooi.jancl:2007} for a general approach for dynamic modalities.

\paragraph*{Action models}

An action model \cite{baltagetal:1998} (or event model) is a structure like a Kripke model but with a precondition function instead of a valuation function. An {\em action model} $E = (\Actions, \arel, \pre)$ consists of a {\em domain} $\Actions$ of {\em actions}, an {\em accessibility function} $\arel: A \imp {\mathcal P}(\Actions \times \Actions)$, where each $\arel_a$ is an accessibility relation, and a {\em precondition function} $\pre: \Actions \imp \lang$, where $\lang$ is a logical language --- which in our case would typically be the epistemic language $\lang(\Dia)$; however, there is also a recursive way to define a language with action model modalities, that we present below. A pointed action model $E_e$, in other words a pair $(E,e)$, where $e \in \Actions$, is an {\em epistemic action}. Just as for Kripke models, we also permit multi-pointed action models $E_F$ where $F \subseteq \Actions$. We will also use `epistemic action' to denote any other information changing update.


Performing an epistemic action that is a pointed action model in a pointed epistemic model involves computing the {\em restricted modal product} of the action model and the epistemic model. This product (also called the update product) encodes the new state of information. It is defined as follows.

Given an epistemic model $M = (S, R, V)$ and an action model $E = (\Actions, \arel, \pre)$, the updated model $(M \otimes E) = (S', R', V')$ is such that \[ \begin{array}{lcl}  S' & \ \ = \ \ & \{ (t,f) \in S \times \Actions \mid M_t \models \pre(f) \} \\ ((t,f),(t',f')) \in R'_a & \text{iff} & R_att' \text{ and } \arel_aff'  \\ (t,f) \in V'(p) & \text{iff} & t \in V(p)
\end{array} \]
In other words: the domain consists of the product domain, but restricted to (state, action) pairs $(t,f)$ such that $M_t \models \pre(f)$, i.e., such that the action can be executed in that state; an agent considers a pair $(t,f)$ possible in the update model if she considered the previous state $t$ possible and the execution of action $f$ in that state; and the valuations do not change after action execution. 

A truthful public announcement $\phi$ is a singleton action model $(\{e\}, \arel,\pre)$, with $\arel_a = \{(e,e)\}$ for all $a \in A$, and $\pre(e)=\phi$.

\begin{example} \label{ex.aml}
Consider a model wherein two agents $a,b$ are uncertain about the value of an atom $p$, and where $p$ is true. Further, consider the epistemic action representing that agent $a$ learns that $p$ is true, but that agent $b$ is uncertain whether that has happened (that is, agent $b$ considers it possible that $a$ has learnt that $p$ is true and also considers it possible that nothing has happened). The epistemic action consists of two actions that have been named after their preconditions $p$ and $\T$. They can be distinguished by agent $a$ but not by agent $b$. 
\begin{center}
\begin{tikzpicture}
\node (00) at (0,0) {$0$};
\node (10) at (2,0) {$\underline{1}$};
\node (t) at (3,0) {\Large$\otimes$};
\node (110) at (4,0) {$\underline{p}$};
\node (120) at (4,2) {$\top$};
\node (e) at (5,0) {\Large$=$};
\node (30) at (8,0) {$\underline{(1,p)}$};
\node (21) at (6,2) {$(0,\top)$};
\node (31) at (8,2) {$(1,\top)$};
\draw (00) to node[above]{$ab$} (10);
\draw (21) to node[above]{$ab$} (31);
\draw (30) to node[left]{$b$} (31);
\draw (110) to node[left]{$b$} (120);
\end{tikzpicture}
\end{center}
The result of executing the action is depicted on the right. The point of the updated model is the one with precondition $p$. It can only be executed in the state wherein $p$ is true. In fact, $a$ is learning that $p$, but $b$ is uncertain between that action and the `trivial' action wherein nothing is learnt. The trivial action has precondition $\T$. The trivial action can be executed in both states of the initial model. Therefore, the updated model contains three states. Its accessibility relations are calculated from the initial epistemic model and the action model. For example, $(0,\top) \sim_b (1,p)$ because $0 \sim_b 1$ and $\top \sim_b p$.  Whereas $(0,\top) \not\sim_a (1,p)$ because $\top \not\sim_a p$, even though $0 \sim_a 1$. Note that we have $(0,\top) \sim_b (1,p)$ although the $b$-link between $(0,\top)$ and $(1,p)$ is not drawn in the figure: it follows from the $b$-link drawn between $(0,\top)$ and $(1,\top)$, the $b$-link drawn between $(1,\top)$ and $(1,p)$, and the assumption of transitivity.

We have that $(1,p) \models \Box_a p \et \neg \Box_b \Box_a p$. Here, $(1,p) \models \neg \Box_b \Box_a p$, because $(1,p) \sim_b (1,\top)$ and $(1,\top) \not\models\Box_a p$; where the latter is because $(1,\top) \sim_a (0,\top)$ and $(0,\top) \not\models p$. Also, $(1,p) \models \Box_a p$, because only $(1,p)$ is $a$-accessible.
\end{example}

The composition of two action models is again an action model. The construction has the same spirit as that of updating epistemic models. Given $E = (\Actions, \arel, \pre)$ and $E' = (\Actions', \arel', \pre')$, their composition $E'' = (\Actions'', \arel'', \pre'')$ is defined as
\[ \begin{array}{lcl} \Actions'' & \ \ = \ \ & \Actions \times \Actions' \\ ((e,e'),(f,f')) \in \arel''_a & \text{iff} & \arel_aef \text{ and } \arel'_ae'f'  \\ \pre''(e,e') & = & \pre(e) \et [E_e]\pre'(e')
\end{array} \]

\begin{example}
Let us give some examples of action model composition. We recall that a sequence of two announcements $\psi$ and $\phi$ is again an announcement, namely that of $\psi\et[\psi]\phi$. For example, if $\psi = p \et \neg \Box_a p$ and $\phi = \Box_a (p \et q)$ we get  $(p \et \neg \Box_a p)\et[p \et \neg \Box_a p]\Box_a (p \et q)$. This is equivalent to $\dia{p \et \neg \Box_a p}\Box_a (p \et q)$. A sequence of two announcements is typically not equivalent to their conjunction! In this case the conjunction is $(p \et \neg \Box_a p)\et \Box_a (p \et q)$, which is inconsistent, as from $\neg \Box_a p$ and $\Box_a p$ we derive a contradiction. Now, consider that in the presence of another agent $b$, agent $a$ was first privately informed that $p \et \neg \Box_a p$ and was then privately informed that $\Box_a (p \et q)$, while in both cases agent $b$ considered possible that nothing happened (so that we get the action model as in Example~\ref{ex.aml}, only with preconditions $p \et \neg \Box_a p$ and $\Box_a (p \et q)$, respectively, instead of $p$). The composition of these two action models is a four-action action model with identity access for $a$ (who knows what it going on) and universal access for $b$ (who considers it possible that no message, one message, or two messages have arrived), and with preconditions: $\top$, $p \et \neg \Box_a p$, $\Box_a (p \et q)$, and $\dia{p \et \neg \Box_a p}\Box_a (p \et q)$.
\end{example}

\paragraph{Language and semantics of AML}

As already noted, the language $\lang(\Dia,\otimes)$ of {\em action model logic} AML contains an additional inductive construct \[ \dia{E_e}\phi \] standing for `there is an execution of (pointed) action model $E_e$ after which $\phi$ (is true)', and where it is required that the domain of $E$ is finite. This clause $\dia{E_e}\phi$ is properly recursive in the formal language definition: it says that, given the (finite number) $|\domain(E)|$ formulas that are the preconditions of $E$ (although we do not see them) and the formula $\phi$ (that we see), also $\dia{E_e}\phi$ is a formula. One should see the action model frame as the operator and all those formulas as its arguments. The set of all action model frames should then be seen as an additional parameter of the logical language, apart from the agents $A$ and the atoms $P$. As the finite pointed action model frames can be enumerated, this is a countable set. 

The definition of modal depth $d(\phi)$ should be extended to the case $\phi = \dia{E_e}\psi$, namely as the sum of $d(\psi)$ and the maximum depth of the preconditions in $E$. Formally: $d(\dia{E_e}\psi) = d(\psi) + \max \{ d(\pre(f)) \mid f \in \Actions \}$, where $E = (\Actions,\arel,\pre)$.

The semantics for epistemic actions is as follows.
\[ \begin{array}{lcl}
M_s \models \dia{E_e} \psi &\mbox{ \ \ iff \ \ } & M_s \models \pre(e) \text{ and } (M \otimes E)_{(s,e)} \models \psi 
\end{array} \] 

\paragraph*{Axiomatization}
Action model logic has a complete Hilbert-style axiomatization with reduction axioms for the action model modalities. This is just as for public announcement logic, which is now a special case. For completeness' sake, the reduction axioms are as follows, where in the final axiom, $E''$ is the composition of $E$ and $E'$. Please note the similarity with the reductions for PAL, with the exception of the axiom for knowledge after action model execution. The axiom for knowledge after public announcement is the special case  where there is only one accessible action. 
\[\begin{array}{lll}
{[E_e}p &\eq &(\pre(e) \imp p) \\
{[E_e]}\neg\phi &\eq & (\pre(e)\imp\neg[E_e]\phi) \\
{[E_e]}(\phi\et\chi) & \eq & ([E_e]\phi\et[E_e]\chi) \\
{[E_e]}\Box_a\phi & \eq & (\pre(e) \imp \Et_{\arel_a ef} \Box_a [E_f]\phi) \\
{[E_e]}[E'_{e'}]\chi & \eq & [E''_{(e,e')}]\chi
\end{array}\]
Similarly to PAL, by way of these reduction axioms, action model logic AML is shown to be as expressive as multi-agent epistemic logic K, given the model class $\mathcal K$. Similarly, given epistemic models and action models where all relations are equivalence relations, AML is as expressive as multi-agent S5 (see e.g.\ \cite{baltagetal:1998,hvdetal.del:2007}).

It is well-known that PAL and AML are {\em bisimulation invariant}, by which we mean that bisimilar epistemic models also satisfy the same formulas in the languages $\lang(\Dia,!)$ and $\lang(\Dia,\otimes)$: if $M_s \bisim M'_{{s'}}$, then for all formulas $\phi$ in those languages, $M_s \models \phi$ iff $M'_{s'} \models \phi$. This invariance also applies to restricted bisimulation and bounded bisimulation, and the respective linguistic restrictions. All this should be considered elementary variations on the standard results from \cite{blackburnetal:2001}. But not all this is explicit for dynamic epistemic logics. For that, see \cite{hvdetal.papal:2020} for PAL and \cite{aware.aamas:2012} for AML. A consequence is that $M_s \bisim M'_{{s'}}$ implies $(M|\phi)_s \bisim (M'|\phi)_{{s'}}$, where $\phi$ is any formula true in $s$: if two models are bisimilar then also their model restrictions to the same formula. Intuitively it should be clear that this must hold, because whenever $\phi$ is true in $M$ there must be a bisimilar state wherein it is true in $M'$ --- and similarly for precondition formulas of action models.

\paragraph{Arrow update logic} Another general framework for dynamic epistemic logic is arrow update logic \cite{kooietal:2011}, of which a generalized version has also been proposed \cite{kooirenne} that is as expressive as action model logic. We will introduce it in Section \ref{sec.au} dedicated to quantification over arrow updates.

\paragraph{More information} Many standard topics in dynamic epistemic logic are not dealt with in this summary technical preview, such as common knowledge, factual change, and change of belief. Comprehensive introductions to dynamic epistemic logic are
\cite{hvdetal.ajl:2005,hvdetal.del:2007,baltagetal.hpi:2008,hvdetal.puzzle:2015,baltagetal.stanford:2016,moss.handbook:2015}, as well as internet resources \url{https://iep.utm.edu/dynamic-epistemic-logic/} and  \url{https://en.wikipedia.org/wiki/Dynamic_epistemic_logic}.

\section{Arbitrary announcement} \label{sec.apal}

\subsection{Language and semantics} 

{\em Arbitrary public announcement logic} APAL \cite{balbianietal:2008} contains a quantifier over announcements. Its language $\lang(\Dia,!,\F)$ is obtained from $\lang(\Dia,!)$ by adding an inductive clause $\F\phi$, where $\dia{!}$ is a (APAL) {\em quantifier}. We call a formula in $\lang(\Dia,!)$ a {\em quantifier-free formula}. Given model $M_s$ and $\dia{!}\phi\in\lang(\Dia,!,\F)$, we then define \[ M_s \models \F \phi \mbox{ \ \ iff \ \ there is a quantifier-free $\psi$ such that } M_s \models \dia{\psi} \phi \]
The formula $\psi$ is called a {\em witness} for quantifier $\dia{!}$. As PAL and K are equally expressive, we may even assume that witness $\psi$ is in $\lang(\Dia)$.

For $\dia{!}\phi$ we read `there is an announcement after which $\phi$ (is true)', and for $[!]\phi$ we read `after any/arbitrary announcement, $\phi$ (is true)'. Implicit in both formulations is that announcements can only be made on condition that they are true. As we see from the above semantics, these paraphrases are slightly imprecise, because the announcements need to be quantifier-free formulas. Without yet other restrictions, the definition would otherwise be circular, as to determine whether $[!]\phi$, we then would have to check the truth of $[[!]\phi]\phi$. 

We have to expand the definition of modal depth with a clause for the quantifier: $d(\dia{!}\phi) := d(\phi)$.

\begin{example}
Given an agent $a$ who is uncertain about the truth of $p$, either a truly informative announcement $p$ or a trivial announcement $\T$ can be made. We have that $M_1 \models \F \Box_a p$ because $M_1 \models \dia{p} \Box_a p$. On the other hand, $M_1 \models \F \neg\Box_a p$ because $M_1 \models \dia{\T} \neg\Box_a p$. Of course we do not have $M_1 \models \F (\Box_a p \et \neg \Box_a p)$.
\begin{center}
\begin{tikzpicture}
\node (0) at (0,0) {$0$};
\node (1) at (2,0) {$\underline{1}$};
\node (2) at (4,0) {$0$};
\node (3) at (6,0) {$\underline{1}$};
\node (4) at (8,0) {$\underline{1}$};
\node (2b) at (4,-.5) {$M$};
\draw[-] (0) -- node[above] {$a$} (1);
\draw[-] (2) -- node[above] {$a$} (3);
\node (12) at (3,0) {$\stackrel {\top} \Pmi$};
\node (34) at (7,0) {$\stackrel {p} \Imp$};
\end{tikzpicture}
\end{center}
\end{example}

\subsection{Validities} \label{sec.apal.know}

An illustrative validity is: $\F (\Box_a p \vel \Box_a\neg p)$. This formalizes that the agent $a$ can always learn the value of an atomic proposition. It is easy to see why this is valid: either $p$ is true, in which case the agent knows it after its announcement (or, in case it already knew that $p$, {\em still} knows it after its announcement), or it is false, in which case the agent knows that it is false after the announcement that $p$ is false. The argument is given in terms of the intuitions for knowledge, on $\mathcal S5$ models, but in fact  applies to any $\mathcal K$ model. The generalization of this validity from atoms $p$ to any formula $\phi$ is also valid, this is discussed in the paragraph on knowability, below.

Some schematic APAL validities of interest are as follows. We note that Church-Rosser only holds on class $\mathcal S5$ (see also Example~\ref{crnok}), and that McKinsey holds on class $\mathcal S5$, where it is unclear if it also holds on class $\mathcal K$. Proofs of these validities can be found in \cite{balbianietal:2008}, except for CR, of which the proof found there is incorrect, and for which a correct proof can be found in \cite{hvdetal.papal:2020}.
\begin{itemize}
\item $\F\F \phi \imp \F \phi$ \quad ({\bf 4}) \\
This expresses that a sequence of two announcements $\psi$ and $\chi$ is again an announcement, namely $\dia{\psi}\chi$ (or $\psi\et [\psi] \chi$).
\item $\G \phi \imp \phi$ \quad ({\bf T}) \\ If $\phi$ is true after any announcement, it is true after the trivial announcement, and therefore it was already true.
\item $\F\G \phi \imp \G\F\phi$  \quad (Church-Rosser) / CR \\
Given $M_s$ and announcements $\psi$ and $\chi$, there are, respectively, two consecutive announcements $\psi'$ and $\chi'$ such that the same (or a bisimilar) pointed epistemic model results: $(M|\psi|\psi')_s \bisim (M|\chi|\chi')_s$.  
\item $\G\F \phi \imp \F\G\phi$  \quad (McKinsey) / MK \\
MK is not first-order definable, but {\bf 4} and MK together are first-order definable by \emph{atomicity}: $\all x \is y (Rxy \, \& \, (\all z Ryz \Imp y=z))$. This is a famous result in correspondence theory \cite{jfak:1985}, see also \cite{blackburnetal:2001}. However, in our setting this $R$ is the update relation between pointed epistemic models, where a pair consisting of $M_s$ and $(M|\psi)_s$ is in the update relation in case $M_s \models\psi$. Atomicity then says that there is a most informative announcement given the uncertainty specified in the above $\phi$ occurring in MK. In fact, the most informative announcement is that of the value of all atoms occurring in $\phi$ \cite{hvdetal.theoria:2012}.
\end{itemize}

The operator $\F$ seems therefore to behave like the modality of the logic S4. However, this is not the case. We recall that PAL and APAL are not normal modal logics, as they do not satisfy uniform substitution. For another example, $p\imp\F\Box_ap$ is valid but $(p\et\neg\Box_ap)\imp\F\Box_a(p\et\neg\Box_ap)$ is invalid (see Example \ref{ex.first}).

\begin{example} \label{crnok}
This counterexample to Church-Rosser on $\mathcal K$ is by Louwe Kuijer. Given a single agent,  consider $M=(S,R,V)$ given by $S=\{s,t,u\}$, $R=\{(s,t),(t,t),(t,u)\}$ and $V(p)=\{s,t\}$. We now have that $M_s \models \dia{!}[!]\Dia\top$, with the witness being the announcement $p$ which makes $s$ and $t$ bisimilar. Yet we also have $M_s \not\models [!]\dia{!}\Dia\top$, since after the announcement $\Box p$ we are left with only the state $s$. As it has no successors, no further announcement can make $\Dia\top$ true. A depiction is:
\begin{center}
\begin{tikzpicture}
\node (0) at (0,0) {$1(s)$};
\node (1) at (2,0) {$1(t)$};
\node (2) at (4,0) {$0(u)$};
\draw[->] (0) --  (1);
\draw[->] (1) --  (2);
\draw[->] (1) edge[loop above,looseness=7] (1); 
\node (r) at (5.5,0) {$\stackrel {p} \Imp$};
\node (0r) at (7,0) {$1(s)$};
\node (1r) at (9,0) {$1(t)$};
\draw[->] (0r) --  (1r);
\draw[->] (1r) edge[loop above,looseness=7] (1r); 
\node (l) at (-1.5,0) {$\stackrel {\Box p} \Pmi$};
\node (0) at (-3,0) {$1(s)$};
\node (rr) at (10.5,0) {$\bisim$};
\node (0rr) at (12,0) {$1(s)$};
\draw[->] (0rr) edge[loop above,looseness=7] (0rr); 
\end{tikzpicture}


\end{center}
\end{example}

\paragraph*{Knowability}
The schema $\F\Box_a\phi$, for `there is an announcement after which the agent knows $\phi$', can be seen as an interpretation of `$\phi$ is {\em knowable}', a suggestion made by Johan van Benthem in \cite{jfak.fitch:2004}. Thus, APAL provides an actual setting to investigate Fitch's knowability paradox \cite{fitch:1963}. Fitch addressed the question whether what is true can become known and in \cite{fitch:1963} he demonstrated that the existence of unknown truths (there is a $\phi$ for which $\phi\et\neg\Box_a\phi$) is inconsistent with the requirement that all truths are knowable (for any $\psi$, $\psi\imp\F\Box_a\psi$). The inconsistency appears by substituting $p\et\neg\Box_ap$ for $\psi$, as follows. Let $M_s$ be a ($\mathcal S5$) model satisfying $p\et\neg\Box_ap$ and towards a contradiction assume that $M_s$ also satisfies $(p\et\neg\Box_ap)\imp\F\Box_a(p\et\neg\Box_ap)$. Then  $M_s\models\F\Box_a(p\et\neg\Box_ap)$. Then there is a formula $\psi$ witnessing this $\F$ quantifier such that $M_s \models \psi$ and $(M|\psi)_s \models \Box_a(p\et\neg\Box_ap)$. Now observe that  $\Box_a(p\et\neg\Box_ap)$ is inconsistent in S5, as we can derive $\Box_a p$ and $\neg \Box_a p$ from it, and therefore $\bot$. It can therefore not be true in $(M|\psi)_s$.

An interesting APAL validity for the Fitch setting is $\F (\Box_a \phi \vel \Box_a \neg\phi)$ \cite{hvdetal.theoria:2012}. This states that in APAL `everything is knowable', meaning that for every formula we can get to know whether it is true or false.

That this is valid, is shown by announcing the value of all variables occurring in $\phi$, and then showing that in any model $M$ with a constant valuation for those variables, for any $\psi$ only containing those variables, $M \models \psi \imp \Box_a \psi$. Therefore, in particular, $M \models \phi \imp \Box_a \phi$ and $M \models \neg\phi \imp \Box_a \neg\phi$, from which follows $M \models \Box_a \phi \vel \Box_a \neg \phi$.

The result is for $\mathcal S5$ models. It is unclear whether the results extends to model class $\mathcal K$.

Such knowability comes at a price: the Moore sentence $p\et\neg\Box_ap$ is knowable, because after being truthfully announced, agent $a$ knows that it is false, as we saw in Example \ref{ex.first}. This seems contrary to the `spirit' of knowability: the idea is, that you can get to know that something \emph{was} true, not that it is true now.

\subsection{Expressivity} \label{sec.apal.exp}

On the class of $\mathcal S5$ models for a single agent, APAL is as expressive as PAL,  whereas for more than one agent, APAL is more expressive than PAL \cite{balbianietal:2008}. On class $\mathcal K$, single-agent APAL is also more expressive than PAL. However, as so often, the proof of that is similar to the proof of the multi-agent case on class $\mathcal S5$. 

There are two different proofs of larger expressivity. We present them in some detail, as both kinds of proof have been used to obtain expressivity results for different versions of APAL.

The APAL quantifier $\dia{!}$ implicitly quantifies over {\em countably many atoms} and also implicitly quantifies over {\em arbitrarily large modal depth} (we recall that the modal depth is the largest number of stacked $\Dia$-operators --- or $\Box$-operators, which does not matter for the count). These properties can be independently used to demonstrate larger expressivity of APAL. Prior to the details, let us first sketch how.

On the assumption that a particular formula $\phi \in \lang(\Dia,!,\dia{!})$ is equivalent to a formula $\psi \in \lang(\Dia,!)$, we use that $\psi$ must contain a finite number of atoms, or that $\psi$ is of finite modal depth. Given that, we can choose two models that have a different value for $\phi$ but that must have the same value for $\psi$, which is a contradiction, so that therefore such a $\psi$ cannot exist. Therefore, PAL is not at least as expressive as APAL. As $\lang(\Dia,!) \subseteq 
\lang(\Dia,!,\dia{!})$, we also trivially have that APAL is at least as expressive as PAL. Therefore, APAL is (strictly) more expressive than PAL. Such expressivity results are often shown with Ehrenfeucht-Fraiss\'e games \cite{ehrenfeucht:1961}. The method above, however, seems suitable for quantified dynamic epistemic logics because it is succinct. And let us now give the details.

\paragraph{Implicit quantification over all atoms}
First, let us use the argument that $\dia{!}$ quantifies over all atoms, found in \cite{balbianietal:2008}. For this argument to work, we need that the set of atoms is countably infinite. 
Consider the formula $\phi = \apaldia (\Box_a p \et \neg \Box_b \Box_a p)$. Let us assume that $\phi \in \lang(\Dia,!,\dia{!})$ is equivalent to $\psi \in \lang(\Dia,!)$. As the formula $\psi$ only contains a finite number of atoms, and as the set of atoms is infinite, there is an atom $q \in P$ such that $q \not\in \var(\psi)$. Consider the following two models for variables $p$ and $q$ (where the value of other variables does not matter, and where in $M$, the value of $q$ does not matter); $10$ means that $p$ is true and $q$ is false, etcetera.

\begin{center}
\begin{tikzpicture}
\node (m) at (-1.5,0) {$M_1$};
\node (0) at (0,0) {$0$};
\node (1) at (2,0) {$\underline{1}$};
\draw[-] (0) -- node[above] {$a$} (1);
\end{tikzpicture}
\quad \hspace{1cm}
\begin{tikzpicture}
\node (m) at (-1.5,0) {$M'_{10}$};
\node (00) at (0,0) {$00$};
\node (10) at (2,0) {$\underline{10}$};
\node (01) at (0,2) {$01$};
\node (11) at (2,2) {$11$};
\draw[-] (00) -- node[above] {$a$} (10);
\draw[-] (01) -- node[above] {$a$} (11);
\draw[-] (00) -- node[right] {$b$} (01);
\draw[-] (10) -- node[right] {$b$} (11);
\node (m) at (3.5,1) {$\stackrel {p \vel q} \Imp $};
\node (l10) at (6.5,0) {$\underline{10}$};
\node (l01) at (4.5,2) {$01$};
\node (l11) at (6.5,2) {$11$};
\draw[-] (l01) -- node[above] {$a$} (l11);
\draw[-] (l10) -- node[right] {$b$} (l11);
\end{tikzpicture}
\end{center}

We now have that 
\[ M_1\not\models\apaldia (\Box_a p \et \neg \Box_b \Box_a p) \]
because there are only two possible model restrictions, the trivial one after which $\Box_a p$ remains false, and the restriction to the $p$-state, after which $\Box_a p$ is true but then also $\Box_b \Box_a p$.

We also have that
\[ M'_{10}\models\apaldia (\Box_a p \et \neg \Box_b \Box_a p) \]
because
\[ M'_{10}\models\dia{p\vel q} (\Box_a p \et \neg \Box_b \Box_a p) \]
as $M'|(p \vel q), 10 \models \Box_a p \et \neg \Box_b \Box_a p$.

On the other hand, $M_1 \bisim^{P\setminus\{q\}} M'_{10}$ and therefore by bisimulation invariance for PAL: \[ M_1 \models \psi \quad \text{iff} \quad M'_{10} \models \psi \]

\paragraph{Implicit quantification over any finite modal depth}
The other proof of the larger expressivity of APAL (originally by Barteld Kooi, but by now also found in, for example, \cite{hvdetal.papal:2020}) uses that $\dia{!}$ quantifies over {\em formulas of  arbitrarily large epistemic depth}. This proof does not need the requirement that there are infinitely many atoms in the logical language.

Again, consider $\phi = \apaldia (\Box_a p \et \neg \Box_b \Box_a p)$, and again, let us assume that $\phi \in \lang(\Dia,!,\dia{!})$ is equivalent to $\psi  \in \lang(\Dia,!)$.  Consider the models depicted below. We abuse notation by having multiple states with the same name, namely of the value of the unique atom $p$. However, we will only evaluate formulas in the (underlined) point of any of such models, thus avoiding ambiguity.

\bigskip

\begin{tikzpicture}
\node (0) at (0,0) {$0$};
\node (b) at (0,-.7) {$\bisim$};
\node (1) at (1.5,0) {$\underline{1}$};
\node (m) at (-1.5,0) {$M_1:$};
\draw[-] (0) -- node[above] {$a$} (1);
\end{tikzpicture} 

\begin{tikzpicture}
\node (0) at (0,0) {$0$};
\node (b) at (0,-.7) {$\bisim$};
\node (1) at (1.5,0) {$\underline{1}$};
\node (2) at (3,0) {$1$};
\node (3) at (4.5,0) {$0$};
\node (m) at (-1.5,0) {\color{white}$M_1:$};
\draw[-] (0) -- node[above] {$a$} (1);
\draw[-] (1) -- node[above] {$b$} (2);
\draw[-] (2) -- node[above] {$a$} (3);
\end{tikzpicture}

\begin{tikzpicture}
\node (0) at (0,0) {$0$};
\node (b) at (0,-.7) {$\bisim$};
\node (1) at (1.5,0) {$\underline{1}$};
\node (2) at (3,0) {$1$};
\node (3) at (4.5,0) {$0$};
\node (4) at (6,0) {$0$};
\node (5) at (7.5,0) {$1$};
\node (5b) at (9,0) {$1$};
\node (6) at (10.5,0) {$0$};
\node (m) at (-1.5,0) {\color{white}$M_1:$};
\draw[-] (0) -- node[above] {$a$} (1);
\draw[-] (1) -- node[above] {$b$} (2);
\draw[-] (2) -- node[above] {$a$} (3);
\draw[-] (3) -- node[above] {$b$} (4);
\draw[-] (4) -- node[above] {$a$} (5);
\draw[-] (5) -- node[above] {$b$} (5b);
\draw[-] (5b) -- node[above] {$a$} (6);
\end{tikzpicture} 

\begin{tikzpicture}
\node (0) at (0,0) {$0$};
\node (1) at (1.5,0) {$\underline{1}$};
\node (2) at (3,0) {$1$};
\node (3) at (4.5,0) {$0$};
\node (4) at (9,0) {$1$};
\node (5) at (10.5,0) {$1$};
\node (5b) at (12,0) {$0$};
\node (m) at (-1.5,0) {$M''_1:$};
\draw[-] (0) -- node[above] {$a$} (1);
\draw[-] (1) -- node[above] {$b$} (2);
\draw[-] (2) -- node[above] {$a$} (3);
\draw[dotted,-] (3) -- node[above] {$> d(\psi)$} (4);
\draw[-] (4) -- node[above] {$b$} (5);
\draw[-] (5) -- node[above] {$a$} (5b);
\end{tikzpicture} 

\begin{tikzpicture}
\node (0) at (0,0) {$0$};
\node (b) at (0,-.7) {$\stackrel{\eta}{\Imp}$};
\node (1) at (1.5,0) {$\underline{1}$};
\node (2) at (3,0) {$1$};
\node (3) at (4.5,0) {$0$};
\node (4) at (9,0) {$1$};
\node (5) at (10.5,0) {$1$};
\node (5b) at (12,0) {$1$};
\node (m) at (-1.5,0) {$M'_1:$};
\draw[-] (0) -- node[above] {$a$} (1);
\draw[-] (1) -- node[above] {$b$} (2);
\draw[-] (2) -- node[above] {$a$} (3);
\draw[dotted,-] (3) -- node[above] {$> d(\psi)$} (4);
\draw[-] (4) -- node[above] {$b$} (5);
\draw[-] (5) -- node[above] {$a$} (5b);
\end{tikzpicture} 

\begin{tikzpicture}
\node (0) at (0,0) {\color{white}$0$};
\node (1) at (1.5,0) {$\underline{1}$};
\node (2) at (3,0) {$1$};
\node (3) at (4.5,0) {$0$};
\node (m) at (-1.2,0) {$(M'|\eta)_1:$};
\draw[-] (1) -- node[above] {$b$} (2);
\draw[-] (2) -- node[above] {$a$} (3);
\end{tikzpicture}

\bigskip

\noindent First observe that $M_1$ is bisimilar to the model depicted below it, which is obtained from taking a mirror-image copy of $M$ and pasting that to $M$ with a $b$-link. Similarly, we can take two copies of that, such that we get the model on the next line, etc., and we get some $M''$ where the rightmost state is more than $d(\psi)$ steps away from the underlined (actual) $p$-state. Let now $M'$ be as this $M''$ except that in the right-most state $p$ is true. This is the only state in $M'$ where $b$ knows that $a$ knows that $p$. Formula $\Box_b \Box_a p$ is a distinguishing formula for that state. Using that, we can find a distinguishing formula for any finite subset of the domain of $M'$ and thus there is an (unspecified but rather complex) announcement $\eta$ after which the three states as depicted remain, such that $\dia{\eta}(\Box_a p \et \neg \Box_b \Box_a p)$ is true in $M'_1$, and thus \[ M'_1 \models \dia{!}(\Box_a p \et \neg \Box_b \Box_a p) \]
Also, as before, \[ M_1 \models \neg\dia{!}(\Box_a p \et \neg \Box_b \Box_a p) \]
On the other hand, $M'_1 \bisim^n M''_1$, i.e., if you cut off the part of the model further away from the actual state $1$ than $n$ steps, the models are bisimilar. Models that are $n$-bisimilar satisfy the same formulas in $\lang(\Dia,!)$ of modal depth at most $n$. Therefore, as $d(\psi) \leq n$, \[ M'_1 \models \psi \quad \text{iff} \quad M''_1 \models \psi \] and thus, as $M''_1 \bisim M_1$, \[ M'_1 \models \psi \quad \text{iff} \quad M_1 \models \psi \]
Again, we obtained a contradiction.

\subsection{Axiomatization and theory} \label{axiomtheory}

The logic APAL has a {\em complete infinitary axiomatization} for the class ${\mathcal S5}$, for which the preferred reference is \cite{balbianietal:2015}.\footnote{This axiomatization is found in \cite{balbianietal:2008} as well, but that publication also contains an incorrect finitary axiomatization. The corrections are discussed in \cite{philippe.corrected:2015,balbianietal:2015}.} The axiomatization is \emph{infinitary} because it contains a derivation rule with infinitely many premisses in order to draw a conclusion. Axiomatizations like that of PAL and AML that contain finitely many axioms (axiom schemes) and finitely many rules with finitely many premisses are \emph{finitary}.

The axiomatization of APAL extends the axiomatization of PAL with one axiom and one derivation rule involving the quantifier. The derivation rule is formulated using the so-called {\em necessity forms} or {\em admissible forms} \cite{goldblatt:1982} $\lang_\sharp$ defined as $\pmb{\phi}(\sharp) := \sharp \mid \phi \imp \pmb{\phi}(\sharp) \mid \Box_a \pmb{\phi}(\sharp) \mid [\phi]\pmb{\phi}(\sharp)$, where $\phi \in \lang(\Dia,!,\dia{!})$, and where we used bold-face font to distinguish $\pmb{\phi}(\sharp) \in \lang_\sharp$ from $\phi \in \lang(\Dia,!,\dia{!})$. Given some $\psi \in  \lang(\Dia,!,\dia{!})$, to obtain $\pmb{\phi}(\psi)$ from $\pmb{\phi}(\sharp)$ we replace the unique symbol $\sharp$ in $\pmb{\phi}(\sharp)$ by $\psi$ and we thus get a formula in $\lang(\Dia,!,\dia{!})$. 

The axiom and rule are:
\[\begin{array}{lll}
{[!]}\phi \imp [\psi]\phi \hspace{3cm} \hfill \text{where } \psi \in \lang(\Dia,!) \\
\left( \ \pmb{\phi}([\psi]\chi) \text{ for all } \psi \in \lang(\Dia,!) \ \right) \text{ implies } \pmb{\phi}([!]\chi)  \hspace{3cm} \hfill \text{where } \pmb{\phi}(\sharp) \in \lang_\sharp
\end{array}\]
It is unknown whether there is a finitary axiomatization as well for APAL. Further, we conjecture that the axiomatization without the $\mathcal S5$ axioms {\bf T}, {\bf 4}, and {\bf 5} is complete for the class ${\mathcal K}$, by an analogous completeness proof. `Conjecture' seems a big word, it is merely that such a proof is not on record.

The logic APAL is not compact \cite{balbianietal:2008}, the satisfiability problem is undecidable \cite{frenchetal:2008} (the usual tiling argument applies), and the model checking problem is PSPACE-complete \cite{agotnesetal.jal:2010}, where this is shown in \cite{agotnesetal.jal:2010} for an APAL variant called GAL that we will present later, but the method also applies to APAL. Also, APAL does not have the {\em finite model property} \cite{hvdetal.nofinite:2021}. A logic has the finite model property (FMP) if any satisfiable formula has a model with a finite domain.

Like APAL, the logic APAL is bisimulation invariant, so adding the quantifier does not change that property. A proof is found in \cite{hvdetal.papal:2020}, and also for a related logic. Note that it is not found in the reference publication  \cite{balbianietal:2008}. So, if two models are bisimilar, they satisfy the same formulas in $\lang(\Dia,!,\dia{!})$. However, unlike PAL, the logic APAL does not preserve restricted bisimulation or bounded bisimulation \cite{hvdetal.papal:2020}. Having just seen the expressivity proofs, it is easy to see why. Because the APAL quantifier implicitly quantifies over all atoms, it does not preserve restricted bisimilarity. For example, in the first expressivity proof, the models are $p$-bisimilar, but the distinguishing formula $\dia{!}(\Box_a p \et \neg \Box_b\Box_a p)$ only contains atom $p$: it is in $\lang(\Dia,!,\dia{!})|p$. Because the APAL quantifier implicitly quantifies over arbitrarily large modal depth, it does not preserve bounded bisimilarity. For example, in the second expressivity proof, the models are $2$-bisimilar, but the (same) distinguishing formula $\dia{!}(\Box_a p \et \neg \Box_b\Box_a p)$ has modal depth $2$. 

\subsection{Variations} \label{sec.variations}

A fair number of variations of APAL have seen the light. Prior to a more detailed presentation, let us see some of this variation.
\[\begin{array}{lll}
M_s \models \F \phi & \text{iff} & \text{there is a quantifier-free}\ \psi \ \text{such that}\ M_s \models \dia{\psi} \phi \hfill \text{ (APAL) \cite{balbianietal:2008}} \\
M_s \models \F \phi & \text{iff} & \text{there is a}\ \psi \ \text{such that}\ M_s \models \dia{\psi} \phi \hfill \text{(full APAL) \cite{hvdetal.aiml:2016}}  \\
M_s \models \F \phi & \text{iff} & \text{there is a positive}\ \psi \ \text{such that}\ M_s \models \dia{\psi} \phi \hspace{2cm} \hfill \text{(APAL$^+$) \cite{hvdetal.papal:2020}} \\ M_s \models \F \phi & \text{iff} & \text{there is a Boolean}\ \psi \ \text{such that}\ M_s \models \dia{\psi} \phi \hfill \text{(BAPAL) \cite{hvdetal.bapal:2022}}
\end{array}\]
But there are many more, that are also reported below. What is the interest? The logic APAL is undecidable, has a high complexity of model checking, and has no known finitary axiomatization. Taming this into decidability, lower complexities, or a finitary axiomatization (or other proof methods such as sequent or tableaux calculi) is a computational advantage for applications such as epistemic planning and security protocols. Although a nice aim, the reality has so far turned out to be fairly tough, as we will see. However, this is, we think, very informative for the reader. Of independent interest is that some variations may provide a better match to address philosophical logical themes such as knowability or AI themes involving agency. Let us proceed.

\subsubsection{Fully arbitrary announcements}

In this variation we really quantify over all announcements, without restriction. It therefore seems to have a large intuitive appeal.
\[ M_s \models \F \phi \text{ \quad iff \quad there is a } \psi \text{ such that } M_s \models \dia{\psi} \phi \]
where $\phi,\psi, \dia{!}\phi, \dia{\psi}\phi$ are all in the same language, so that in particular $\psi$ is really an arbitrary formula in that language. As already mentioned, if this were a definition, it would be circular, as one of those $\psi$ may be the formula $\apaldia\phi$. However, in \cite{hvdetal.aiml:2016} a logic is proposed with quantification over announcements where the above is a \emph{property} of the logic, instead of the semantics of the quantifier. The language $\lang_{Ord}(\Dia,!,\dia{!})$, where $Ord$ are the ordinals, has additionally to the $\dia{!}$ quantifier also quantifiers $\dia{!_\alpha}$ for any ordinal $\alpha$. Let now $\lang_\alpha$ be the sublanguage only containing quantifiers $\dia{!_\beta}$ for $\beta < \alpha$, then we can define:
 \[ \begin{array}{lll} M_s \models \dia{!_\alpha} \phi & \text{iff} & \text{there is } \psi \in \lang_\alpha \text{ such that } M_s \models \dia{\psi} \phi \\
M_s \models \dia{!} \phi & \text{iff} & \text{there is } \alpha \in Ord \text{ such that } M_s \models \dia{!_\alpha} \phi
\end{array}\] 
The axiomatization of the logic is unclear, as the collection of $\dia{!_\alpha}$ in the logical language is a proper class. In other words, although it is fun (and maybe even elegant) to have such a logic wherein arbitrary is really arbitrary, it may not be of much practical value and does not help us with pushing down complexities. An open question is whether $\dia{!}\phi$ is equivalent to $\dia{!_\omega}\phi$, that is, whether the hierarchy stabilizes at $\lang_\alpha = \lang_\omega$.\footnote{Originally suggested by Alexandru Baltag, and also by a reviewer.} If so, we can then simply define that: $M_s \models \dia{!} \phi$ iff there is a $n \in \Naturals$ such that $M_s \models \dia{!_n} \phi$, and it might bring the axiomatization closer.

\subsubsection{Positive announcements}
In this variation we only allow quantification over positive formulas. It is conjectured to bring decidability of satisfiability.
\[ M_s \models \F \phi \quad \text{iff} \quad \text{there is a positive}\ \psi \ \text{such that}\ M_s \models \dia{\psi} \phi \]
In a positive formula, an epistemic modality is never bound by a negation. The \emph{positive formulas} are defined as:
\[ \phi ::= p | \neg p| \phi \vel \phi | \phi \et \phi | \Box_a \phi \]
The positive formulas correspond to the universal fragment in the first-order translation (essentially, where negations do not bind universal quantifiers \cite{VanBenthem1984}).\footnote{The above positive formulas are to be distinguished from their extension \[ \phi ::= p | \neg p| \phi \vel \phi | \phi \et \phi | \Box_a \phi | [\neg\phi]\phi | \apalbox \phi \] that is also known as the positive formulas, and that figure in, for example, \cite{hvdetal.synthese:2006,balbianietal:2008}, where the $[!]$ in the inductive case $[!]\phi$ is the APAL quantifier. The negation in $[\neg\phi]\psi$ may appear strange at first sight but this is a consequence of the semantics of public announcement: note that $M_s \models [\neg\phi]\psi$ \ iff \ ($M_s \models \neg \phi$ implies $(M|\neg\phi)_s \models \psi$) \ iff \ ($M_s \models \phi$ or $(M|\neg\phi)_s \models \psi$). An interesting question seems whether, if the $[!]$ in the inductive case $[!]\phi$ were the APAL$^+$ quantifier instead of the APAL quantifier, every extended positive formula would be equivalent to a positive formula.}
  They have the property that they are \emph{preserved} after an informative update, such as an announcement. They are therefore also \emph{successful}: they remain true after their announcement. The logic with this quantifier is called APAL$^+$, for {\em positive} arbitrary public announcement logic. All results reported here are for class $\mathcal S5$. Note that the language of APAL$^+$ contains  the usual public announcement, so that any formula may be announced, also formulas that are not positive. The restriction is only for witnesses of the quantifier $\dia{!}$.

From a more semantic perspective we can observe that positive formulas are also preserved after a {\em refinement} of a given model (see Section \ref{sec.prelims}). Given a model $M=(S,R,V)$ and $T \subseteq S$, $T$ is called {\em closed under refinements} iff for all $s,t \in S$ such that $M_s \lumis M_t$, if $s \in T$ then $t \in T$. The denotation of a positive announcement is closed under refinements, and dually, on finite models a subset of the domain that is closed under refinements is the result of a positive announcement. This is a powerful tool to investigate APAL$^+$. Refinements also have a much more general use in dynamic epistemic logics. Section \ref{sec.rml} is dedicated to refinements.

The logic APAL$^+$ is incomparable in expressivity to APAL. This result is not obvious, as we are comparing two logics with different quantifiers. Both directions of the incomparability result have non-trivial proofs that cannot be easily sketched in this survey, and that use closure under refinements as a proof tool. Just as for APAL, the complexity of model checking in APAL$^+$ is PSPACE-complete. The (known) axiomatization of APAL$^+$ is much like that of APAL: in the axiom and in the derivation rule we merely replace the witness for a positive witness. For details on expressivity, model checking and axiomatization we refer to \cite{hvdetal.papal:2020}.

We conjecture that the logic APAL$^+$ is decidable.\footnote{Work in progress by Tim French et al.\ employ a technique combining the S5 modalities for knowledge with the KD4-like modalities for public announcements --- we can alternatively see the model transformation induced by a public announcement in dynamic epistemic logic as a pair in the accessibility relation associated to that announcement, in a larger model that is constructed from the direct sum of all model restrictions of a given model, as in \cite{WC13}.} Decidability of APAL$^+$ is interesting for potential applications. In many scenarios only positive information change plays a role, given some initial information state containing ignorance, such as security protocols with mutual acknowledgements by principals of messages with factual content, or scenarios with sensing in order to reduce uncertainty and provide diagnostics.

\subsubsection{Boolean announcements}

In this variation we only allow quantification over Booleans. We now get a finitary axiomatization and decidability.
\[ M_s \models \F \phi \quad \text{iff} \quad \text{there is a Boolean}\ \psi \ \text{such that}\ M_s \models \dia{\psi} \phi \]
A {\em Boolean} or a formula in {\em propositional logic} is a formula in $\lang(\emptyset)$. Boolean announcements result in model restrictions, and, as public announcements only change the information about the true value of propositional variables but not their actual value, such announced Booleans remain true after (truthful!) announcement: Boolean announcements are \emph{successful}. Also, Booleans are positive formulas. The resulting logic BAPAL is decidable \cite{bapaldec}, and has a finitary axiomatization \cite{hvdetal.bapal:2022}. Still, it does not have the finite model property \cite{bapaldec}. The results reported here are for class $\mathcal S5$. The relevant axiom and derivation rule are:
        \[ \begin{array}{lll}
{[!]}\phi \imp [\psi_0] \phi \hspace{3cm} \hfill \text{ where } \psi_0 \in \lang(\emptyset) & \\
\psi \imp [\chi][p]\phi \text{ implies } \psi \imp [\chi][!]\phi \hfill \hspace{1cm} \text{ where } p \not\in\var(\psi),\var(\chi),\var(\phi)
\end{array}\]
The atom $p$ not occurring in any of the formulas occurring in the derivation rule is called \emph{fresh}. It has to be fresh in order to avoid that its value depends on occurrences of atoms in these formulas (which would permit invalid rule applications). Note that $\psi \imp [\chi][p]\phi$ and $\psi \imp [\chi][!]\phi$ are instantiations of the necessity form $\psi \imp [\chi]\sharp$. The APAL derivation rule ``$\pmb{\psi}([\eta]\phi)$ for all quantifier-free $\eta$, implies $\pmb{\psi}([!]\phi)$'', where  $\pmb{\psi}(\sharp)$ may be any necessity form, has an infinity of premisses, namely one for every formula $\eta$, whereas the BAPAL derivation rule has one premiss only, namely for a fresh variable $p$. This makes the axiomatization finitary.

The logic BAPAL is more expressive than PAL (which is shown as for APAL), and BAPAL is not at least as expressive as APAL \cite{hvdetal.bapal:2022} (on class $\mathcal S5$). However, it is unknown if APAL is not at least as expressive as BAPAL. To prove that, one would somehow have to show that the Boolean quantifier can be `simulated' by an an APAL quantifier that is properly entrenched in preconditions and postconditions relative to $\phi$. This seems quite hard. The proof that BAPAL is not at least as expressive as APAL uses, not surprisingly, that the BAPAL quantifier does not quantify over formulas of arbitrarily large modal depth, as the modal depth of Booleans is zero. The proof is instructive. We therefore sketch it. 

\medskip

The proof that BAPAL is not at least as expressive as APAL is very similar to the proof that PAL is less expressive than APAL in Section~\ref{sec.apal.exp}. We recall that the proof that PAL is less expressive than APAL cansisted of two parts. One part was that (trivially) APAL is at least as expressive as PAL, because the language of PAL is a fragment of the language of APAL. This part no longer holds for BAPAL. The other part consisted in showing that PAL is not at least as expressive as APAL. We can reuse that part for BAPAL. In the version of the proof for BAPAL we again use bounded bisimilarity, but slightly differently. To emphasize the similarity in method we will use the same epistemic models as in Section~\ref{sec.apal.exp} and not the models used in the proof in \cite{hvdetal.bapal:2022}, that are a bit different.

Consider again the APAL formula $\dia{!} (\Box_a p \et \neg \Box_b \Box_a p)$. Let us suppose that there exists an equivalent formula $\psi$ in the same logical language but interpreted according to the BAPAL semantics (the languages of APAL and BAPAL are the same in this survey, so we need to distinguish APAL and BAPAL by their semantics). Note that this is a difference with the previous proof, where $\psi$ was a quantifier-free formula, in the language of PAL. Now recall the models $M_1$, $M'_1$ and $M''_1$ of the previous proof, and that $M_1 \bisim M''_1$ and $M'_1 \bisim^{d(\psi)} M''_1$. 

Unlike APAL, the logic BAPAL has the property that bounded bisimulation implies bounded modal equivalence, i.e., the same truth value for formulas of modal depth up to that bound \cite{hvdetal.bapal:2022}. Therefore, $M'_1 \bisim^{d(\psi)} M''_1$, so that models $M'_1$ and $M''_1$  both make $\psi$ true or both make $\psi$ false. And as $M_1 \bisim M''_1$, also $M_1$ and $M'_1$ have the same value for $\psi$. On the other hand, as before, $M_1 \not\models\dia{!} (\Box_a p \et \neg \Box_b \Box_a p)$ whereas $M'_1 \models \dia{!} (\Box_a p \et \neg \Box_b \Box_a p)$. Contradiction, and end of proof.

Informally, it is  also easy to see that Boolean announcements cannot achieve much to distinguish $M'_1$ from $M''_1$. One can either announce $p$ or $\top$ ($\neg p$ is ruled out as this is false in the point $1$). The first announcement trivializes $M'_1$ as well as $M''_1$ to a singleton where $p$ is common knowledge, whereas the second announcement is not informative in either model. 

\subsubsection{APAL with memory} \label{sec.memory}

In this variation we keep the original value of propositions in memory. We can express that after an announcement a formula \emph{was} true. It has a finitary axiomatization.

In {\em APAL with memory} (APALM) \cite{BaltagOS18,baltagetal:2022}, instead of models $M=(S,R,V)$ we have models $M=(S,T,R,V)$, where $T \subseteq S$, and where $(S,S,R,V)$ is an {\em initial model} (and where all relations $R_a$ are equivalence relations --- the results are for class $\mathcal S5$). The domain $S$ is the initial domain, whereas the domain $T$ is the current domain, the current model restriction. This additional structural information makes it possible to refer to what was true in the past, as well as to what is true now. The logical language has additional operators to access this historical information, namely $\phi^0$ (for `$\phi$ was initially true'), a constant $0$ that is true when the model is initial, and the universal modality $U$. The semantics for $\phi^0$ is (where $s \in T$):
\[ (S,T,R,V)_s \models \phi^0 \text{ iff } (S,S,R,V)_s \models \phi \] The semantics for the quantifier are as in APAL. Namely \[ M_s \models \F \phi \quad \text{iff} \quad \text{there is a quantifier-free}\ \psi \ \text{such that}\ M_s \models \dia{\psi} \phi \]
However, the semantics for public announcement are not as in APAL. Different from APAL, and from PAL, the inductive construct $\dia{\phi}\psi$ now has the restriction that $\phi$ must be quantifier-free. This restriction on what can be publicly announced makes it harder to compare APAL to APALM. It is unclear whether in the logic APAL, for all $\phi,\psi\in\lang(\Dia,!,\dia{!})$ there is a $\chi \in\lang(\Dia,!)$ such that $\dia{\phi}\psi$ is equivalent to $\dia{\chi}\psi$. One unanswered question is then whether the set of APAL validities is the same as the set of APALM validities restricted to the language $\lang(\Dia,!,\dia{!})$.

There are also relations between APALM and the logic with modalities for what is true \emph{before} an announcement  by \cite{BalbianiDH16}. Like APALM, that logic is interpreted on structures where apart from the current model we keep the initial model in memory.

Let us now look into the difference between APAL and APALM. The difference has to do with the phenomenon in PAL that non-bisimilar states may become bisimilar after an announcement, as in the following example. The power of APALM is that it can distinguish between such (standardly) bisimilar states because the model still remembers that that the states {\em were} different.

\begin{example} Consider states $t$ and $u$ in the following model $M$, and in its restriction $M|p$.

\begin{center}
\begin{tikzpicture}
\node (0) at (0,0) {$0(s)$};
\node (1) at (2,0) {$\underline{1}(t)$};
\node (2) at (4,0) {$1(u)$};
\draw[-] (0) -- node[above] {$b$} (1);
\draw[-] (1) -- node[above] {$a$} (2);
\node (m) at (5.5,0) {$\stackrel{p}{\Imp}$};
\node (1) at (7,0) {$\underline{1}(t)$};
\node (2) at (9,0) {$1(u)$};
\draw[-] (1) -- node[above] {$a$} (2);
\end{tikzpicture}
\end{center}
In $M$, the states $t$ and $u$ have different knowledge properties, for example, $M_t \models \neg \Box_b p$ whereas $M_u \models \Box_b p$, and they are not bisimilar. However, after the public announcement of $p$ they have become indistinguishable: $(M|p)_t$ and $(M|p)_u$ are bisimilar and therefore satisfy the same formulas. In APALM, we can still distinguish the state $t$ and $u$ in $M|p$, because $(M|p)_t \models (\neg \Box_b p)^0$ whereas $(M|p)_u \models (\Box_b p)^0$.
\end{example}

The logic APALM has a finitary axiomatization. This is possible because of the memory feature of APALM \cite[Section~3]{BaltagOS18}. The derivation rule introducing the quantifier is similar to the one in BAPAL. It is not known whether APALM is decidable. 

Also, APALM may have an interesting take on knowability. We recall from Section~\ref{sec.apal.know} that in APAL everything is knowable in the sense that $\dia{!}\Box_a\phi\vel\dia{!}\Box_a\neg\phi$ is valid \cite{hvdetal.theoria:2012}, but at the price of changing the value of propositions. You can formalize that you can get to know that a proposition is true (now), but you cannot formalize that you can get to know that a proposition \emph{was} true. But in APALM you can, namely as $\phi \imp \dia{!}\Box_a\phi^0$. And $\dia{!}\Box_a\phi^0 \vel \dia{!}\Box_a\neg\phi^0$ then formalizes that for every proposition you can get to know whether it was true or false. Are these APALM validities?

\subsubsection{Almost Arbitrary Announcements}

In this variation we quantify over propositions that are stronger or weaker than a given formula. This permits a notion of dynamic consequence with novel substructural properties.
\[ M_s \models \dia{\chi^\downarrow} \phi \quad \text{iff} \quad \text{there is a quantifier-free $\psi$ 
\emph{implying}} \ \chi \text{ such that } M_s \models \dia{\psi} \phi \]
where $\chi$ is also quantifier-free, and where by ``$\psi$ implies $\chi$'' we mean that $\psi\imp\chi$ is valid (in PAL). So we then get: \begin{center} $M_s \models \dia{\chi^\downarrow} \phi$ \quad iff \quad there is a quantifier-free $\psi$ such that $\models\psi\imp\chi$  and $M_s \models \dia{\psi} \phi$. \end{center} This variation of APAL semantics was initially suggested by Igor Sedl\'ar, for a more general setting of substructural logics. Results have been published in \cite{DitmarschLKS20} and subsequently in \cite{hvdetal.almost:2022} (the expanded journal version of the former). The resulting logic is called IPAL$^\downarrow$ (where the I stands for `implying'). As in APAL, the quantification in IPAL$^\downarrow$ is over quantifier-free formulas, however, only over those formulas $\psi$ that are at least as strong as a given formula $\chi$. Structurally, given some model $M$, we quantify over all restrictions of $M$ that are \emph{contained in} $M|\chi$. In the semantics, the following are interchangeable (see \cite[Prop.\ 4]{DitmarschLKS20}):
\[ \begin{array}{ll}
(i) & \models \psi\imp\chi \text{ and } M_s \models \dia{\psi}\phi \\
(ii) & M \models \psi\imp\chi \text{ and } M_s \models \dia{\psi}\phi \\
(iii) & M_s \models \dia{\psi\et\chi}\phi
\end{array}\]
where in formulation $(iii)$ not $\psi$ but $\psi\et\chi$ is the formula implying $\chi$.

A dual `at most as strong as' quantifier $ \dia{\chi^\uparrow}$ has semantics
\[ M_s \models \dia{\chi^\uparrow} \phi \quad \text{iff} \quad \text{there is a quantifier-free $\psi$  \emph{implied by} } \chi \text{ such that } M_s \models \dia{\psi} \phi \]
In other words, now we quantify over all restrictions of $M$ that \emph{contain} $M|\chi$. This begets the logic IPAL$^\downarrow$; IPAL is the logic containing both operators.

The axiomatization of IPAL$^\downarrow$ is straightforward, by replacing the APAL axiom and derivation rule by the following, where $\pmb{\eta}(\sharp)$ is a necessity form and where we recall the above paraphrase (iii) of the IPAL$^\downarrow$ semantics:
\[\begin{array}{ll}
{[\chi^\downarrow]}\phi \imp [\psi\et\chi]\phi, \text{ for all quantifier-free } \psi \\
\text{From } \pmb{\eta}([\psi\et\chi]\phi) \text{ for all quantifier-free } \psi, \text{ infer } \pmb{\eta}([\chi^\downarrow]\phi) \hfill 
\end{array}\]
The logic is also undecidable, where the proof method of \cite{agotnesetal:2016} applies, as $[\top^\downarrow] \phi$ is equivalent to $\apalbox\phi$. So we can build a tiling of the plane as usual.

One can show that IPAL is less expressive than APAL, by a technically involved argument not resembling anything shown so far in this survey and therefore omitted (see \cite{hvdetal.almost:2022}).

The relation between IPAL$^\downarrow$ and substructural logic is that we can view $\models [\psi_1^\downarrow]\dots[\psi_n^\downarrow]\phi$ as so-called \emph{dynamic consequence} $\psi_1\dots\psi_n\Imp\phi$, as an alternative to Van Benthem's dynamic consequence \cite{jfak.ajl:2008,Aucher16a} defined as $\models [\psi_1]\dots[\psi_n]\phi$. This may be of proof theoretical interest as this alternative dynamic consequence relation satisfies more substructural properties. For example, unlike \cite{jfak.ajl:2008}, it satisfies `strong weakening', defined as ``from $\psi_1\dots\psi_n\Imp\phi$, infer $\psi_1\dots\psi\dots\psi_n\Imp\phi$'' (insert $\psi$ anywhere in the sequence $\psi_1\dots\psi_n$). This novel notion of dynamic consequence clearly deserves closer exploration.

A possibly very different logic seems to result if in $\dia{\chi^\downarrow}\phi$ and $\dia{\chi^\uparrow}\phi$ the formula $\chi$ is not required to be quantifier-free but may also contain quantifiers. The resulting logic QIPAL is also succinctly presented in \cite{hvdetal.almost:2022}. It should be noted that these semantics are well-defined. As observed by Sa\'ul Fern\'andez Gonz\'alez, this is not immediately clear, as to determine the truth of $\dia{\chi^\downarrow}\phi$ we need to establish the validity of $\psi\imp\chi$ in QIPAL. However, in QIPAL the paraphrased semantics $(iii)$ above is also equivalent to the semantics with $(i)$, from which well-definedness easily follows by induction on the number of quantifiers.

\subsubsection{Knowability logic}

In this variation we directly interpret the APAL quantifier on epistemic models and omit the public announcements from the logical language. So let now $\dia{!}\phi\in\lang(\Dia,\dia{!})$.
\[ M_s \models \F \phi \text{ \ iff \ there is a quantifier-free } \psi \text{ such that } M_s \models \psi \text{ and } (M|\psi)_s \models \phi \]
We recall that the APAL semantics of $\dia{!}\phi\in\lang(\Dia,!,\dia{!})$ use public announcement. \[ M_s \models \F \phi \mbox{ \ \ iff \ \ there is a quantifier-free $\psi$ such that } M_s \models \dia{\psi} \phi \] The semantics are equivalent. But for the former we do not need public announcement in the logical language. Given that formulas of shape $\apaldia \Box_a \phi$ represent ``$\phi$ is knowable'', the logic with language $\lang(\Dia,\F)$ without announcements is called {\em knowability logic}~\cite{hvd.wollic:2012}. It is not known what its axiomatization is. Wiebe van der Hoek mentioned that, in some way, we `know' what its axiomatization is, because it consists of the validities of APAL not containing public announcements. But it is unknown how to get these systematically. The axiom and rule for $\F$ in APAL need to be replaced by something not using announcements. 

Another such logic for knowability has been proposed in~\cite{wenetal:2011}, now based on a relation-restricting update, instead of a domain-restricting update.

A logic for knowability called LK (as well as a number of variants) was proposed in \cite{mojie:2021}, of which the logical language is the fragment of $\lang(\Dia,!,\dia{!})$ wherein  the two modalities $\dia{!}$ and $\Box_a$ are bundled/packed. The language of this logic is $\phi := p \mid \neg \phi \mid (\phi \et \phi) \mid \Dia_a \phi \mid \dia{\phi}\phi '\mid \dia{!}\Box_a \phi$, where $\dia{!}\Box_a\phi$ stands for ``$\phi$ is knowable by $a$.'' Such bundled logics often have, obviously, different axiomatizations, complexity (or decidability) of decision problems, etcetera \cite{PadmanabhaRW18,wang:2018}. But not in this case: the axiomatization of the logic LK is again similar to that of APAL, and LK is also undecidable. We conjecture that LK is less expressive than APAL.

\subsubsection{Arbitrary announcement as effort}

In this variation we interpret the APAL quantifier on epistemic models with more, topological, structure.
\[ M_s \models \dia{!} \phi \quad \text{iff} \quad \text{there is a quantifier-free {\bf observable} $\psi$ such that } M_s \models \dia{\psi} \phi \]
Here, `observable' is a topological notion defined further below. In this setting an arbitrary announcement quantifier is interpreted as a kind of `effort' modality: if you can get to know a proposition after incorporating the informative consequences of an announcement, this is like getting to know a proposition after making some {\em effort}. Instead of quantifying over submodels reached by executing announcements, we can alternatively quantify over submodels of a given model reached by an alternative semantics for such announcements on models that have more structure than the epistemic models central in this survey: only some subsets of the domain that are denotations of formulas are \emph{permitted}.  Examples of such richer structures are the subset spaces in the work by Parikh and collaborators \cite{DabrowskiMP96}, or more topological, or topologically inspired, structures in publications like \cite{Ozgun17,bjorndahl:2018,DitmarschKO19,aybuke19}. In the above semantics of the quantifier, a formula $\psi$ is \emph{observable} if the topological interior of its denotation is true: in other words, `permitted' then means `open'. We refer to the cited works for details.

\subsubsection{Quantifying over formulas for subsets of atoms}

In this variation we only allow quantification over formulas containing a subset of atoms.
\[ M_s \models \dia{!} \phi \text{ iff there is a } \psi \in \lang(\Dia,!)|P(\phi) \text{ such that } M_s \models \dia{\psi} \phi \]
Here, $P(\phi)\subseteq P$ is the set of atoms occurring in $\phi$, and $\lang(\Dia,!)|P(\phi)$ is the language of the quantifier-free formulas only containing atoms in $P(\phi)$. The logic with this quantifier is called SCAPAL and it is reported in \cite{hvdetal.almost:2022}.  Barteld Kooi already observed many years before \cite{hvdetal.almost:2022} that such a modality does not distribute over disjunction, i.e., $\dia{!}\phi\vel\dia{!}\psi$ is not equivalent to $\dia{!}(\phi\vel\psi)$. Dually, $[!]\phi\et [!]\psi$ is not equivalent to $[!](\phi\et\psi)$. 

\begin{example}
Consider again the formula $\apaldia (\Box_a p \et \neg \Box_b \Box_a p)$ used in the expressivity proofs for APAL and the same models and announcement (see Section \ref{sec.apal.exp}). Let $M$ be the model on the left.
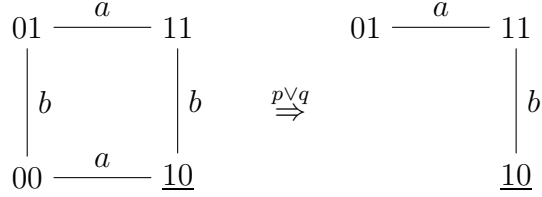
\begin{center}
\begin{tikzpicture}
\node (00) at (0,0) {$00$};
\node (10) at (2,0) {$\underline{10}$};
\node (01) at (0,2) {$01$};
\node (11) at (2,2) {$11$};
\draw[-] (00) -- node[above] {$a$} (10);
\draw[-] (01) -- node[above] {$a$} (11);
\draw[-] (00) -- node[right] {$b$} (01);
\draw[-] (10) -- node[right] {$b$} (11);
\node (m) at (3.5,1) {$\stackrel {p \vel q} \Imp $};
\node (l10) at (6.5,0) {$\underline{10}$};
\node (l01) at (4.5,2) {$01$};
\node (l11) at (6.5,2) {$11$};
\draw[-] (l01) -- node[above] {$a$} (l11);
\draw[-] (l10) -- node[right] {$b$} (l11);
\end{tikzpicture}
\end{center}
We now have that 
\[ \begin{array}{llll}
M_{10}\models\apaldia ((\Box_a p \et\neg \Box_b \Box_a p) \vel q) & \hspace{3cm} (i) \\
M_{10}\not\models\apaldia (\Box_a p \et\neg \Box_b \Box_a p) & \hfill (ii)  \\
M_{10}\not\models\apaldia q & \hfill (iii) 
\end{array}\]
The first is true because, as depicted:
\[\begin{array}{l} M_{10}\models\dia{p \vel q} (\Box_a p \et \neg \Box_b \Box_a p), \text{ so} \\ M_{10}\models\dia{p \vel q} ((\Box_a p \et \neg \Box_b \Box_a p) \vel q), \text{ and therefore} \\ M_{10}\models\apaldia ((\Box_a p \et \neg \Box_b \Box_a p) \vel q) \end{array}\] where we note that announcement $p\vel q$ is indeed in the language containing atoms occurring in $P((\Box_a p \et \neg \Box_b \Box_a p) \vel q) = \{p,q\}$. The second is false because the only model restrictions containing $10$ that we can obtain with formulas only involving $p$ are $\{10,11\}$ (which is $M|p$) and $\{10,11,00,01\}$ (which is $M$ itself). The third is false because $q$ is false in state $10$.

Therefore, $\dia{!}\phi\vel \dia{!}\psi$ is not equivalent to $\dia{!}(\phi\vel\psi)$.
\end{example}

A similar yet slightly different version is the logic with, for each $Q \subseteq P$, quantifier 
\[ M_s \models \dia{!_Q} \phi \text{ iff there is a } \psi \in \lang(\Dia,!)|Q \text{ such that } M_s \models \dia{\psi} \phi \] which results in the logic called SAPAL (S for `subset'). If we require $Q$ to be finite we get FSAPAL. These logics are also reported in \cite{hvdetal.almost:2022}. Although all these logics have slightly different expressivity, their axiomatizations are not very different from APAL and they are also undecidable.

\subsubsection{Quantifying over formulas of bounded modal depth}

In this variation we only allow quantification over formulas of bounded modal depth. This is a novel proposal. We are unaware of a publication or work in progress.
\[ M_s \models \dia{!^n} \phi \text{ iff there is a } \psi \in \lang(\Dia,!) \text{ with } d(\psi)\leq n \text{ and such that } M_s \models \dia{\psi} \phi \]
This quantification over quantifier-free formulas of bounded modal depth has been suggested by Tim French. We have already seen a bit of this variation: the logic BAPAL has quantifiers $\dia{!^0}$ in this sense. That may not be immediately obvious. The quantifier-free formulas of modal depth $0$ may contain announcement modalities but do not contain any epistemic modalities $\Dia$. But from such formulas we can eliminate the announcements. It is easy to see that for $\phi,\psi\in\lang(!)$, $[\phi]\psi$ is equivalent to $\phi\imp\psi$. Using that we can prove by induction on the number of announcements in a formula that any $\phi\in\lang(!)$ is equivalent to a $\phi'\in\lang(\emptyset)$, that is, to a Boolean.

It seems interesting to investigate whether, like BAPAL, this logic has a finitary axiomatization and is decidable.

\subsection{Open problems and further research}

There are many more relevant works, that we do not discuss in detail. How logics like APAL can be employed to formalize knowability is currently actively discussed in the formal epistemology community \cite{Xu2021-XUEIL,Holliday2018-HOLKTA}. The tendency is towards more explicit temporal perspectives, as in APALM (Section~\ref{sec.memory}). A version of APAL to reason about asynchronous information change is presented in \cite{BalbianiDG20}. A so-called `mental program' arbitrary public announcement logic is presented in \cite{charrieretal:2015}. It is a decidable version of APAL, however, the basic (epistemic-like) modalities and the structures on which such modalities are interpreted, are already very different from the epistemic models in this survey.

Open problems about APAL and its variants include:
\begin{itemize}
\item Whether APAL has a finitary axiomatization \cite{balbianietal:2015}
\item Whether APAL$^+$ is decidable \cite{hvdetal.papal:2020}
\item Is APAL not at least as expressive as BAPAL \cite{hvdetal.bapal:2022}?
\item Does the Full APAL hierarchy stabilize at $\omega$ \cite{hvdetal.aiml:2016}?
\item The axiomatization of the logic of knowability \cite{hvdetal.theoria:2012}
\item Formalizing knowability in APALM \cite{baltagetal:2022}
\item Is LK less expressive than APAL \cite{hvdetal.almost:2022}?
\item The substructural properties of dynamic consequence based on IPAL$^\downarrow$ \cite{hvdetal.almost:2022}
\end{itemize}

\section{Group announcement and coalition announcement}  \label{sec.galcal}

\subsection{Announcements by agents}

This section is devoted to group announcement logic GAL and coalition announcement logic CAL. These are clearly also variations of APAL, but we deal with them separately in this survey, as they are more targeted towards modeling aspects of agency and therefore have received quite a bit of attention from that community. Extensions of GAL and CAL with common knowledge and distributed knowledge are discussed in Section~\ref{sec.iterate}.

\paragraph*{How to model an announcement by an agent} We recall that a truthful public announcement made by an outside observer is a true public announcement. This is information coming into the system from a trusted outside source. A truthful public announcement made by an agent whose knowledge or beliefs are modelled in the system is not necessarily true. Now there is a difference between truthful and true. Announcements made by agents can be truthful but not true, and can be true but not truthful. An announcement  made by an agent is truthful if it is believed by that agent. If agent $a$ truthfully says $\phi$ we therefore model this as the announcement $\Box_a \phi$. 

We now have to distinguish our usual setting in this survey, the treatment of knowledge and of updates preserving $\mathcal S5$ models, from the treatment of belief and of other updates. 
\begin{quote}
{\em On $\mathcal S5$ models, the truthful announcement $\phi$ by agent $a$ is true.}
\end{quote}
This is, because $\Box_a \phi$ implies $\phi$. It is standard in dynamic epistemic logics to model `agent $a$ makes a truthful public announcement $\phi$' in this way as `a truthful public announcement $\Box_a\phi$ is made', where the first occurrence of truthful implies true, and the second occurrence of truthful means true.

It is often said that modelling an announcement $\phi$ by an agent $a$ as $\Box_a\phi$ is a way to have a notion of agency in dynamic epistemic logic. This is true to the extent that, for example, a multi-agent systems modeller can informally let $\Box_a\phi$ always mean that agent $a$ observes (announces) $\phi$. Formally we cannot distinguish the announcement by agent $a$ that $\phi$, modelled in this way, from the announcement by the outside observer that $a$ knows $\phi$.
\begin{quote}
\mbox{\em On not-$\mathcal S5$ models, the announcement $\phi$ by agent $a$ may be truthful but not true.}
\end{quote}
When agent $a$ says $\phi$ and $\neg\phi\et\Box_a \phi$ is true, agent $a$ is truthful because $\Box_a\phi$ is true. This now represents that agent $a$ believes $\phi$. This belief is mistaken (incorrect), because $\phi$ is false. Formula $\neg\phi\et\Box_a \phi$ formalizes that agent $a$ has an incorrect belief in $\phi$. 
\begin{quote}
{\em On not-$\mathcal S5$ models, the announcement $\phi$ by agent $a$ may be true but not truthful.}
\end{quote}
There are now two options. When agent $a$ says $\phi$ and $\phi\et\Box_a \neg\phi$ is true, agent $a$ is \emph{lying} because she believes $\phi$ to be false. But in fact she is mistakenly saying the truth. Whereas when agent $a$ says $\phi$ and $\phi\et\neg(\Box_a \phi \vel \Box_a\neg\phi)$ is true, agent $a$ is \emph{bluffing} because she is uncertain whether $\phi$ is true, although confidently announcing $\phi$. In fact she is unknowingly saying the truth, because $\phi$ is true. See \cite{mahon.stanford:2008}. Speech acts such as lying and bluffing can easily be modelled as action models \cite{hvd.lying:2014}, and they can also be executed on $\mathcal S5$ models, but the updated model is then no longer $\mathcal S5$ \cite{hvdetal.del:2007}. 

Belief, lying, and bluffing are outside our scope. In this section we restrict ourselves to truthful public announcements by agents on models for knowledge.

\paragraph*{What is a simultaneous announcement?} Let us also elaborate on the simultaneity of announcements made by groups of agents. When three agents $a,b,c$ respectively say $\psi_a$, $\psi_b$, and $\psi_c$ simultaneously, we model this as the announcement $\Box_a \psi_a \et \Box_b \psi_b \et \Box_c \psi_c$. This is a simultaneous announcement by $a$, $b$, and $c$ consisting of three conjuncts. We do not model this as three successive announcements. The difference matters, because the truth value of $\Box_b \psi_b$ before the announcement of $\Box_a\psi_a$ may be different from the truth value of $\Box_b \psi_b$ after the announcement of $\Box_a\psi_a$. Just imagine that $a$ says that $p$ is true and that $b$ says that he is uncertain whether $p$; $b$ can say it at the same time as $a$ but not afterwards. We need not imagine simultaneous announcements as everybody talking at the same time. Public announcement is a misnomer for a \emph{public event}, such as `the three children who know that they are muddy now step forward' in the muddy children problem \cite{mosesetal:1986,hvdetal.puzzle:2015}. This can be observed when done simultaneously.  Or, back to $a,b,c$ talking, we can instead imagine $a,b,c$ writing, respectively, $\psi_a$, $\psi_b$, and $\psi_c$ on a piece of paper put into three envelopes after which the external observer or authority publicly opens the envelopes and reads their contents aloud in whatever order. It would be more accurate to say that we model such agents' actions as \emph{independent} events, as usual in concurrency. And do the three muddy children really step forward simultaneously, within a split microsecond? Of course not. It is rather that their decisions to step forward were made independently, pertaining to the same perceived previous state of information.

\subsection{Group announcement logic}

{\em Group Announcement Logic} GAL is a variation of APAL with, instead of quantifiers $\dia{!}$, quantifiers $\dia{!_G}\phi$, for any group of agents $G \subseteq A$, meaning `$\phi$ is true after a simultaneous truthful public announcement by the agents in $G$' \cite{agotnesetal:2008,agotnesetal.jal:2010}. 

The semantics are as follows, where a set of formulas is quantifier-free if all formulas in the set are quantifier-free (so the members of $\{\psi_a \mid a \in G\}$ below are in $\lang(\Dia,!)$). \[ M_s \models \dia{!_G}\phi \mbox{ \ \ iff \ \ there is a quantifier-free } \{\psi_a \mid a \in G\} \text{ such that } M_s \models \dia{\Et_{a \in G} \Box_a\psi_a} \phi \]
Given $\dia{!_G}\phi$, we imagine the remaining agents in $A\setminus{G}$ staying silent. Therefore, $\dia{!_\emptyset}\phi$ is equivalent to $\phi$. Whatever the $G$ in $\dia{!_G}$, all agents in $A$ are always listening: whatever public announcement is made, it is always observed by all agents, and in the same way.

\begin{example} \label{ex.three}
Given are two agents $a,b$ such that $a$ knows whether $p$ and $b$ knows whether $q$ (and this is common knowledge), and let in fact $p$ be true and $q$ be false (the underlined state). Anne $(a)$ can achieve that Bill knows whether $p$, namely by informing him of the value of $p$, and Bill $(b)$ can similarly achieve that Anne knows whether $q$. Neither agent can achieve both outcomes at the same time, but together they can achieve that. 
\begin{center}
\begin{tikzpicture}
\node (0) at (0,0) {$\underline{10}$};
\node (1) at (0,2) {$11$};
\node (a) at (1.5,1) {$\stackrel {\Box_a p} \Pmi $};
\draw[-] (0) -- node[right] {$a$} (1);
\end{tikzpicture}
\quad
\begin{tikzpicture}
\node (00) at (0,0) {$00$};
\node (10) at (2,0) {$\underline{10}$};
\node (01) at (0,2) {$01$};
\node (11) at (2,2) {$11$};
\node (a) at (3.5,0) {$\stackrel {\Box_b \neg q} \Imp$};
\draw[-] (00) -- node[above] {$b$} (10);
\draw[-] (01) -- node[above] {$b$} (11);
\draw[-] (00) -- node[right] {$a$} (01);
\draw[-] (10) -- node[right] {$a$} (11);
\end{tikzpicture}
\quad
\begin{tikzpicture}
\node (0) at (0,0) {$00$};
\node (1) at (2,0) {$\underline{10}$};
\draw[-] (0) -- node[above] {$b$} (1);
\node (a) at (0,2) {$\stackrel {\Box_a p \et \Box_b \neg q} \Imp$};
\node (1) at (3,2) {$\underline{10}$};
\end{tikzpicture}
\end{center}
We can evaluate in the square model $M$ in the middle that
\begin{itemize}
\item $M_{10} \models \dia{!_a} \Box_b p$ but $M_{10} \not\models \dia{!_b} \Box_b p$
\item $M_{10} \models \dia{!_b} \Box_a \neg q$ but $M_{10} \not\models \dia{!_a} \Box_a \neg q$
\item $M_{10} \models \dia{!_{ab}} (\Box_b p \et \Box_a \neg q)$ but $M_{10} \not\models \dia{!_a} (\Box_b p \et \Box_a \neg q)$ and $M_{10} \not\models \dia{!_b} (\Box_b p \et \Box_a \neg q)$
\end{itemize}
Whatever the actual state, $a$ and $b$ can get to know it on this model by collaborating in the $\dia{!_{ab}}$ sense.
\end{example}

A number of GAL validities are as follows, where $G,H \subseteq A$.
\begin{itemize}
\item $\dia{!_G} \dia{!_H} \phi \imp \dia{!_{G \union H}}\phi$ \\ If an announcement by group $G$ is followed by an announcement by group $H$, then $G$ and $H$ could have made a joint announcement with the same informative content.
\item $\dia{!_G}\dia{!_G}\phi \imp \dia{!_G}\phi$ \\ A corollary of the previous.
\end{itemize}
%
The logic GAL shares various properties with APAL. For example, the infinitary axiomatization and the method to prove its completeness are similar, the model checking complexity is also PSPACE-complete, GAL is also undecidable \cite{agotnesetal:2014,agotnesetal:2016}, and it also does not have the finite model property \cite{hvdetal.nofinite:2021}. Expressivity is discussed later.

\subsection{Coalition announcement logic}

Another variation on APAL with aspects of agency is {\em coalition announcement logic} CAL. In group announcement logic GAL we investigate the consequences of the simultaneous announcement by $G$. The agents not in $G$ do not take part in the action. In CAL we quantify over what the agents in $G$ can achieve by their joint announcement, no matter what the other agents simultaneously announce. 

In the logical language we replace the GAL quantifier $\dia{!_G}$ by the CAL quantifier $\caldia{!_G}$, where $G \subseteq A$. The notation is supposed to suggest its $\is\all$-semantics, and that of the dual $\calbox{!_G}$ its $\all\is$-semantics. The semantics of coalition announcement are then as follows. 
\[\begin{array}{lcl} M_s \models \caldia{!_G}\phi & \text{ \ iff \ } & \text{there is a quantifier-free } \{ \psi_a \mid a \in G\} \text{ such that} \\ &&  \text{for all quantifier-free } \{ \psi_a \mid a \in A\setminus G\} \\ && M_s \models \Et_{a \in G} \Box_a\psi_a \text{ and } M_s \models [\Et_{a \in A} \Box_a\psi_a]\phi \end{array}\]

\begin{example} \label{ex.four}
Consider again the four-state model $M$ of Example \ref{ex.three}. Although we have that $M_{10} \models \dia{!_a} \Dia_a q$ (namely by $a$ simply doing nothing / announcing $\T$), it is {\bf not} the case that $M_{10} \models \caldia{!_a} \Dia_a q$: agent $b$ can prevent $a$ from remaining ignorant about $q$ by announcing $\neg q$. It then does not matter whether $a$ announces $p$ or $\T$, either way $a$ will learn $\neg q$.
\begin{center}
\begin{tikzpicture}
\node (0) at (0,0) {$\underline{10}$};
\node (a) at (1,1) {$\stackrel {\Box_a p \et \Box_b\neg q} \Pmi $};
\end{tikzpicture}
\ 
\begin{tikzpicture}
\node (00) at (0,0) {$00$};
\node (10) at (2,0) {$\underline{10}$};
\node (01) at (0,2) {$01$};
\node (11) at (2,2) {$11$};
\node (a) at (3.8,1) {$\stackrel {\Box_a \top \et \Box_b \neg q} \Imp$};
\draw[-] (00) -- node[above] {$b$} (10);
\draw[-] (01) -- node[above] {$b$} (11);
\draw[-] (00) -- node[right] {$a$} (01);
\draw[-] (10) -- node[right] {$a$} (11);
\end{tikzpicture}
\ 
\begin{tikzpicture}
\node (0) at (0,0) {$00$};
\node (1) at (2,0) {$\underline{10}$};
\draw[-] (0) -- node[above] {$b$} (1);
\end{tikzpicture}
\end{center}
We still have that $M_{10} \models \caldia{!_{ab}} (\Box_b p \et \Box_a \neg q)$ --- but this is trivial, as there are no other agents around to say something to prevent it. The power of group announcements by all agents is the same as that of an announcement by that `grand coalition': $\dia{!_A}\phi \eq \caldia{!_A}\phi$. 
\end{example}
The logic CAL was proposed and summarily discussed in \cite{agotnesetal:2008}. No axiomatization is given. Publication \cite{agotnesetal:2008} also succinctly presents an original neighbourhood semantics for the CAL quantifier. Given the $\is\all$ character of its semantics this is not surprising --- we illustrate this by Example~\ref{ex.five}, below. Further below we  relate CAL to coalition logic \cite{Pauly2002}, that has a similar neighbourhood semantics.

\begin{example} \label{ex.five}
Consider a different representation of the three models featuring in Example~\ref{ex.four}, namely as a single model $\mathcal M$ with point $10$ as underlined (without giving details, we can see it as the direct sum of the three models $M$, $M|(\Box_a p \et \Box_b\neg q)$, and $M|(\Box_a \top \et \Box_b\neg q)$). To distinguish states with similar names in different parts of the model we have given them different names.
\begin{center}
\begin{tikzpicture}
\node (0) at (0,0) {$10x${\color{white}y}};
\end{tikzpicture}
\ 
\begin{tikzpicture}
\node (00) at (0,0) {$00$};
\node (10) at (2,0) {$\underline{10}$};
\node (01) at (0,2) {$01$};
\node (11) at (2,2) {$11$};
\draw[-] (00) -- node[above] {$b$} (10);
\draw[-] (01) -- node[above] {$b$} (11);
\draw[-] (00) -- node[right] {$a$} (01);
\draw[-] (10) -- node[right] {$a$} (11);
\end{tikzpicture}
\ 
\begin{tikzpicture}
\node (0) at (0,0) {$00y$};
\node (1) at (2,0) {$10y$};
\draw[-] (0) -- node[above] {$b$} (1);
\end{tikzpicture}
\end{center}
A {\em neighbourhood} of a state $s$ in the domain of $\mathcal M$ is a set of subsets of the domain of $\mathcal M$. Any such subset in the neighbourhood is a set of {\em neighbours}. Let the two sets of neighbours of state $10$ be $X = \{10x\}$ and $Y = \{10,10y\}$. So the set $\{X,Y\}$ is the neighbourhood of $10$. 

We now define, ad hoc (see \cite{agotnesetal:2008} for a proper definition), that for $\phi\in\lang(\Dia)$, $\mathcal M_s \models \phi$ if ($M_s \models \phi$ or $(M|(\Box_a p \et \Box_b\neg q))_s \models \phi$ or $(M|(\Box_a \top \et \Box_b \neg q))_s \models \phi$). We also consider a novel modality $\dia{a}$, that is supposed to be reminiscent of $\caldia{!_a}$, and for which we define, as usual in neighbourhood semantics, that $\mathcal M_s \models \dia{a}\psi$ iff there is a set of neighbours of $s$ in $\mathcal M$ (in other words, a set of states in the neighbourhood of $s$) where $\phi$ is true in all of its states. 

Let us now replay the previously observed $M_{10} \not\models \caldia{!_a} \Dia_a q$ in the current setting and show that $\mathcal M_{10} \not\models \dia{a} \Dia_a q$. Then we have that $\mathcal M_{10} \not\models \dia{a}\Dia_a q$ because $\Dia_a q$ is not true in $X$ and $\Dia_a q$ is not true in $Y$, where the latter fails because $\Dia_a q$ is not true in both states of $Y$. In other words it is not the case that there is  ($\exists$) a set of neighbours in the neighbourhood of state $10$ such that for all ($\all$) states in that set $\Dia_a q$ is true.

Note that the non-monotonic logician would write $[a]\phi$ for $\dia{a}\phi$ \cite{helleetc}, focussing on the $\all$ of $\exists\all$, not $\dia{a}\phi$ as we do, focussing on the $\exists$ of $\exists\all$ and in that way on our storyline for CAL.
\end{example}

The logic CAL was a main focus of investigation in various related works involving Rustam Galimullin \cite{Galimullin19,frenchetal:2019,GalimullinA17,GalimullinAA19}. They contain an axiomatization for CAL with an additional auxiliary operator bundling an announcement by agents in $G$ with a quantification over a announcement by agents not in $G$. This technical device then facilitates proving completeness for the axiomatization. 

The axiomatization of CAL with only the coalition quantifiers $\caldia{!_G}$ remains an open question.

The undecidability (of satisfiability) of CAL is shown in \cite{agotnesetal:2016}, in an integrated setting also including APAL and GAL (and that simplifies prior proofs of undecidability for APAL \cite{frenchetal:2008} and GAL \cite{agotnesetal:2014}). The PSPACE complexity of model checking is shown in \cite{GalimullinAD18}, and its lack of the finite model property in \cite{hvdetal.nofinite:2021}.

\subsection{Expressivity of GAL and CAL}

The logics GAL and CAL are related in interesting ways. As mentioned in Example \ref{ex.four}, if all agents $A$ together can achieve $\phi$ in GAL, then obviously as well in CAL, as there are no remaining agents to counteract it: \[ \dia{!_A} \phi \imp \caldia{!_A} \phi \] On the other hand, if the empty coalition $\emptyset$ can achieve $\phi$ in CAL, then $\phi$ will be true after any GAL announcement: \[ \caldia{!_\emptyset} \phi \imp [!_A] \phi \]
Continuing along that line, as this is equivalent to $\caldia{!_\emptyset} \phi \imp \dia{!_\emptyset}[!_A] \phi$, it was conjectured \cite{hvd.wollic:2012} that the CAL quantifier might be definable in terms of the GAL quantifier as follows.
\[ \caldia{!_G}\phi \eq \dia{!_G}[!_{A \backslash G}] \phi \]
This schema can be paraphrased as follows: \begin{quote} {\em There is something that the agents in $G$ can say no matter what the agents not in $G$ say simultaneously and after which $\phi$, iff there is something that the agents in $G$ can say no matter what the agents not in $G$ say subsequently and after which $\phi$.} \end{quote}
If the conjecture had been true, GAL would have been at least as expressive as CAL,  because we can then see the language of GAL as a fragment of the language of CAL. Then, showing that GAL is as expressive as CAL would have seemed within reach. However, the conjecture is false \cite{GalimullinAA19}. It is false because non-bisimilar states can become bisimilar after an announcement (we recall the discussion on APAL with memory \cite{BaltagOS18} in Section \ref{sec.memory} on page \pageref{sec.memory}). The agents not involved in that announcement are then no longer able to distinguish between those states in a subsequent announcement. However, they might have been able to prevent such states becoming indistinguishable if they could have made a simultaneous announcement, as in the following example.

\begin{example} \label{ex.galcal}
Consider the models below for a single variable $p$. As before, $p$ is false in the $0$ states and true in the $1$ states. As usual we assume transitivity of the accessibility relation. Note that in model $M$ formula $\Box_a \neg p$ is distinguishing for state $v$. 

\bigskip

\begin{tikzpicture}
\node (m) at (-1.7,0) {$M:$};
\node (0) at (0,0) {$1$};
\node (1) at (2,0) {$0$};
\node (2) at (4,0) {$1$};
\node (3) at (6,0) {$0$};
\node (4) at (8,0) {$1$};
\node (5) at (10,0) {$0$};
\node (0a) at (0,0.4) {$u'$};
\node (1a) at (2,0.4) {$t'$};
\node (2a) at (4,0.4) {$s$};
\node (3a) at (6,0.4) {$t$};
\node (4a) at (8,0.4) {$u$};
\node (5a) at (10,0.4) {$v$};
\draw[-] (0) -- node[above] {$ac$} (1);
\draw[-] (1) -- node[above] {$a$} (2);
\draw[-] (2) -- node[above] {$a$} (3);
\draw[-] (3) -- node[above] {$ac$} (4);
\draw[-] (4) -- node[above] {$b$} (5);
\end{tikzpicture}

\medskip

\begin{tikzpicture}
\node (n) at (-1.7,0) {$N:$};
\node (0) at (0,0) {\color{white}$1$};
\node (1) at (2,0) {$0$};
\node (2) at (4,0) {$1$};
\node (3) at (6,0) {$0$};
\node (4) at (8,0) {$1$};
\node (5) at (10,0) {\color{white}$0$};
\draw[-] (1) -- node[above] {$a$} (2);
\draw[-] (2) -- node[above] {$a$} (3);
\draw[-] (3) -- node[above] {$ac$} (4);
\end{tikzpicture}

\medskip

\begin{tikzpicture}
\node (m) at (-1.3,0) {$M|\Dia_a p:$};
\node (0) at (0,0) {$1$};
\node (1) at (2,0) {$0$};
\node (2) at (4,0) {$1$};
\node (3) at (6,0) {$0$};
\node (4) at (8,0) {$1$};
\draw[-] (0) -- node[above] {$ac$} (1);
\draw[-] (1) -- node[above] {$a$} (2);
\draw[-] (2) -- node[above] {$a$} (3);
\draw[-] (3) -- node[above] {$ac$} (4);
\end{tikzpicture}

\medskip

\begin{tikzpicture}
\node (m) at (-1.3,0) {\color{white}$M|\Dia_a p:$};
\node (2) at (4,0) {$1$};
\node (3) at (6,0) {$0$};
\node (4) at (8,0) {$1$};
\draw[-] (2) -- node[above] {$a$} (3);
\draw[-] (3) -- node[above] {$ac$} (4);
\end{tikzpicture}

\bigskip

We will now sketch the proof that
\[ \begin{array}{lcl}
M_s \not\models & [!_a]\dia{!_{bc}} & (\Dia_a\Box_c \neg p \et \Dia_a\Box_c p \et \Dia_a\neg(\Box_c p \vel \Box_c \neg p)) \\
M_s \models & \calbox{!_a} & (\Dia_a\Box_c \neg p \et \Dia_a\Box_c p \et \Dia_a\neg(\Box_c p \vel \Box_c \neg p))
\end{array}\]
from which follows that $\caldia{!_G}\phi \eq \dia{!_G}[!_{A \backslash G}] \phi$ is not valid for all $\phi$, $A$ and $G$.

The first is the case because if $a$ announces $\Dia_a p$ (as depicted), in model $M|\Dia_a p$ states $t$ and $t'$, and $u$ and $u'$, are bisimilar. The bisimulation contraction is shown below it. After that announcement by $a$, agents $b$ and $c$ can no longer make the formula $\Dia_a\Box_c \neg p \et \Dia_a\Box_c p \et \Dia_a\neg(\Box_c p \vel \Box_c \neg p)$ true. That formula describes that $a$ considers possible three $c$-equivalence classes, including one where $c$ is uncertain and that therefore requires at least two states. Altogether we therefore need a model with at least four states. But the contraction of $M|\Dia_a p$ contains three states, which is not enough.

Whereas if they can make their announcement simultaneously with $a$, they can still use that  $u$ and $u'$ are different. If $a$ again were to announce $\Dia_a p$, $b$  can simultaneously announce $\phi_b := \neg \Box_b (p \et \neg (\Box_c p \vel \Box_c \neg p))$ ($\phi_b$ is only false in $M_{u'}$) and $c$ can simultaneously announce $\top$ (i.e., $\Box_c \top$). From this joint announcement by $a,b,c$ model $N$ results, in which $\Dia_a\Box_c \neg p \et \Dia_a\Box_c p \et \Dia_a\neg(\Box_c p \vel \Box_c \neg p)$ is true. The only other announcement $a$ can initially make in state $s$ is $\top$, and then $b$ and $c$ can also enforce $N$, by respective announcements $\phi_b$ and $\neg \Box_c \Box_a \neg p$ (of which the negation is only true in $M_v$).
\end{example}

Example \ref{ex.galcal} is a simple demonstration of the differences between GAL and CAL. But it is not an expressivity proof. In \cite{frenchetal:2019} it is shown that GAL is not at least as expressive as CAL. That proof is not so simple. It is unknown whether CAL is at least as expressive as GAL. A corollary in \cite{frenchetal:2019} is that GAL is not at least as expressive as APAL. In \cite{agotnesetal.jal:2010} it was already shown that APAL is not at least as expressive as GAL. Therefore, GAL and APAL are incomparable \cite[Theorem 5.6]{frenchetal:2019}. 

Finally, we recall the structures with `memory' of the initial information state that allow to continue to distinguish states that have become bisimilar after announcement. Section \ref{sec.memory} presented the logic APALM with that memory feature. In the journal version \cite{baltagetal:2022} of \cite{BaltagOS18} and in the journal version \cite{frenchetal:2022} of \cite{frenchetal:2019} the logics GAL and CAL are extended with memory as in APALM, resulting in GALM and CALM, and it is then shown that the CAL quantifier is now after all definable with the GAL quantifier as $\caldia{!_G}\phi \eq \dia{!_G}[!_{A \backslash G}] \phi$.

\subsection{Applications} \label{sec.appgalcal}

\paragraph*{Agency} The expression $\dia{!_G} \phi$ has the smell of `group of agents $G$ is able to achieve $\phi$', such that, taking a single agent, $\Box_a \dia{!_a} \phi$ (on ${\mathcal S5}$ models) seems to formalize that agent $a$ knows that she is able to achieve $\phi$, as in logics combining agency and knowledge \cite{Jamroga2004,agotnes:synthese-06,vdhoeketal:2002c}. These are tricky issues in the setting for group announcements. In dynamic epistemic logics it has been proposed to define `knowledge de re' as `there is an announcement which $a$ knows will achieve the goal', and `knowledge de dicto' as `$a$ knows that there is an announcement which will achieve the goal' \cite{agotnesetal.jal:2010}. As required for such de re / de dicto distinctions, in the former a specific witness formula is required, whereas in the latter the witness formula may be different for each state that agent $a$ considers possible. The intuitive descriptions match the GAL formalizations $\dia{!_a}\Box_a\phi$ respectively $\Box_a \dia{!_a}\phi$ \cite[Prop.~26]{agotnesetal.jal:2010}. Given this identification the following results can then be obtained (see \cite{agotnesetal:2008} and \cite[Prop.~27]{agotnesetal.jal:2010}):
\begin{itemize}
\item $\Box_a \dia{!_a} \phi \imp \dia{!_a} \phi$ is valid. \\ `If you know that you can do something, you can do it.'
\item $\dia{!_a} \Box_a \phi \imp \Box_a \dia{!_a} \phi$ is  valid. \\  `Knowledge de re implies knowledge de dicto'. Also valid is $\dia{!_a} \Box_a \phi \imp \Box_a\dia{!_a} \Box_a \phi$.
\item $\Box_a \dia{!_a} \phi \imp \dia{!_a} \Box_a \phi$ is not valid. \\
`Knowledge de dicto does not imply knowledge de re'. \\
In different states, different announcements may be required to make $\phi$ true. As you do not know what the actual state is, you therefore do not know what announcement makes $\phi$ true in the actual state. You only know that in all states that you consider possible there is an announcement that makes $\phi$ true. For example, in state $s$ formula $\phi$ is true after you announce $p$ but not after you announce $q$; in indistinguishable state $t$ formula $\phi$ is true after you announce $q$ but not after you announce $p$. Should you announce $p$ or should you announce $q$? You are not able to achieve $\phi$!
\end{itemize}

\paragraph{Security}
Given that $\dia{!_G}\dia{!_G}\phi \imp \dia{!_G} \phi$ is valid in GAL (see above), and that we can iterate this result for any finite sequence of announcements by group $G$, we can formalize communication protocols consisting of finite executions in GAL. This includes communication protocols where agents take turns in saying something. If $a$ announces $\phi$ and subsequently $b$ announces $\psi$, we can also see this as: $a$ announces $\phi$ and  simultaneously $b$ announces $\T$ (keeps quiet), and subsequently, $a$ announces $\T$ and simultaneously $b$ announces $\psi$. Of particular interest seems the formalization of security protocols executed by two \emph{principals} Alice $a$ and Bob $b$ in the presence of an \emph{eavesdropper} Eve $e$ --- in such protocols the agents sending each other messages are the principals, and an eavesdropper is, not surprisingly, an agent intercepting (`hearing') a message. If $\phi$ is some epistemic goal (information goal) and $\psi$ is a security constraint that needs to be preserved throughout the execution, a finite protocol for $a$ and $b$ to learn the secret safely should observe \[ \psi \imp \dia{!_{ab}} (\phi\et\psi) \hspace{3cm}  (i) \]
A typical example is whether in cards cryptography, given a commonly known deck of cards dealt over three players (where each player can only see her own cards), players Alice and Bob can learn the card deal by sending public messages (announcements) to each other, without Eve learning the card deal \cite{hvd.studlog:2003}. Part of the problem specification would then be, where $0_a$ stands for `Alice holds card $0$': \[ 0_a \imp \dia{!_{ab}} (\Box_b 0_a \et \neg (\Box_c 0_a \vel \Box_c \neg 0_a))  \hspace{3cm}  (ii) \] As GAL and APAL have incomparable expressivity, we conjecture that $(i)$ is not formalizable in APAL. The shadow of doubt is that the incomparability proof in \cite{frenchetal:2019} uses three agents. It therefore holds for more agents, but maybe not for less, and $(i)$ only requires two communicating agents. Then again, what does safety for two agents mean anyway without the presence of a third agent who is the eavesdropper, as in $(ii)$? 

\paragraph{Coalition logic, and playability}

The following axioms of coalition logic \cite{Pauly2002} are validities of coalition announcement logic CAL \cite{agotnesetal:2008}. \[ \begin{array}{ll}  & (\caldia{!_G} \phi \et \calbox{!_H} \phi') \imp \caldia{!_{G\union H}} (\phi \et \phi') \hspace{1cm} \text{if } G\inter H = \emptyset \\ 
& \calbox{!_\emptyset} \phi \imp \caldia{!_A} \phi
\end{array} \]
The second already featured in this section. The first formalizes that if two disjoint groups can independently bring about $\phi$ respectively $\phi'$, then  they  can jointly bring about $\phi\et\phi'$. If the groups are not disjoint, that may no longer be the case, as an agent in the intersection may have do $x$ to bring about $\phi$ but may have to do $y \neq x$ to bring about $\phi'$.

There are also relations to {\em game logic} \cite{parikh:1985,PaulyP03a} and also to the {\em forcing} operator $\{G\}$ proposed in \cite{jfak.bulletin:2001} that quantifies over strategies in extensive games (where a strategy is a sequence of moves); see also \cite{jfak.logicingames:2014}. For a single agent $i$, $\{i\}\phi$ is true if ``player $i$ has a strategy for playing [the] game which forces a set of outcomes all of which satisfy $\phi$'' \cite{jfak.bulletin:2001}. As $\caldia{!_G}$ models what coalition $G$ can achieve/force in any finite sequence of moves, this is akin to the extensive game setting in \cite{jfak.bulletin:2001}. However, there are issues. The schema $\{G\}\{G\}\phi \imp \{G\}\phi$ is valid for the forcing operator. But although $\dia{!_G}\dia{!_G}\phi \imp \dia{!_G}\phi$ is valid in GAL, it is unknown whether $\caldia{!_G}\caldia{!_G}\phi \imp \caldia{!_G} \phi$ is valid in CAL \cite[page 45]{Galimullin19}.

\subsection{Open problems and further research}

The relation of GAL and CAL to (epistemic extensions of) logics of agency such as ATL \cite{Alur2002} and STIT \cite{horty:2001} seems worth investigating. A logic much like CAL but in an ATL-setting with therefore an explicit notion of agency, including a complete axiomatization, was proposed by de Lima in \cite{delima:2011,Lima14,delima:2019}. 

More recent investigations building forth upon the logics GAL and CAL, in areas as diverse as abstract argumentation and strategic voting, include \cite{Galimullin21,ParmannA21,XiongA20,sedlar:2022,eide:2019}. A group announcement logic for ignorance (with ignorance modalities instead of knowledge modalities as epistemic primitives) is presented in \cite{jiefan:2016}.

For comparisons with logics as STIT and ATL we should keep in mind that GAL and CAL do not have factual change, which makes such a comparison restricted. Then again, also in dynamic epistemic logics one can model factual change and quantify over that, which we will address in Section \ref{sec.iterate} on epistemic planning.

Open problems about GAL and CAL include:
\begin{itemize}
\item What is the complete axiomatization of CAL? \cite{GalimullinA17}
\item Is GAL at least as expressive as CAL? \cite{frenchetal:2019,frenchetal:2022}
\item How does CAL with public factual change relate to epistemic extensions of logics of agency?
\end{itemize}

\section{Quantifying over action models} \label{sec.am} 

In this section we are moving from quantifying over public announcements to quantifying over arbitrary epistemic actions, such as private announcements. We recall that already for PAL the term `announcement' is somewhat questionable. And even more so for non-public information change. However, a constant throughout dynamic epistemic logics is that a model update is induced by an epistemic action, and that an epistemic action needs to satisfy a precondition formula. We simply continue to call this formula the announcement. But it can then be private instead of public. Whether there is an epistemic action after which a proposition is true, is then still the same as what is \emph{to be announced} in order to make that proposition true.

\subsection{Arbitrary action model logic} \label{sec.aaml}

In the logic APAL we quantify over announcements, i.e., public events. We can also quantify over events that may not be public, where the obvious suspect is to quantify over action models. This was observed in \cite[Section 5]{balbianietal:2008}, including the obvious semantics, but only truly realized in various works involving James Hales \cite{hales2013arbitrary,frenchetal:2014,hales:2016}, where our presentation follows \cite{hales2013arbitrary}.

We expand the language $\lang(\Dia,\otimes)$ of action model logic with a quantifier $\dia{\otimes}$ with the following semantics, where $E_e$ is an epistemic action, and where $M$ is a $\mathcal K$ model (the results in this section are typically not for $\mathcal S5$ models).
\[ \begin{array}{lcl} M_s \models \dia{\otimes} \phi & \text{ iff } & \text{there is a } E_e \text{ with quantifier-free preconditions such that } M_s \models \dia{E_e} \phi \end{array} \] 
We call the resulting logic AAML, for {\em arbitrary action model logic}. We recall that in APAL the quantifier over announcements adds expressivity and causes undecidability of satisfiability. It may therefore come as a surprise that in AAML the quantifier does not add expressivity and does not cause undecidability. The logic AAML is as expressive as action model logic AML (and therefore as expressive as the logic K), because the quantifier $\dia{\otimes}$ can be eliminated by reduction axioms. We recall from Section~\ref{sec.prelims} that such a reduction axiom has the shape of an equivalence between two formulas $\phi_1$ and $\phi_2$ such that a subformula of $\phi_1$ bound by a quantifier is (strictly) more complex than any subformula of $\phi_2$ bound by that quantifier, or such that that quantifier no longer occurs in $\phi_2$. Every formula with quantifiers can then be reduced (rewritten) to an equivalent formula without quantifiers. The technique to reduce AAML is borrowed from another logic, refinement modal logic, and therefore we will only discuss it section \ref{sec.rml} devoted to that logic. 

Because AAML is as expressive as AML by way of a reduction, this logic AAML also allows  {\em synthesis}. This we will now discuss.

\paragraph*{Synthesis} Logics with quantifiers over information change sometimes allow {\em synthesis}. There are (at least) two kinds of synthesis in a dynamic epistemic logic. The logic need not have quantifiers over information change. In the \emph{local synthesis} problem, given an epistemic model $M_s$ and a formula $\phi$,  the problem is whether there exists an epistemic action $\alpha$ (such as a pointed action model $E_e$, but we will later also consider other epistemic actions) such that $M_s \models \dia{\alpha}\phi$.  In the \emph{global synthesis} problem, given a formula $\phi$, the problem is whether there exists an epistemic action $\alpha$ such that whenever $M_s \models \dia{\beta}\phi$ for some epistemic action $\beta$, then also $M_s \models \dia{\alpha}\phi$. In a logic with a quantifier $\dia{\bullet}$ for information change due to such $\beta$s, global synthesis  amounts to the validity of the schema $\dia{\bullet}\phi \eq \dia{\alpha}\phi$. From here on, by synthesis we mean global synthesis. Not all  dynamic epistemic logics allow synthesis, for example, PAL does not. The somewhat trivial reason for this failure is that we cannot make a typical disjunction of epistemic parts, such as $\Box_a p \vel \Box_a \neg p$, \emph{always} true: we would have to choose between the announcement $p$ and the announcement $\neg p$. Although correct, this is somewhat trivial, because allowing the `test' of the value of $p$, that is, non-deterministic choice between the announcements $p$ and $\neg p$, would solve the problem for this example. Such a program would amount to a two-pointed action model, with preconditions $p$ respectively $\neg p$, and where all agents can distinguish the two actions.

Clearly, a dynamic epistemic logic with quantifiers can only allow synthesis if the quantifier can be eliminated by a reduction, so that any formula with quantifiers is equivalent to a formula without. So for APAL, we need not even think about synthesis, as APAL is (strictly) more expressive than PAL (Section~\ref{sec.apal}). 

The logic AAML allows synthesis \cite{hales2013arbitrary}. More specifically, given $\dia{\otimes}\phi$, where $\phi$ is quantifier-free, there is a multi-pointed action model $E_T$ (where $T \subseteq \domain(E)$) such that $\dia{\otimes}\phi \eq \dia{E_T}\phi$ is valid: given any formula $\phi \in \lang(\Dia,\otimes)$ we can \emph{synthesize} an action model $E_T$, that is finite, and with quantifier-free preconditions, such that, whenever there is an action model such that after its execution $\phi$ is true, then $E_T$ can also be executed to make $\phi$ true. We cannot always synthesize a single-pointed action model $E_t$. It may have to be multi-pointed, as for the disjunctive epistemic goal $\phi = \Box_a p \vel \Box_a \neg p$ above. By an easy induction over the number of quantifiers in a formula we can also lift the restriction that $\phi$ be quantifier-free. It can be any formula in $\lang(\Dia,\otimes,\dia{\otimes})$. 


Because there is synthesis for AAML, and because in an actual state $s$ some $e \in T$ of the synthesized $E_T$ will execute, the logic satisfies that
\[ \begin{array}{lcl} M_s \models \dia{\otimes} \phi & \text{ iff } & \text{there is a } E_e \text{ such that } M_s \models \dia{E_e} \phi \hspace{2cm} (i) \end{array} \] 
without any language restriction. Note that  $(i)$  is a property of the semantics and does not define the semantics of $\dia{\otimes}$, as in the beginning of this subsection. We cannot have $(i)$ be the semantics of the quantifier, as $\dia{\otimes}\phi$ itself can then be a precondition formula of action model $E$. We recall from Section \ref{sec.variations} a similar result for fully arbitrary public announcement logic \cite{hvdetal.aiml:2016}.

\paragraph*{Axiomatization} 
The axiomatization of AAML in \cite{hales2013arbitrary} is somewhat indirect, as it uses a translation into refinement modal logic \cite{bozzellietal.inf:2014}. This translation involves replacing the AAML quantifier by the refinement quantifier, and some additional arguments around action models. Also, the \cite{hales2013arbitrary} presentation has parameterized quantifiers $\dia{\otimes_G}$ for $G\subseteq A$ instead of a single quantifier $\dia{\otimes}$, for technical reasons to make the translation work. We therefore do not give the axiomatization in this section but only in Section \ref{sec.rml} on refinement modal logic. We do not know any results on the complexity of satisfiability of AAML.

\subsection{Variations}

We describe two variations on arbitrary action model logic. They outline ideas for future research rather than survey published work.

\paragraph{Quantifying over actions with a fixed precondition}
A public announcement $\phi$ is an event identically observed by all agents, corresponding to a singleton action model with precondition $\phi$ and with the single action accessible for all agents (so the relation is reflexive, and the action model in class $\mathcal S5$). Now consider {\em any} epistemic action with precondition $\phi$. For example, any way that Anne can inform Bill of $p$, no matter whether this is private or public, whether a third agent Cath learns about it or not, and so on. Consider an additional inductive language construct $\dia{\phi^\otimes}\psi$ for `after any action informing the agents of $\phi$, $\psi$ (is true)'. Its semantics therefore are
\[ M_s \models \dia{\phi^\otimes} \psi \quad \text{iff} \quad M_s \models \phi \et \dia{\otimes} \psi \]
or, alternatively but equivalently,
\[ \begin{array}{lll} M_s \models \dia{\phi^\otimes} \psi & \text{iff} & \text{there is a } E_e \text{ with quantifier-free preconditions such that }  \\ && \pre(e)=\phi \text{ and } M_s \models \dia{E_e} \psi \end{array}\]
Such a quantification was suggested in \cite{aucher.planning:2012}. This was based on the similar \cite{aucher:2010}, that proposed instead of action models an {\em action language}, on a par with the static language. Another somewhat similar quantification is found in \cite{Lima14}, but for coalitional actions. The settings of \cite{aucher.planning:2012,aucher:2010,Lima14} use a different language than here, but inspired this presentation.

We now show that $\dia{\otimes}$ and  $\dia{\phi^\otimes}$ are interdefinable.

It is clear that $\dia{\phi^\otimes}$ is definable in the language $\lang(\Dia,\otimes,\dia{\otimes})$. Its semantics provide the definition:
\[ \dia{\phi^\otimes} \psi \quad \text{:=} \quad \phi \et \dia{\otimes} \psi \]

It is not so clear that $\dia{\otimes}$ is also definable in the language $\lang(\Dia,\otimes)$ expanded with the $\dia{\phi^\otimes}$ modalities, as these modalities appear to have an aspect of bundling/packing \cite{wang:2018}. Semantically, we now want that:  $\dia{\otimes}\psi$ iff (there is a $\phi$ such that $\dia{\phi^\otimes}\psi$). This is where (global) synthesis comes in handy: the AAML validity $\dia{\otimes}\psi\eq\dia{E_T}\psi$,  wherein multi-pointed action model $E_T$ is synthesized from the given formula $\psi$. Consider the following argument:

If $M_s \models  \dia{\otimes} \psi$, then by global synthesis $M_s \models \dia{E_T}\psi$, so that there is a $t \in T$ for which $M_s \models \dia{E_t}\psi$. From this we obtain that $M_s \models \pre(e)$ and also that $M_s \models \dia{\otimes}\psi$, and from both we obtain $M_s \models  \pre(t)\et\dia{\otimes}\psi$, that is, $M_s \models \dia{\pre(t)^\otimes} \psi$. Therefore $M_s \models \Vel_{t \in T} \dia{\pre(t)^\otimes} \psi$.

Vice versa, if $M_s \models \Vel_{t \in T} \dia{\pre(t)^\otimes} \psi$, then there is a $t \in T$ such that $M_s \models \dia{\pre(t)^\otimes} \psi$, and therefore, by the definability of $\dia{\phi^\otimes}\psi$, we get $M_s \models  \pre(t) \et \dia{\otimes} \psi$, from which it follows that $M_s \models  \dia{\otimes} \psi$.

We can therefore define $\dia{\otimes}\psi$ as

\[ \dia{\otimes} \psi \quad \text{:=} \quad \Vel_{t \in T} \dia{\pre(t)^\otimes} \psi \]  

It might be worthwhile to explore in depth such an action model logic for epistemic actions with given precondition.

\paragraph{Quantifying over private group announcements} 
We recall group announcement logic GAL where we quantify over known announcements simultaneously made by the agents in group $G \subseteq A$, and coalition announcement logic CAL where we quantify over known announcements simultaneously made by the agents in group $G \subseteq A$ no matter what (known) announcements are also simultaneously made by the agents not in $G$. We can also consider such quantifications over any informative action `involving agents in $G$'. That is, announcements only made by the agents in $G$ and that may not be public. Under such conditions of partial observability we require that agents make the same announcement in for them indistinguishable actions. This then represents that they know what they say.  Analogously to the GAL and CAL quantifiers over joint public announcements by agents, we could then consider quantifiers over $\mathcal S5$ action models satisfying that constraint. Let us merely give an example of such an epistemic action.

\begin{example}
Consider four agents Anne, Bill, Cath, and Dave. Anne and Bill have joint access to a bank account (or a nuclear storage facility, whatever \dots) by a secret code. Anne's secret code $p$ is only known to her and Bill's secret code $q$ is only known to him. They want to transfer their codes to Cath and Dave, respectively, without making them public. Clearly, the way to do this is that Anne privately informs Cath, and Bill privately informs Dave, and such that these actions are commonly known to all. Let $\Kw_i \phi$ (for `agent $i$ knows whether $\phi$') abbreviate $\Box_i\phi\vel\Box_i \neg\phi$. Then the specification of the problem is that \[ (\Kw_a p \et \neg \Kw_b p \et \Kw_b q \et \neg \Kw_a q) \imp \dia{\otimes_{ab}} (\Kw_c p \et \Kw_d q \et \neg \Kw_a q\et \neg \Kw_c q \et \neg \Kw_b p \et \neg \Kw_d p) \] The $\dia{\otimes_{ab}}$ quantifier informally represents that $a,b$ can simultaneously (that is, independently) make a private announcement, similarly to the $\dia{!_{ab}}$ quantifier in GAL wherein $a,b$ simultaneously make a public announcement. An epistemic action witnessing this $\dia{\otimes_{ab}}$ quantifier, that satisfies that Anne and Bill know what they say, is:
\begin{center}
\begin{tikzpicture}
\node (00) at (0,0) {$\Box_a \neg p \et \Box_b \neg q$};
\node (01) at (0,2) {$\Box_a \neg p \et \Box_b q$};
\node (10) at (4,0) {$\Box_a p \et \Box_b \neg q$};
\node (11) at (4,2) {$\Box_a p \et \Box_b q$};
\path[-] (00) edge node[above] {$bd$} (10);
\path[-] (01) edge node[above] {$bd$} (11);
\path[-] (00) edge node[left] {$ac$} (01);
\path[-] (10) edge node[left] {$ac$} (11);
\end{tikzpicture}
\end{center}
\end{example}
We omit the semantics for such action models and quantifiers from the survey. We conjecture that such logics with quantifiers over possibly private actions by groups of agents have complete axiomatizations with reduction axioms eliminating all dynamic features.\footnote{Namely by combining features of the axiomatizations of refinement epistemic  logic \cite{halesetal:2011} (see Section~\ref{sec.rml}, later), of AAML (see above), and of certain arrow update logics \cite{hvdetal.aus:2020} (see Section~\ref{sec.au}, next).} 

\subsection{Open problems and further research} \label{open.aaml}

The list of open problems for APAL and its variations, and for GAL and CAL, is longer and much more specific than the list below. This is more a feature than a problem: the logics in this section all permit reduction, so that they are decidable and as expressive as the base logic (K or S5). However, they presumably have widely varying complexities for satisfiability, model checking, synthesis, or planning. It may also be that the intuitions for action model dynamics are harder to grasp than for announcements, despite quantifying over the former being easier than quantifying over the latter.
\begin{itemize}
\item Synthesis for action model logic, for $\mathcal S5$ epistemic models and $\mathcal S5$ action models \cite{hales2013arbitrary}.
\item The axiomatization of the logic with quantification over private group announcements.
\end{itemize}

\section{Arrow updates} \label{sec.au}

In public announcement logic an updated model has a restricted domain, and this is not different in action model logic, where the update is a domain-restricted modal product.  Instead of domain restrictions we can also consider updating models by restricting accessibility relations. Already in the 1990s, relational restrictions have been the basis for dynamic epistemic logics \cite{gerbrandyetal:1997}. We focus on the more recent {\em arrow update logic} AUL \cite{kooietal:2011} (see also \cite{Kuijer15,kuijer:2015}) and {\em arrow update model logic} AUML \cite{kooirenne}, for both of which versions with quantification over information change have been proposed \cite{hvdetal.aij:2017,hvdetal.undecidable:2017,hvdetal.aus:2020}. 

Such relation-changing logics are more focussed on belief and belief change than on knowledge and knowledge change (you believe what holds in all accessible states, which may exclude the actual state). But we will see that arrow update logics also model  knowledge change. Arrow updates may then be more succinct than action models with the same update effect  \cite{kooietal:2011}.

For domain as well as relational restrictions we principally have those in mind that are with respect to modally definable subsets, i.e., denotations of formulas in $\lang(\Dia)$. Other restrictions are addressed in Section \ref{sec.sabotage}.

\subsection{Arrow update logics}

A pair in the accessibility relation of an epistemic model is also called an \emph{arrow}.  Indeed it is common to denote $(s,t) \in R$ (or $Rst$) as $s \imp t$, where the `$\imp$' symbol resembles an arrow. State $s$ is the \emph{source} of the arrow and $t$ is the \emph{target} of the arrow. In {\em arrow update logic} (AUL) \cite{kooietal:2011} one specifies which arrows are preserved, by way of specifiying what formulas should be satisfied at the source (state) of the arrow and the target (state) of the arrow. This determines the model transformation. It is called an {\em arrow update}. The logic AUL contains modalities for arrow updates.

While restricting the relations, in AUL the domain remains the same. There is also a version of arrow update logic wherein updated models can have larger modal complexity and in particular larger domains (representing higher-order uncertainty among the agents over the information change): just as action model logic is a generalization of public announcement logic, generalized arrow update logic or \emph{arrow update model logic} AUML has been proposed in \cite{kooirenne} as a generalization of AUL. It contains modalities for {\em arrow update models}. 

We will present both logics simultaneously by way of presenting arrow update model logic and defining arrow update logic as a special case. 

\paragraph{Arrow update model} 
Given a logical language $\lang$, an {\em arrow update model} $\aumodel$ is a pair $(\austates,\aufunction)$ where $\austates$ is a non-empty domain (set) of {\em outcomes} (also denoted $\Domain(\aumodel)$) and where $\aufunction$ is an {\em arrow relation} $\aufunction: A \imp \powerset((\austates\times\lang) \times (\austates \times\lang))$. 

For each agent $a$, the arrow relation links (outcome, formula) pairs to each other.  We write $\aufunction_a$ for $\aufunction(a)$, and we write $(o,\phi)\imp_a(o',\phi')$ for $((\austate,\phi),(\austate',\phi'))\in \aufunction_a$, or even, if the outcomes are unambiguous, $\phi\imp_a\phi'$. Formula $\phi$ is the {\em source condition} and formula $\phi'$ is the {\em target condition} of the $a$-labelled {\em arrow} from {\em source} $o$ to {\em target} $o'$. The arrow updates of AUL \cite{kooietal:2011} correspond to singleton pointed arrow update models. So in that case we can indeed write $\phi\imp_a\phi'$, as there is just one outcome.

Given an epistemic model $M = (S,R,V)$ and an arrow update model $(\austates,\aufunction)$, the updated model $M * \aumodel = (S',R',V')$ is defined as follows, where  $p \in P$, $a \in A$, $s,s'\in S$, and $o,o'\in O$:
\[ \begin{array}{lll}
S' & = & S \times \austates \\ 
((s,\austate), (s',\austate')) \in R'_a & \text{iff} & \text{there are } \phi,\phi'\in\lang \text{ such that:} \\ && (s,s') \in R_a, (\austate,\phi)\imp_a(\austate',\phi'), M_s \models \phi, \text{and } M_{s'} \models \phi' \\  V'(p) & = & V(p) \times \austates
\end{array} \]
Example \ref{ex.aul} demonstrates how to execute an arrow update for a singleton arrow update model with three arrows. 
\begin{example} \label{ex.aul}
Given initial uncertainty about $p$ for two agents Anne and Bill, a typical arrow update is the action wherein Anne ($a$) opens an envelope containing the truth about $p$ while Bill ($b$) observes Anne reading the contents of the letter. We preserve all arrows satisfying one of $p \imp_a p, \neg p \imp_a \neg p$, and $\T \imp_b \T$.\footnote{Formally, given arrow update model $(\{o\},\aufunction)$, we have that $\aufunction(a) = \{ ((o,p),(o,p)), ((o,\neg p),(o,\neg p)) \}$ and $\aufunction(b) = \{((o,\T), (o,\T)) \}$.} Therefore, only two arrows disappear, $\neg p \imp_a p$ and $p \imp_a \neg p$. It is depicted below, including the equivalent simplified visualization, for which it is harder to determine the update as the arrows are not explicit.

\begin{center}
\begin{tikzpicture}
\node (0) at (0,0) {$0$};
\node (1) at (2,0) {$\underline{1}$};
\draw[<->] (0) -- node[above] {$ab$} (1);
\draw[->] (0) edge[loop above,looseness=15] node[above] {$ab$} (0); 
\draw[->] (1) edge[loop above,looseness=9] node[above] {$ab$} (1); 
\node (t) at (3,0) {\large $\Imp$};
\node (0t) at (4,0) {$0$};
\node (1t) at (6,0) {$\underline{1}$};
\draw[<->] (0t) -- node[above] {$b$} (1t);
\draw[->] (0t) edge[loop above,looseness=15] node[above] {$ab$} (0t); 
\draw[->] (1t) edge[loop above,looseness=9] node[above] {$ab$} (1t); 
\end{tikzpicture}
\quad\quad\quad\quad\quad\quad
\begin{tikzpicture}
\node (0) at (0,0) {$0$};
\node (1) at (2,0) {$\underline{1}$};
\draw[-] (0) -- node[above] {$ab$} (1);
\node (t) at (3,0) {\large $\Imp$};
\node (0t) at (4,0) {$0$};
\node (1t) at (6,0) {$\underline{1}$};
\draw[-] (0t) -- node[above] {$b$} (1t);
\end{tikzpicture}
\end{center}
\end{example} 

Just as in action model logic, we wish to have a logical language and semantics with modalities for arrow updates, which requires their inclusion in an inductive definition of the logical language, subject to certain constraints in order to keep the language well-defined.

\paragraph{Syntax and semantics}
The language $\lang(\Dia,\uparrow)$ of {\em arrow update model logic} AUML has an additional inductive construct $\dia{U_o}\phi$ where $U = (O, \aufunction)$ with $o \in \austates$ is an arrow update model with $O$ finite and with $\aufunction_a$ finite for all $a \in A$, and with source and target conditions that have already inductively been defined as formulas. We read $\dia{U_o}\phi$ as `after executing arrow update $U_o$, $\phi$ (is true)'. 

In order to distinguish arrow update logic AUL from arrow update model logic AUML the language for singleton arrow update models is called $\lang(\Dia,\uparrow_1)$.

Given a pointed epistemic model $M_s$ and a pointed arrow update $U_o$, the semantics for the arrow update modality are as follows.
\[ \begin{array}{lcl} 
M_s \models \dia{\aumodel_\austate} \phi & \text{ iff } & (M * \aumodel)_{(s,\austate)} \models \phi \text{ where $M * \aumodel$ is defined as above} 
\end{array} \]
The arrow update modality is self-dual: $\dia{\aumodel_\austate} \phi \eq [\aumodel_\austate]\phi$. AUML is as expressive as the base modal logic K, because every formula in $\lang(\Dia,\uparrow)$ is equivalent to a formula in $\lang(\Dia)$, by a reduction. The crucial reduction axiom is as follows. The conjunction is over all `arrows' for agent $a$ between outcomes $o$ and $o'$ in the domain of $U$. We recall that there can be multiple arrows for each agent, such as the two arrows for agent $a$ in Example \ref{ex.aul}.
\[ {[U_o]}\Box_a \phi \eq \Et_{(o,\psi) \imp_a (o',\psi')} (\psi \imp \Box_a (\psi' \imp [U_{o'}]\phi)) \]
As an arrow update is a singleton arrow update model, this is also the reduction axiom for AUL. And just like AUML, also AUL is as expressive as K. A relevant detail concerning the respective reductions showing this, is that the composition of two arrow update models is again an arrow update model, which is fairly obvious (\cite{kooirenne}, details omitted), and that the composition of two arrow updates (singleton arrow update models) is again an arrow update, which is less obvious (\cite[Def.\ 3.1]{kooietal:2011}). Let us at least sketch the idea: if $\phi \imp_a \psi$ is in the first arrow update, called $U_o$, and $\phi' \imp_a \psi'$ is in the second arrow update, then $\phi\et[U_o]\phi' \imp_a \psi'\et[U_o]\psi'$ is an arrow in their composition. We note that the arrow in the composition is again with respect to the same, unique, outcome $o$.

Results for arrow update logics are for class $\mathcal K$ and not for class $\mathcal S5$. It is unknown how to obtain complete axiomatizations when restricting oneself to updates between models with equivalence relations. However, the $\mathcal S5$ version of refinement modal logic, called refinement epistemic logic, that will be presented in Section~\ref{rel}, gives reason for optimism. Arrow update logic AUL is more succinct than action model logic AML in the somewhat informal sense that certain arrow updates are exponentially smaller than action models with the same update effect \cite{kooietal:2011}. We call it informal because a (typically technically quite involved) proof of the result is lacking, although we conjecture that it holds.

Intuitively, AUML and AML come close, they are interchangeable formalisms for modelling multi-agent system dynamics, where it is hard to say where one `should be' preferred over the other. When restricted to $\mathcal S5$ models, the arrow updates of the logic AUL come intuitively very close to what are known as {\em semi-public announcements} (also known as {\em semi-private} --- is a glass half empty or is it half full?), such as when Alice shows her card to Bob only, with Cath watching \cite{hvd.jolli:2002}.
 
\subsection{Arbitrary arrow update logics} \label{sec.aaul}

A quantifier-free arrow update model is an arrow update model with all source and target conditions in $\lang(\Dia,\uparrow)$. We obtain {\em arbitrary arrow update logic} AAUL by adding an additional inductive construct $\dia{\AAUL_1}$ to language $\lang(\Dia,\uparrow)$, thus obtaining language $\lang(\Dia,\uparrow, \dia{\uparrow_1})$, and with semantics
\[ \begin{array}{l} 
M_s \models \dia{\AAUL_1} \phi \quad \text{ iff } \quad \text{ there is a quantifier-free singleton $U_o$ such that } M_s \models \dia{U_o} \phi
\end{array} \]
Similarly, we obtain {\em arbitrary arrow update model logic} AAUML by adding an additional inductive construct $\dia{\AAUL}$ to language $\lang(\Dia,\uparrow)$, thus obtaining language $\lang(\Dia,\uparrow, \dia{\uparrow})$, and with semantics
\[ \begin{array}{l} 
M_s \models \dia{\AAUL} \phi \quad \text{ iff } \quad \text{ there is a quantifier-free $U_o$ such that } M_s \models \dia{U_o} \phi
\end{array} \]
The logic AAUL has a complete axiomatization, is more expressive than the logic K, is incomparable in expressivity to APAL \cite{hvdetal.aij:2017}, and it is undecidable \cite{hvdetal.undecidable:2017}. Also, the model checking problem is PSPACE-complete, as for APAL. 

The logic AAUML also has a complete axiomatization, is as expressive as the logic K (by a reduction), and it is therefore decidable \cite{hvdetal.aus:2020}. The complexity of satisfiability and of model checking are unknown. The crucial reduction axiom of AAUML is as follows, where for each agent $a$, $\Phi_a$ is a set of formulas in $\lang(\Dia,!,\dia{\uparrow})$ with typical members $\phi_a$.\footnote{These formulas are also required to be in a particular form called `disjunctive negation normal form', DNNF, such that negations are only binding atoms. It can be shown that each formula in $\lang(\Dia,!,\dia{\uparrow})$ is equivalent to a formula in DNNF, so this is not restrictive. See \cite{hvdetal.aus:2020} for details.}
\[ \dia{\AAUL}\bigwedge_{a\in A}(\bigwedge_{\phi_a \in \Phi_a}\Dia_a\phi_a\wedge \square_a\psi_a) \quad \leftrightarrow \quad \bigwedge_{a\in A}\bigwedge_{\phi_a \in \Phi_a}\Dia_a\dia{\AAUL}(\phi_a\wedge\psi_a) \hspace{3cm}  (*) \]
If the quantifier binds a conjunction containing a Boolean formula, one uses that facts do not change in this logic, that is, that $\dia{\AAUL}\phi_0 \eq \phi_0$ is valid for Booleans $\phi_0$. Altogether we can treat any conjunction bound by a quantifier and again we have a reduction. A different incarnation of  axiom $(*)$ will feature in the next Section~\ref{sec.rml} on refinement modal logic.

\paragraph{Synthesis}
The logic AAUML permits (global) synthesis: given $\phi\in\lang(\Dia,\uparrow)$, there is an arrow update model $U_o$ (with a single point) such that $\dia{\AAUL}\phi\eq\dia{U_o}\phi$ is valid. We recall from the previous section that AAML permits synthesis as well, with the difference that one can only synthesize a multi-pointed action model from a given formula, not necessarily a single-pointed one.

\paragraph{Other arrow updates} A precursor of arbitrary arrow update logic AAUL is the logical semantics for knowability logic of \cite{wenetal:2011}. The quantification in \cite{wenetal:2011} is also over (equivalence) relational restrictions in a given model with constant domain. However, their quantifier is not bisimulation invariant, as the restriction is not parameterized by logical conditions such as the source and target formulas of arrow update logic. The authors discuss alternative solutions to the knowability paradox in their semantics. 

Finally, the dual of arrow preservation in arrow update logics is arrow removal, as in sabotage logic \cite{jfak.sabotage:2005} and various other relation-changing logics \cite{ArecesDFS17} presented in Section~\ref{sec.sabotage}.

\subsection{Open problems and further research}

Arrow updates often compare well intuitively with action models, and seem to permit succinct descriptions of knowledge dynamics. We merely list some topics for further research.

\begin{itemize}
\item Axiomatization of arrow update logic on $\mathcal S5$ models \cite{hvdetal.aij:2017}
\item Axiomatization of arrow update model logic on $\mathcal S5$ models \cite{kooirenne,hvdetal.aus:2020}
\item Synthesis for arrow update model logic on $\mathcal S5$ models \cite{hvdetal.aus:2020}
\end{itemize}

\section{Refinement modal logics}  \label{sec.rml}

Refinement modal logic (RML) \cite{bozzellietal.inf:2014}\footnote{But also \cite{hvdetal.loft:2009,hvdetal.felax:2010,halesetal:2011,hales:2011} predating the focal publication \cite{bozzellietal.inf:2014}, as well as \cite{hales:2016, bozzellietal.tcs:2014,achilleosetal:2013,xingetal:2019} coming later.} is a stranger in our midst, unless you get to know her better. Please allow us to make the proper introductions. Refinement modal logic does not contain announcements, nor (modalities for) action models or similar, but it contains quantifiers over information change. We recall knowability logic, wherein you can say `there is an announcement after which $\phi$' but without announcements in the logical language (Section~\ref{sec.apal.know}). That is intuitively appealing but theoretically problematic, as no axiomatization is known. In refinement modal logic you can (intuitively but not formally) say `there is an action model after which $\phi$', but without modalities for action models in the logical language. So this is as in arbitrary action model logic (Section~\ref{sec.aaml}), but without action models in the logical language. How to get rid of those? This is not problematic, but we need a detour through modal logic. Please read on.

\subsection{Refinement modal logic}

\paragraph{Refinement} In arbitrary public announcement logic we quantify over announcements, in arbitrary action model logic we quantify over action models, and in arbitrary arrow update (model) logic we quantify over arrow updates. All these quantifications have in common that the logical language contains modal operators for specific model transformations corresponding to the execution of epistemic actions, and the quantification is defined in terms of these modalities. In {\em refinement modal logic} RML \cite{bozzellietal.inf:2014} we quantify over the refinements of a given model. We recall (from Section~\ref{sec.prelims}) that, from the requirements {\bf atoms}, {\bf forth} and {\bf back} of a bisimulation, a refinement relation only needs to satisfy {\bf atoms} and {\bf back}; and that $M \succeq M'$ means that model $M'$ is a refinement of model $M$ (that is, that there is a refinement relation between the models $M$ and $M'$). The semantics of the refinement quantifier directly use this refinement relation. The language of refinement modal logic does not have modalities for the execution of epistemic actions. Although the refinement quantifier therefore does not quantify over epistemic actions, the relation to other quantifiers is so close that this warrants detailed discussion of RML in this survey. 

We start with some elementary examples illustrating the refinement relation.

\begin{example} \label{ex.refex}
Given the properties of the refinement relation, a refinement can be seen as a restriction of a bisimilar copy of a given model. Consider the models depicted below.
\begin{center}
\begin{tikzpicture}[->]
\node (m) at (5,0) {$M$};
\node (10) at (1,0) {$\bullet$};
\node (20) at (2,0) {$\bullet$};
\node (30) at (3,0) {$\bullet$};
\node (40) at (4,0) {$\bullet$};
\draw (10) -> (20);
\draw (20) -> (30);
\draw (30) -> (40);
\node (mp) at (5,-.6) {$M'$};
\node (10b) at (1,-.6) {$\bullet$};
\node (20b) at (2,-.6) {$\bullet$};
\node (30b) at (3,-.6) {$\bullet$};
\draw (10b) -> (20b);
\draw (20b) -> (30b);
\node (mpp) at (5,-1.2) {\pmb{$M''$}};
\node (00bb) at (0,-1.2) {$\bullet$};
\node (10bb) at (1,-1.2) {$\bullet$};
\node (20bb) at (2,-1.2) {$\bullet$};
\node (30bb) at (3,-1.2) {$\bullet$};
\draw[very thick] (10bb) -> (00bb);
\draw[very thick] (10bb) -> (20bb);
\draw[very thick] (20bb) -> (30bb);
\node (mppp) at (5,-1.8) {$M'''$};
\node (10bbb) at (1,-1.8) {$\bullet$};
\node (20bbb) at (2,-1.8) {$\bullet$};
\node (30bbb) at (3,-1.8) {$\bullet$};
\node (40bbb) at (4,-1.8) {$\bullet$};
\node (20n) at (0,-1.8) {$\bullet$};
\node (30n) at (-1,-1.8) {$\bullet$};
\node (40n) at (-2,-1.8) {$\bullet$};
\draw (10bbb) -> (20bbb);
\draw (20bbb) -> (30bbb);
\draw (30bbb) -> (40bbb);
\draw (10bbb) -> (20n);
\draw (20n) -> (30n);
\draw (30n) -> (40n);
\node (00x) at (1,-2.8) {$0$};
\node (10x) at (2,-2.8) {$1$};
\node (20x) at (3,-2.8) {$2$};
\node (30x) at (4,-2.8) {$3$};
\node (10nx) at (0,-2.8) {$-1$};
\node (20nx) at (-1,-2.8) {$-2$};
\node (30nx) at (-2,-2.8) {$-3$};

\end{tikzpicture}
\end{center}
We note that:
\begin{itemize}
\item $M'$ is a submodel (model restriction) of $M$.
\item $M''$ is a refinement of $M$ (arrows in bold font).
\item $M'''$ is a bisimilar copy of $M$.
\end{itemize}
Model $M''$ is a submodel of $M'''$ that is a bisimilar copy of $M$. Model $M''$ is not a submodel of $M$. A refinement is a model restriction of a bisimilar copy of a model. 

In fact, all of $M',M'', M'''$ are refinements of $M$. Let us give the refinement relations linking $M$ to its respective refinements. Consider the numbers below the three models. Let the names of the states be those numbers, possibly primed, or doubly or triply primed. Assume that all states in all models have the same valuation of atoms. Let $R^\sharp$ be the accessibility relation in a model $M^\sharp$, where $\sharp$ is one of $'$, $''$, $'''$, and consider the relation $Z^\sharp$ between $M$ and $M^\sharp$ that relates any state $i$ to $i^\sharp$ in case $i^\sharp$ exists and also to $-i^\sharp$ in  case it exists. Then all such $Z^\sharp$ are refinement relations. Relation $Z'''$ is also a bisimulation. We merely illustrate why the relation $Z''$ between $M$ and $M''$ defined as above is a refinement relation. Note that $Z'' = \{(0,0''),(1,1''), (1,-1''), (2,2'')\}$. This relation is a refinement, because it satisfies the requirements {\bf atoms} and {\bf back}. Crucially, $Z''$ it is not a bisimulation, because it does not satisfy {\bf forth}: observe that $(2,2'') \in Z''$ and $(2,3) \in R$, but there is no state $t$ in $M''$ such that $(3,t) \in Z''$ and $(2'',t) \in R''$.
\end{example}

\begin{example}
Two models can be each other's refinement, but not bisimilar. Consider the models $M$ and $N$ depicted below. The dashed relation is a refinement between $N$ and $M$, where we recall the notation $N \lumis M$, and the dotted relation is a refinement between $M$ and $N$, that is, $M \lumis N$. Still, $M$ and $N$ are not bisimilar.
\begin{center}
\begin{tikzpicture}
\node (m) at (-1,0) {$M:$};
\node (10) at (0,0) {$7$};
\node (20) at (2,0) {$4$};
\node (30) at (4,0) {$5$};
\node (40) at (6,0) {$6$};
\node (n) at (-1,-2) {$N:$};
\node (2m) at (2,-2) {$1$};
\node (3m) at (4,-2) {$2$};
\node (4m) at (6,-2) {$3$};
\draw[<-] (10) -> (20);
\draw[->] (20) -> (30);
\draw[->] (30) -> (40);
\draw[->] (2m) -> (3m);
\draw[->] (3m) -> (4m);
\draw[dashed,bend right = 15,<-] (10) to (3m);
\draw[dotted,bend left,->] (20) to (2m);
\draw[dashed,bend right,<-] (20) to (2m);
\draw[dotted,bend left,->] (30) to (3m);
\draw[dashed,bend right,<-] (30) to (3m);
\draw[dotted,bend left,->] (40) to (4m);
\draw[dashed,bend right,<-] (40) to (4m);
\end{tikzpicture}
\end{center}
\end{example}

In \cite{bozzellietal.inf:2014} and most related works on refinement modal logic, instead of the refinement relation as above (for the set of all agents) the refinement relation has a parameter $G \subseteq A$, and apart from {\bf atoms} and {\bf back} (for all agents), additionally, {\bf forth} is required for the agents in $A \setminus G$. As this was partly for technical reasons, in this survey we only consider refinement for the set $A$ of all agents, which simplifies the presentation. 

\paragraph*{Language and semantics of refinement modal logic}
The language $\lang(\Dia,\rmldia)$ of refinement modal logic is the language $\lang(\Dia)$ of multi-agent modal logic extended with an inductive construct $\rmldia\phi$, representing `there is a refinement after which $\phi$'. It has the following semantics. 
\[ M_s \models \rmldia \phi \quad \mbox{iff} \quad  \text{there is } M'_{s'} \text{ such that } M_s \lumis M'_{s'} \text{ and } M'_{s'} \models \phi \]
As the refinement relation is transitive, reflexive, and confluent (for all $x,y,z$, if $x \succeq y$ and $x \succeq z$, there is $w$ such that $y \succeq w$ and $z \succeq w$), we  have for the refinement modality the corresponding validities:
\begin{itemize}
\item $\rmlbox \phi \imp \phi$ \quad ({\bf T})
\item $\rmldia \rmldia \phi \imp \rmldia \phi$ \quad ({\bf 4}) 
\item $\rmldia \rmlbox \phi \imp \rmlbox \rmldia \phi$ \quad (Church-Rosser) / CR
\end{itemize}
The refinement quantifier also satisfies: \begin{itemize} \item $\rmlbox \rmldia \phi \imp \rmldia \rmlbox \phi$ \quad (McKinsey) / MK \end{itemize} As this has not been observed in the literature, it requires proof. We recall the relational property of \emph{atomicity} corresponding to {\bf 4} plus MK, and that it is valid in APAL (Section \ref{sec.apal.know}). The proof that MK is valid for RML is simpler. 

Towards a contradiction assume that MK does not hold. Then there is an epistemic model $M_s$ and a $\phi$ such that $M_s \models \rmlbox \rmldia \phi$ as well as $M_s \models \rmlbox \rmldia \neg\phi$, where $M  = (S,R,V)$ and $s \in S$. Now consider $M^\emptyset= (S, R^\emptyset, V)$ that is as $M$ except that for all $a \in A$, $R^\emptyset_a = \emptyset$: we remove all pairs from all accessibility relations. We note that $M \succeq M^\emptyset$ and in particular $M_s \succeq M^\emptyset_s$. As all relations are empty, $M^\emptyset_s \models \rmldia \phi$ iff $M^\emptyset_s \models \phi$ : the unique refinement of an empty relation is the empty relation. Now $M_s \models \rmlbox \rmldia \phi$ and $M_s \succeq M^\emptyset_s$ imply $M^\emptyset_s \models \rmldia \phi$, and in view of the above, $M^\emptyset_s \models \phi$. Also, $M_s \models \rmlbox \rmldia \neg\phi$ and $M_s \succeq M^\emptyset_s$ imply $M^\emptyset_s \models \rmldia \neg\phi$, and therefore $M^\emptyset_s \models \neg\phi$. We have obtained a contradiction.

Just as a refinement of a given model is a restriction of a bisimilar copy of that model, refinement quantification is bisimulation quantification followed by relativization. In {\em bisimulation quantified} logics we have explicit quantifiers $\all p$ over propositional variables $p$ \cite{french:2006,visser:1996}. {\em Relativization} $\phi^p$ of a formula $\phi$ to an atomic proposition $p$ is a syntactic way to describe model restrictions, such as the consequences of a public announcement: $(M|p)_s \models \phi$ iff $M_s \models \phi^p$, which of course in PAL is the same as $M_s \models \dia{p}\phi$, see \cite{milleretal:2005,jfak.book:2011}. In refinement modal logic, given $\phi \in \lang(\Dia)$, we have \cite[Sect.\ 4.3]{bozzellietal.inf:2014}: \[ \rmldia\phi \text{ is equivalent to } \is p \dia{p}\phi \] 
Example~\ref{ex.refex} was an illustration: given $M$, we first take its bisimilar copy $M'''$ and make an atom $p \in P$ true (only) in states $-1''',0''',1''',2'''$. Then announce $p$. Model $M'$ now results.

This equivalence is a non-trivial result, because announcing an atom is far more restrictive than announcing an arbitrary formula, which in its turn is far more restrictive than executing an arbitrary epistemic action, such as a pointed action model. We will now discuss that.

\paragraph{Relation between action models and refinement} 
\begin{quote} {\em Executing an action model produces a refinement of the initial model, and for every refinement of a finite model there is an action model producing it.} \end{quote}
We sketch the proof found in \cite{bozzellietal.inf:2014}.

Given pointed model $M_s$ and epistemic action $E_e$, the resulting $(M \otimes E)_{(s,e)}$ is a refinement of $M_s$ by way of the relation $Z$ consisting of all pairs $(t,(t,f))$ such that $M_t \models \pre(f)$. Some states of the original model may get lost in the modal product, namely if there is no action whose precondition can be executed there. But all `surviving' (state,action)-pairs simply can be traced back to their first argument: clearly a refinement.

For the other direction, let refinement $N_t$ of a {\em finite} model $M_s$ be given, and let $Z$ be the refinement relation. Consider action model $E$ that is isomorphic to $N$ (for convenience of the presentation we let the names of actions and states correspond), and wherein the valuation of propositional variables in a state $t'$ of model $N$ is replaced by a precondition $\pre(t') \in \lang(\Dia)$ of action $t'$ that is the distinguishing formula of the set of states $s'$ in $M$ such that $(s',t') \in Z$. As $M$ is finite, this distinguishing formula exists \cite{browneetal:1987,jfak.odds:1998}. We now have that $M_{s'} \models \pre(t')$ iff $(s',t') \in Z$, and that $(M  \otimes E)_{(s,t)}$ is bisimilar to $N_t$. 

\begin{example} \label{ex.actrml}
Recalling Example \ref{ex.aml} on page \pageref{ex.aml}, the picture below shows once more  the execution of an action model, now including the refinement relation (visualized as dashed arrows) between the initial model and the updated model.

\begin{center}
\begin{tikzpicture}
\node (00) at (0,0) {$0$};
\node (10) at (2,0) {$\underline{1}$};
\node (t) at (2.75,0) {\Large$\otimes$};
\node (110) at (3.5,0) {$\underline{p}$};
\node (120) at (3.5,2) {$\top$};
\node (e) at (4.5,0) {\Large$=$};
\node (30) at (7,0) {$\underline{(1,p)}$};
\node (21) at (5,2) {$(0,\top)$};
\node (31) at (7,2) {$(1,\top)$};
\draw (00) to node[above]{$ab$} (10);
\draw (21) to node[above]{$ab$} (31);
\draw (30) to node[left]{$b$} (31);
\draw (110) to node[left]{$b$} (120);
\end{tikzpicture}
\quad\quad\quad
\begin{tikzpicture}
\node (00) at (0,0) {$0$};
\node (10) at (2,0) {$\underline{1}$};
\node (30) at (6,0) {$\underline{1}$};
\node (21) at (4,2) {$0$};
\node (31) at (6,2) {$1$};
\draw (00) to node[above]{$ab$} (10);
\draw (21) to node[above]{$ab$} (31);
\draw (30) to node[left]{$b$} (31);
\draw[->,dashed,bend left =10] (00) to  (21);
\draw[->,dashed,bend left =10] (10) to  (31);
\draw[->,dashed,bend left =10] (20) to  (30);
\end{tikzpicture}
\end{center}
\end{example}

The restriction to finite models is important and the result does not hold on all models. For a counterexample, consider the model $M$ consisting of all (infinitely many) valuations of a countably infinite set of atoms $P$ and that are all indistinguishable by a single agent $a$. The accessibility relation for $a$ is the universal relation on that domain. Now consider model $M^=$ that is like $M$ except that the accessibility relation for $a$ is the identity relation. Note that $M^=$ is a refinement of the model $M$. You can get from $M$ to $M^=$ by the epistemic action of announcing the valuation, in whatever state you are. But for that you have to announce the value of infinitely many atoms, and an infinite conjunction is not a formula. It cannot be a precondition of an action. There is no action model producing $M^=$. This appears to make the relation with action model logic problematic. But in fact it is rather straightforward once you take the perspective of a given formula. A given $\phi \in \lang(\Dia)$ bound by the refinement modality only contains a finite number of atoms. Therefore you can announce the values of all the atoms occurring in $\phi$, in the form of an action model or an arrow update. With additional manipulations similar to those above one can then prove that \cite[Theorem 28]{hvdetal.aus:2020}:
\[ \begin{array}{lll}
\rmldia\phi & \text{is equivalent to} & \dia{\otimes}\phi; \\
\rmldia\phi & \text{is equivalent to} & \dia{\uparrow}\phi; \\
\dia{\uparrow}\phi & \text{is equivalent to} & \dia{\otimes}\phi. 
\end{array}\]
It is also reported there that this result generalizes to arbitrary formulas $\phi$ in the respective languages, by a proper inductively defined translation. Therefore, these three quantifiers are interchangeable in any formula, by an argument involving natural induction over quantifier occurrence, and reduction axioms. It is interesting to observe that, even though the three logics AAML, AAUML, and RML are all as expressive as the logic K, they differ in so-called {\em update expressivity}. Given the model class of interpretation, this is the relation between pointed models induced by the respective updates (a typical update that can only be expressed in RML is the one producing $M \succeq M^=$ above). For such matters see \cite{Kuijer15} and \cite[Section 5]{hvdetal.aus:2020}.

\paragraph*{Axiomatization}

The axiomatization of RML reported in \cite{bozzellietal.inf:2014} is in terms of parametrized refinement quantifiers $\dia{\lumis_a}$ for $a \in A$ (interpreting $a$-refinement relations $\lumis_a$, where we recall that in this survey we only consider $\lumis$). There is an alternative axiomatization for RML with only refinement quantifiers $\lumis$ for the set $A$ of all agents. It consists of the axiomatization of arbitrary arrow update model logic in \cite[Section 4, Theorem 19]{hvdetal.aus:2020} (see the previous Section~\ref{sec.aaul}), wherein we replace all $\dia{\uparrow}$ quantifiers by $\dia{\succeq}$ quantifiers, and remove all reduction axioms involving arrow update models. This is justified by recalling the above-mentioned equivalence of $\dia{\uparrow}\phi$ and $\dia{\succeq}\phi$ and its generalization to arbitrary formulas. It is simpler than the axiomatization in \cite{bozzellietal.inf:2014}. We only give the crucial reduction axiom for the $\rmldia$ quantifier. It is obtained as indicated above from the axiom $(*)$ of AAUML in Section~\ref{sec.aaul}.
\[\begin{array}{ll}
\langle \lumis\rangle\bigwedge_{a\in A}(\bigwedge_{\phi_a \in \Phi_a}\Dia_a\phi_a\wedge \square_a\psi_a)\leftrightarrow  \bigwedge_{a\in A}\bigwedge_{\phi_a \in \Phi_a}\Dia_a\langle \lumis\rangle(\phi_a\wedge\psi_a) \hspace{3cm} \hfill (**) 
\end{array} \]
%
Given such reductions, the logic RML is therefore as expressive as the multi-agent modal logic K, and it is therefore also decidable, unlike APAL. 

The complexity of RML is AEXP$_{pol}$ for single-agent version \cite{bozzellietal.tcs:2014}. AEXP$_{pol}$ is the class of problems solvable by alternating Turing machines running in single exponential time but only with a polynomial number of alternations. The complexity is conjectured to be non-elementary for multi-agent RML. In \cite{achilleosetal:2013} a PSPACE upper bound is given for complexity of the \emph{existential fragment} of (single-agent) RML: this is a negation normal form except that only $\dia{\succeq}$ and not $[\succeq]$ may occur. Apart from refinement modal logic RML, there is also refinement $\mu$-calculus, for various results on this $\mu$-RML see \cite{bozzellietal.inf:2014}. For example, the complexity of satisfiability of $\mu$-RML is non-elementary.

\subsection{Refinement epistemic logic} \label{rel}
The semantics of the refinement quantifier depend on the class of models it quantifies over. In particular, quantifying over models from the class ${\mathcal K}$ is different from quantifying over models from the class $\mathcal S5$, a class of interest as the $\Box_a$ modalities then represent knowledge. The semantics are straightforward. Let $M$ be an $\mathcal S5$ model, $s \in \Domain(M)$ and $\phi \in \lang(\Dia,\lumis)$, then \[ M_s \models \rmldia \phi \quad \mbox{iff} \quad  \text{there is \pmb{an $\mathcal S5$ model} } M'_{s'} \text{ such that } M_s \lumis M'_{s'} \text{ and } M'_{s'} \models \phi \]
But the axiomatization is not straightforward. The axiomatization of refinement modal logic on models with particular frame properties does not extend the axiomatization with respect to $\mathcal K$. Indeed, the axiom $(**)$ above is unsound on class $\mathcal S5$.

\begin{example} \label{next}
In axiom $(**)$, let $\Phi_a = \{\Box_a p, \neg \Box_a p\}$ and $\psi_a = \top$, and consider a two-state model $N$ wherein agent $a$ does not know $p$ and $p$ is actually true. Then \[ \Dia_a\rmldia\Box_ap \et \Dia_a\rmldia\neg\Box_ap \text{ is true:} \] Agent $a$ considers it possible that after having been informed of $p$ she knows that $p$, and she also considers it possible that after not having been informed of anything she still does not know $p$. But  \[ \dia{\lumis}(\Dia_a\Box_a p \et \Dia_a \neg\Box_a p) \text{ is false:} \] Consider towards a contradiction a hypothetical model that is a refinement of $N$ and wherein $\Dia_a\Box_ap \et \Dia_a\neg\Box_ap$ is true. The {\bf S5} validity $\Box_a\neg\Box_a \phi \eq \neg\Box_a \phi$, that is, $\Box_a \phi \eq \Dia_a\Box_a \phi$, tells us that whenever you see $\Dia_a\Box_a$ you can delete $\Dia_a$. The {\bf S5} validity $\Box_a\Box_a \phi \eq \Box_a \phi$ tells us that whenever you see $\Box_a\Box_a$ you can delete a $\Box_a$. Further note that $\Dia_a\neg\Box_ap$ is equivalent to $\neg\Box_a\Box_ap$. Because of all that, $\Dia_a\Box_ap \et \Dia_a\neg\Box_ap$ is equivalent to $\Box_ap\et\neg\Box_ap$, which is inconsistent.
\end{example}

However, refinement epistemic logic \emph{has} been axiomatized, and the axiomatization is again a reduction system, namely to the logic S5. Axiomatizations for refinement modal logic on a large variety of model classes, such as the above for $\mathcal S5$, can be found in \cite{halesetal:2011, hales:2016}. A welcome member there, called {\em refinement doxastic logic}, is an axiomatization for RML on class $\mathcal KD45$, the serial, transitive, and Euclidean frames, that are typically used to model belief. There are some surprises and open questions: {\em transitive} refinement modal logic (RML on class $\mathcal K4$, the transitive frames) is (strictly) more expressive than the modal logic K4 that it extends, and its axiomatization is an open question. This can be found in an unpublished manuscript by Tim French.

\subsection{Simulation modal logic}

Instead of the language $\lang(\Dia,\rmldia)$ of refinement modal logic with the semantics \[ M_s \models \rmldia \phi \quad \mbox{iff} \quad  \text{there is } M'_{s'} \text{ such that } M_s \lumis M'_{s'} \text{ and } M'_{s'} \models \phi \]
now consider the language $\lang(\Dia,\langle\preceq\rangle)$ of {\em simulation modal logic} with the semantics
\[ M_s \models \dia{\preceq} \phi \quad \mbox{iff} \quad  \text{there is } M'_{s'} \text{ such that } M_s \simul M'_{s'} \text{ and } M'_{s'} \models \phi \]
This operator quantifies over the simulations of a given model instead of over its refinements. This logic has been investigated and axiomatized in \cite{xingetal:2019}, in a language and semantics with quantifiers that combine refinement and simulation for subsets of the set of all agents. To highlight how their results compare to RML we only consider simulation for the set of all agents. We now recall the $(**)$ axiom of RML (and the $(*)$ axiom of AAUML), juxtaposed to the corresponding reduction $(***)$ for simulation that is extrapolated from \cite[Axiom {\bf CCRKco2}]{xingetal:2019} to emphasize their resemblance. 
\[\langle \lumis\rangle\bigwedge_{a\in A}(\bigwedge_{\phi_a \in \Phi_a}\Dia_a\phi_a\wedge \square_a\psi_a)\leftrightarrow  \bigwedge_{a\in A}\bigwedge_{\phi_a \in \Phi_a}\Dia_a\langle \lumis\rangle(\phi_a\wedge\psi_a) \hspace{3cm} (**)  \]
\[ \langle \simul\rangle\bigwedge_{a\in A}(\bigwedge_{\phi_a \in \Phi_a}\Dia_a\phi_a\wedge \square_a\psi_a)\leftrightarrow  \bigwedge_{a\in A}\Box_a \Vel_{\phi_a \in \Phi_a}\langle \simul\rangle(\phi_a\wedge\psi_a) \hspace{2.5cm} (***) \]
Simulation modal logic quantifies over models with more uncertainty than the current one. Refinement preserves  the positive formulas: hard knowledge. Whereas simulation preserves ignorance. It seems worthwhile to investigate simulation modal logic in depth, in particular on the class $\mathcal S5$: \emph{simulation epistemic logic}, and to obtain synthesis for that.

\subsection{Open problems and further research}

Refinement modal logic and its variations have not received as much attention from the research community as APAL and its variations, despite theoretical properties that compare well (decidability, reduction, \dots). We recall the similar story about arbitrary action model logic spelled out in Section~\ref{open.aaml}. There, we conjectured that this has to do with non-public dynamics being harder to grasp than public dynamics, despite quantifying over non-public dynamics being so much more elegant than quantifying over public dynamics. Maybe this gap can be closed for RML by considering yet other variations, or applications: dynamic modalities that are interpreted as pruning tree-representations of models might be suitable for various other structures than epistemic models and for extending various other logics. Given refinement modal $\mu$-calculus ($\mu$-RML) presented in \cite{bozzellietal.tcs:2014}, two such candidates for extension with refinement modalities are CTL \cite{EmersonC82,EmersonH85}, where a relation with QCTL, quantified CTL, is to be expected \cite{LaroussinieM14}, and PDL \cite{fisheretal:1979,hareletal:2000}.

Open problems and further research include:
\begin{itemize}
\item Investigating refinement CTL and refinement PDL, and their expressivity \cite{bozzellietal.tcs:2014}
\item The axiomatization of simulation modal logic for a single simulation quantifier for all agents \cite{xingetal:2019,bozzellietal.tcs:2014}
\item The axiomatization of simulation epistemic logic on class $\mathcal S5$ \cite{xingetal:2019,bozzellietal.tcs:2014}
\item Synthesis for different refinement and simulation modal logics \cite{hales2013arbitrary,hvdetal.aus:2020}
\end{itemize}

\section{Iteration, factual change, group epistemics, and planning} \label{sec.iterate}

The final two sections of this survey are more exploratory in nature. They are less a survey on work done, and more about further research, because little work has been done in dynamic epistemic logic on those subjects involving quantification over information change. In Section~\ref{sec.iterate} we focus on topics central in dynamic epistemic logic but for which extensions with quantifiers have not or rarely been considered. In Section~\ref{sec.sabotage} we focus on update logics (model changing logics) with quantifiers but that are not dynamic epistemic logics. 

\subsection{Iterated updates} \label{sec.itt}

\paragraph*{Standard program operations on updates}

So far, we always considered dynamic modalities $\dia{\pi}$, where $\pi$ is an \emph{update}, inducing transformation of (`updating') an epistemic model, not necessarily deterministically. We can see the update as a program. In public announcement logic the programs are the announcements, in action model logic the programs are the pointed finite action models, and so on. We can then wonder about standard program operations, as in PDL: test, choice, sequential execution, and arbitrary iteration (`Kleene star') \cite{hareletal:2000}. Except for the arbitrary iteration, the main topic of this section, such program operations are syntactically definable in dynamic epistemic logics, and (maybe therefore) not commonly considered.
\[\begin{array}{lll}
\dia{?\psi}\phi & \eq & \psi \et \phi \\
\dia{\pi \cup \pi'}\phi & \eq & \dia{\pi}\phi \vel \dia{\pi'}\phi \\
\dia{\pi \ ; \ \pi'}\phi & \eq & \dia{\pi}\dia{\pi'}\phi
\end{array}\]

\paragraph*{Test}
The semantics for the modality for a test $?\psi$ contrasts well with the semantics for public announcement of $\psi$ that we already know:
\[ \begin{array}{lll}
M_s \models \dia{?\psi}\phi & \text{iff} & M_s \models \psi \text{ and } M_s \models \phi \\
M_s \models \dia{\psi}\phi & \text{iff} & M_s \models \psi \text{ and } (M|\psi)_s \models \phi 
\end{array}\]
If a test $?\psi$ can be executed, $\psi$ must have been true, and as the resulting model is the same as the original one, any postcondition $\phi$ is true if $\phi$ already was true. This explains that $\dia{?\psi}\phi \eq (\psi \et \phi)$ is valid, and that a test is definable in dynamic epistemic logic. A public announcement is a test on $\psi$ {\em that is observed by all agents} and thus resulting in a different model in which to interpret $\phi$, whereas the test $?\psi$ is not observed by any agent, and thus resulting in the same model in which to interpret $\phi$.

\paragraph*{Choice}
Non-determistic choice between programs does not commonly feature in dynamic epistemic logics because it is definable as $\dia{\pi \cup \pi'}\phi \ \eq \ \dia{\pi}\phi \vel \dia{\pi'}\phi$. The box-version of that equivalence, $[\pi \cup \pi']\phi \ \eq \ [\pi]\phi \et [\pi']\phi$, is considered less intuitive because the choice $\cup$ becomes a conjunction $\et$. For a finite set $\Pi$ of programs we thus get that $\dia{\Union_\Pi \pi}\phi \eq \Vel_\Pi \dia{\pi}\phi$. Indirectly we have already seen this with the multi-pointed action models: $[E_F]\phi \eq \Et_{e \in F} [E_e]\phi$ and in particular $[E]\phi \eq \Et_{e \in \domain(E)} [E_e]\phi$. For an infinite set $\Pi$ of programs, such as `all announcements' ($\Pi = \lang(\Dia,!)$) or `all finite pointed action models', we can still consider $\Union_\Pi \pi$ as a semantic object, but $\Vel_\Pi \dia{\pi}$ would be an infinite disjunction, which is not in the logical language. However, for both finite and countably infinite $\Pi$  we can consider $\dia{\Union_\Pi \pi}\phi$ as a quantification $\dia{\bullet}\phi$. If $\Pi = \lang(\Dia,!)$ we get $\dia{\bullet}=\dia{!}$ and if $\Pi$ is the set of all finite pointed action models we get $\dia{\bullet}=\dia{\otimes}$.

\paragraph*{Sequential update}

Given that $\dia{\pi \ ; \ \pi'}\phi \eq \dia{\pi}\dia{\pi'}\phi$, if $\pi=\pi'$ we get $\dia{\pi \ ; \ \pi}\phi \eq \dia{\pi}\dia{\pi}\phi$, in other words, $\dia{\pi^2}\phi \eq \dia{\pi}^2\phi$. Similarly, for any $n \in \Naturals$, $\dia{\pi^n}\phi \eq \dia{\pi}^n\phi$, where $\dia{\pi^0}\phi = \dia{\pi}^0\phi := \phi$. In particular, for iteration of public announcements we have $\dia{\phi}^n\psi \eq \dia{\phi^n}\psi$, and for iteration of epistemic actions $E_e$ we get that $\dia{E_e}^n\psi \eq \dia{E_e^n}\psi$. This brings us to arbitrary iteration.

\paragraph*{Arbitrary iteration}
Let us now define \begin{center} $\dia{\pi^*}\phi$ \qquad iff \qquad there is $n \in \Naturals$ such that $\dia{\pi^n}\phi$.\end{center} In view of the above we write $\dia{\pi^*}\phi$ and $\dia{\pi}^*\phi$ interchangeably. We get a very different form of quantification over information change if we merely consider arbitrary iteration of a given update $\pi$, and dynamic epistemic languages with so-called infinitary (dynamic) modalities for such arbitrary iteration.  Such iterated updates have rarely been investigated as part of the dynamic epistemic logical language. The exception is \cite{milleretal:2005}, wherein it is shown that already for very simple settings ($\mathcal K$ models, single-agent) the logic become undecidable. Axiomatizations of such logics are also unknown. Possibly for such reasons, the dynamic epistemic logic community has stayed away from arbitrarily iterated updates in the language.\footnote{Note that the arbitrary iteration in epistemic PDL of \cite{jfaketal.lcc:2006}, often presented as a dynamic epistemic logic dialect, is not over the updates. The iteration is over tests $?\phi$ and epistemically interpreted modalities, in order to formalize relativized common knowledge. Whereas the extension of that language with updates, that begets the {\em Logic of Communication and Change} of \cite{jfaketal.lcc:2006}, does not allow iteration of updates.}

On the other hand, iterated updates have received much attention in dynamic epistemic logics in purely semantic investigations. These are `off-line' (meta-level) analyses of the consequences of iterated dynamics on given epistemic models described with the language $\lang(\Dia)$. The more interesting of such proposals involve Moorean phenomena wherein the truth value of knowledge or ignorance propositions keeps flipping back and forth. Let us name a few. 

\begin{itemize}
\item 
A most publishable unpublished work is \cite{sadzik:2006}, wherein updated epistemic models that are the result of executing arbitrarily iterated finite action models display fascinating oscillations between stable points reached during such iterations. 

\item 
In \cite{baltagetal.tark:2009}, belief dynamics under iterated revision is explored (belief revision is modelled as a particular form of update on enriched epistemic models). This developed further into, or rather merged with, {\em learning theory} investigating `asymptotic' knowledge properties of epistemic models after arbitrary iteration of action models \cite{BaltagGS19,BolanderG18}. 

\item 
Iterated sincere and insincere updates (lying) are modelled in \cite{agotnesetal.truelie:2017}, and it is shown that you can keep changing your opinion on some matter of interest in arbitrarily haphazard and irregular ways up to infinity (not very encouraging for everyday politics). 

\item 
In {\em gossip protocols} the informative updates represent telephone calls, where the denotation of such a protocol is a set of possibly infinite call sequences. However, given the restricted setting, the informative consequences of call sequences are bounded: stable information points again, in execution trees generated by such protocols \cite{AptW18,logicofgossiping:2020}. Such analysis are PDL-style analyses, and although such calls can be modelled as action models \cite{gattinger.phd:2018}, it is not entirely clear  how to associate a gossip protocol with a `dynamic epistemic logic style' update modality, except in the case of protocols where all executions are finite.

\item Meta-level iterated updates are also a main focus of epistemic planning in dynamic epistemic logic, to which a separate discussion is devoted later in this section.

\end{itemize}

\paragraph*{Iterated arbitrary updates}
Let now $\dia{\bullet}$ be any quantifier.

Most of our quantifiers satisfy $\phi\imp\dia{\bullet}\phi$, as there is a trivial, non-informative, update. This is the announcement of the always true formula $\top$, for which there also is a corresponding action model, arrow update, group announcement, etcetera. For this trivial $\pi$ the equivalence $\phi\eq\dia{\pi}\phi$ is valid. With implication $\dia{\pi}\phi\imp\dia{\bullet}\phi$ that gives us $\phi \imp \dia{\bullet}\phi$. 

Most of our quantifiers also satisfy $\dia{\bullet}\phi\imp\dia{\bullet}\dia{\bullet}\phi$, which can be obtained from  $\phi \imp \dia{\bullet}\phi$ by elementary further manipulations. 

And most quantifiers also satisfy  $\dia{\bullet}\dia{\bullet}\phi\imp\dia{\bullet}\phi$. This holds when the composition of two updates is again an update (of the same kind). For most of the quantifiers we have seen, this is indeed the case:

A sequence of two announcements $\phi$ and $\psi$ is again an announcement, namely that of $\dia{\phi}\psi$. We recall that a PAL validity is $\dia{\phi}\dia{\psi}\chi \eq \dia{\dia{\phi}\psi}\chi$ \cite{hvdetal.del:2007}. 
A sequence of two action models is again an action model, namely the composition of those two action models \cite{baltagetal:1998} (see Section \ref{sec.prelims}). A sequence of two arrow updates is an arrow update \cite{kooirenne}, and the relational composition of two refinements or of two simulations is again a refinement respectively a simulation \cite{bozzellietal.inf:2014} (pruning a tree, and pruning it again, is still pruning a tree; similarly, growing a tree and letting it grow more, is still growing a tree).

From $\dia{\bullet}\phi\imp\dia{\bullet}\dia{\bullet}\phi$ and $\dia{\bullet}\dia{\bullet}\phi\imp\dia{\bullet}\phi$ follows  $\dia{\bullet}\phi\eq\dia{\bullet}\dia{\bullet}\phi$, and therefore $\dia{\bullet}\phi\eq\dia{\bullet}^n\phi$ for any positive $n \in \mathbb N$. This gets us $\dia{\bullet}\phi\eq\dia{\bullet}\dia{\bullet}^*\phi$, which is written as $\dia{\bullet}\phi\eq\dia{\bullet}^+\phi$.

We also have that $\dia{\bullet}(\phi \vel \psi) \ \eq \ \dia{\bullet}\phi \vel \dia{\bullet}\psi$. 

Considering program operations on modal quantifiers therefore does not have added value. In contrast, the investigation of dynamic epistemic logics with arbitrary iteration of epistemic actions and with additionally quantifiers over such epistemic actions seems worthwhile to pursue.

\subsection{Factual change} \label{factual}

Dynamic epistemic logic is typically a logic wherein, in a given unchanging world, we model changes of knowledge due to observations by agents. There is no factual change (also known as ontic change, literally: change of the world). But there can be. Factual change  can be public --- when a solar eclipse obscures the sun to all alike --- but also private --- when you change the pincode of your bankcard. Integrated ontic and epistemic change can be straightforwardly described as action models \cite{jfaketal.lcc:2006,hvdetal.world:2008}, and extending action model logic with quantifiers over such epistemic actions is also straightforward. We refrain from details.

However, concerning quantification there is a big `if'. Unrestricted quantification over action models with factual change is conceptually problematic, because in that case `anything goes'. Given agents $A$ and atoms $P$, any finite epistemic model $M_s$ can be transformed into any other finite epistemic model $N_t$ by an action model with factual change \cite{hvdetal.world:2008}. The procedure is elementary. First announce the distinguishing formula of $s$ in $M$. Note that an announcement is a singleton action model. The result is the singleton epistemic model with point $s$. Then execute an action model with the shape of $N$, trivial preconditions, and assigning the values to the atoms that they have in model $N_t$. Now take the composition of these two action models. This action model updates $M_s$ into (a model isomorphic to) $N_t$. We merely illustrate this by a three-agent example as below, where $p \et \neg q$ is the distinguishing formula of state $10$.
\begin{center}
\begin{tikzpicture}
\node (00) at (0,0) {$00$};
\node (10) at (2,0) {$\underline{10}$};
\node (01) at (0,2) {$01$};
\node (11) at (2,2) {$11$};
\draw[-] (00) -- node[above] {$a$} (10);
\draw[-] (01) -- node[above] {$ac$} (11);
\draw[-] (00) -- node[right] {$b$} (01);
\draw[-] (10) -- node[right] {$b$} (11);
\node (m) at (3.5,1) {$\stackrel {p \et \neg q} \Imp $};
\node (l10) at (5,0) {$\underline{10}$};
\node (m2) at (6,1) {\large $\otimes$};
\node (ll10) at (8,0) {\footnotesize{\underline{make $p$ and $q$ true}}};
\node (ll11) at (8,2) {\footnotesize{make $p$ and $q$ false}};
\draw[-] (ll10) -- node[right] {$bc$} (ll11);
\node (m3) at (10,1) {\large $=$};
\node (lll10) at (11.5,0) {$\underline{11}$};
\node (lll11) at (11.5,2) {$00$};
\draw[-] (lll10) -- node[right] {$bc$} (lll11);
\end{tikzpicture}
\end{center}
Let now $\dia{\otimes}$ be a modality quantifying over such action models with factual change. Using the above procedure, we can show that on the class of finite $\mathcal S5$ models $\dia{\otimes}\phi$ is valid for any consistent $\phi \in \lang(\Dia)$. We merely rephrase the proof in \cite{hvdetal.world:2008} (that holds for a more general class of models): Let $N_t$ be an epistemic model satisfying $\phi$. Let $M_s$ be arbitrary. Let now $E_e$ be the action model that is the composition of two such as above, so that $(M\otimes E)_{(s,e)}$ is bisimilar to $N_t$. Then $M_s \models \dia{E_e}\phi$. Therefore $M_s \models \dia{\otimes}\phi$. End of proof. 

We conjecture that the restriction to finite models is not necessary to obtain the validity of $\dia{\otimes}\phi$, by using the synthesis results for action model logic reported in \cite{hales2013arbitrary} and Section~\ref{sec.aaml}. Which brings us to the morale of this story: a logic wherein we can realize every consistent formula $\phi$ by an update may not be of much practical interest. However, much more restricted quantification over epistemic and ontic actions is the common way to proceed in epistemic planning. This will be discussed in Section~\ref{planning}.

\subsection{Group epistemic operators}

Let us now consider logics with group epistemic operators, for a group $G \subseteq A$ of agents. The three best known types of group epistemic operators are mutual, common, and distributed knowledge. `It is {\em mutual knowledge} between the agents in $G$ that $\phi$' (`everybody in $G$ knows $\phi$'), notation $E_G \phi$, is defined as the conjunction of $\Box_a \phi$ for all $a \in G$. `It is {\em common knowledge} between the agents in $G$ that $\phi$', notation $C_G \phi$, is syntactically defined as the arbitrary iteration of everybody knows $\phi$, or semantically defined as the truth of $\phi$ in all states reachable from a given state by the relation $(\Union_{a \in G} R_a)^*$ (the finite $G$-paths). `It is {\em distributed knowledge} between the agents in $G$ that $\phi$', notation $D_G \phi$, is defined as truth of $\phi$ in all states reachable from a given state by the intersection relation $\Inter_{a \in G} R_a$. For references, see \cite{faginetal:1995,handbookintro:2015}.

When adding common knowledge to public announcement logic, the public announcement modality can no longer be eliminated from the language by reduction axioms, and the axiomatization is non-trivial \cite{baltagetal:1998,hvdetal.del:2007}, which has led to the proposal of relativized common knowledge \cite{jfaketal.lcc:2006}. Adding distributed knowledge to public announcement logic is also non-trivial for the axiomatization, because proving completeness requires adjusting the standard canonical model technique \cite{WangA13}, and it also needs a different notion of bisimulation \cite{Roelofsen07}. Given the often surprising and unexpected expressivity results for logics wherein epistemic and dynamic modalities interact, and also for logics wherein these interact with quantifiers over dynamic modalities, the expressivity landscape is more complex in the presence of group epistemic operators. Few proposals have been made to extend such logics with group epistemic modalities. We can roughly distinguish contributions adding group epistemic modalities to quantifying over public change, to quantifying over semi-public change, and to quantifying over private change (or yet other epistemic actions). 

\paragraph{Group epistemics and quantifying over public change}

GAL with distributed knowledge is investigated in \cite{GalimullinAA19} and the extended journal version \cite{rustametal.2021}, with results on axiomatization and expressivity. 

APAL and GAL with common knowledge are investigated in \cite{GalimullinA21}. Different semantics for the APAL quantifier $\dia{!}\phi$ are proposed, and similarly for the CAL and GAL quantifiers. Let $\lang(\dots,C)$ be the extension with common knowledge of a language $\lang(\dots)$. The three investigated APAL variants are as follows, where $\dia{!} \phi \in\lang(\Dia,!,\dia{!},C)$. Note that in all cases the quantification is over quantifier-free formulas. 
\[ \begin{array}{lcll}
M_s \models \dia{!} \phi & \text{iff} & \text{there is a $\psi\in\lang(\Dia)$ such that } M_s \models \dia{\psi} \phi \\ 
M_s \models \dia{!} \phi & \text{iff} & \text{there is a $\psi\in\lang(\Dia,C)$ such that } M_s \models \dia{\psi} \phi \\ 
M_s \models \dia{!} \phi & \text{iff} & \text{there is a $\psi\in\lang(\Dia,!,C)$ such that } M_s \models \dia{\psi} \phi 
\end{array} \]
There are results on expressivity and on axiomatization. 

\paragraph{Group epistemics and quantifying over semi-public change}

In \cite[Section~4.3]{hvdetal.aij:2017} arbitary arrow update logic (without common knowledge) is shown to be incomparable in expressivity to epistemic logic with common knowledge, on class $\mathcal S5$. 

In \cite{Kuijer17} it is shown that arbitrary arrow update logic with common knowledge does not have a finitary axiomatization. This could in principle throw further light on the open question whether APAL (without common knowledge) has a finitary axiomatization. 

\paragraph{Group epistemics and quantifying over private change}

Extensions of arbitrary action model logic AAML and refinement modal logic RML with common knowledge or with distributed knowledge seem more manageable than extensions of APAL, GAL and AAUL with common knowledge, because the quantifiers can be eliminated from AAML and RML. It seems worth investigating if they can still be eliminated from such extensions. Such extensions would be relevant for epistemic planning and for information-based security protocols (Section \ref{sec.appgalcal}), wherein distributedly known initial problem configurations and commonly known epistemic goals are important.  We do not know of any publications.

\subsection{Epistemic planning} \label{planning}

Let us assume multi-agent $\mathcal S5$ models. 
An epistemic planning problem consists of an initial state $s$ (where we should think of this as a pointed epistemic model $M_s$) or a set of initial states, a finite set $\Sigma$ of epistemic actions $\sigma$, and a goal formula $\gamma \in \lang(\Dia)$ \cite{bolanderetal:2011,Bolander17,BolanderEHMN19}. Instead of an initial state we may have a property of the initial state, a formula $\delta \in \lang(\Dia)$. In that case, we wish to find plans for any state satisfying $\delta$ (any configuration where block $C$ is on block $B$, any model wherein agent $a$ is ignorant of $p$, \dots). Solving an epistemic planning problem then becomes the question whether \[ \delta \et \dia{(\Union_\Sigma \sigma)^*} \gamma \] is satisfiable. If all $\sigma\in\Sigma$ are deterministic (that is, if the updates are partial functions), some sequence $\sigma_1 \dots \sigma_n$ in $(\Union_\Sigma \sigma)^*$ such that $M_s \models \delta$ and $M_s \models \dia{\sigma_1}\dots\dia{\sigma_n} \gamma$, is called a \emph{plan} to achieve the goal $\gamma$ in $M_s$.\footnote{For non-deterministic epistemic actions it is common in the planning community to define by abbreviation $(\sigma)\phi = \dia{\sigma}\phi \et [\sigma]\phi$ \cite{BolanderEHMN19}. We then require $M_s \models (\sigma_1)\dots(\sigma_n) \gamma$.} As the set of epistemic actions $\Sigma$ for the given planning problem is finite, we can view $\Union_\Sigma \sigma$ as a single non-deterministic program $\pi$. The problem whether $\delta \et \dia{\pi^*} \gamma$ is satisfiable is often undecidable \cite{AucherB13}, and the target in the epistemic planning community is to find decidable planning problems. The undecidability resides in the part $\dia{\pi^*}$. In cases where $\dia{\pi^*}$ can be seen as a quantifier $\dia{\bullet}$ (such as APAL or AAML, see Section~\ref{sec.itt}), this epistemic planning problem is decidable if the logic with language $\lang(\Dia,\bullet,\dia{\bullet})$ is decidable. The search for decidable logics with quantification over information change is therefore related to the search for decidable planning problems and to epistemic synthesis. 

The complexities may still be different. Expression $\delta \et \dia{\bullet} \gamma$ contains a single quantifier, but an arbitrary formula in $\lang(\Dia,\bullet,\dia{\bullet})$ may contain multiple quantifiers. So, even when decidable, the complexity of the problem whether any formula $\phi$ is satisfiable may be higher than the complexity of the problem whether the particular formula $\delta \et \dia{\bullet} \gamma$ is satisfiable.

There is also a difference in focus. Planning is typically about factual change (put block $B$ on top of block $C$), and epistemic planning in dynamic epistemic logic is about combinations of epistemic and factual change \cite{Bolander17,BolanderCPS20}. The quantification is then over a larger set of epistemic actions (see Section~\ref{factual}). Epistemic planning is an excellent vehicle to bring applications of dynamic epistemic logic within reach, such as robot planning \cite{BolanderDH21} (in a  language with common knowledge, factual change, and action models) and cognitive planning \cite{DavilaLLM21} (involving modalities for intentions and attitudes).

\subsection{Open problems and further research}

This is clearly a list of topics for further research and not a list of open problems:
\begin{itemize}
\item Decidable dynamic epistemic logics with arbitrary iteration of updates \cite{milleretal:2005}
\item Axiomatization of RML with common knowledge
\item A dynamic epistemic logic of gossip \cite{gattinger.phd:2018}
\item Quantifying over epistemic and ontic actions
\end{itemize}

\section{Changing models with modalities} \label{sec.sabotage}

\subsection{Sabotage logic and relation-changing logics}

We have dwelled a long time within the confinement of dynamic epistemic logics, considering the various ways to quantify over epistemic actions interpreted as model updates with corresponding dynamic modalities, with the intention to model arbitrary information change. There is a different tradition in modal logic with update modalities that are quantifiers and that affect the domain of a model or change the relations of the model, including investigations involving modal $\mu$-calculus and determination of first-order definability of modal logical fragments \cite{Lavalette04,BenthemI08}. Such logics are not intended to model change of knowledge and are not considered dynamic epistemic logics. Let us call them {\em modal update logics}. We do not present such modal update logics in detail, but point out how dynamic epistemic logics can benefit from results obtained in modal update logics. 

Modal update logics, with dynamic modalities for updates, are all very different and display a far greater variety than dynamic epistemic logics, but we can still roughly compare them with dynamic epistemic logics on a number of features. These features are not necessarily independent. Note that these features are informal, tentative, and (unless cited) not presented elsewhere.
\begin{itemize}
\item modal update logics permit local updates unlike the global updates in dynamic epistemic logics
\item modal update logics permit semantic updates unlike the syntactic updates in dynamic epistemic logics
\item modal update logics permit particular updates unlike the universal updates in dynamic epistemic logics
\end{itemize}

\paragraph{Local versus global} The update in a dynamic epistemic logic is a model transformer and not a pointed model transformer. It is a function transforming a model into another model. We call this {\em global}. This induces a relation between points in the initial model and points in the updated model. 

But the update is not a function of a pointed model. If it is, we call it {\em local}. 

For example, a public announcement $\phi$ induces a relation between a model $M$ and the restriction $M|\phi$ of that model. This induces a relation between \emph{pointed} epistemic models, for example, between $M_s$ and $(M|\phi)_s$. But although $s$ plays a role in the relation, because after all $\phi$ needs to be satisfied there, the state $s$ that is the point of $M$ does not determine the relation: any state $t$ in the domain of $M$ that also satisfies $\phi$ and that is not the designated point is treated the same as $s$. 

An update of $M_s$ removing a state $t$ different from $s$, or removing a pair $(s,t)$ from the accessibility relation, would be local.

This distinction between local and global was proposed in \cite{aucheretal:2009}.

\paragraph{Semantic versus syntactic}
The model transformations in dynamic epistemic logic are almost always with respect to the extension of \emph{formulas} evaluated in the initial model, such as the announcement $\phi$ (or the preconditions of an action model, or the source and target conditions of an arrow update). In that sense the update is \emph{syntactic} (formula dependent). 

If instead of a restricted model $M|\phi$ we were to consider a restricted model $M|T$ where $T$ is a subset of the domain of $M$, that may not be the extension of a formula, we call the update \emph{semantic}. 

Syntactic updates are typically less expressive than semantic updates, because the updated model is then a function of logical properties of the initial model. Indeed, public announcement logic is as expressive as epistemic logic. 

\paragraph{Particular versus universal}
Instead of restricting the model to \emph{all} states satisfying $\phi$, or to \emph{all} arrows (pairs in the accessibility relation) satisfying $\phi$ at the source and $\psi$ at the target, as in dynamic epistemic logics, we now restrict the model by removing \emph{a} state (that is, one state) in the domain, or \emph{a} pair in the relation, as in sabotage logic \cite{jfak.sabotage:2005}. 

\paragraph{} Let us see a few of such model transforming logics  \cite{Lavalette04,jfak.sabotage:2005,BenthemI08,aucheretal:2009,arecesetal:2012,fervari:2014, ArecesFHM18,ArecesFHM17,AucherBG18,GrossiR19}.

\paragraph{Sabotage, bridge and swap}
In Johan van Benthem's {\em sabotage games} \cite{jfak.sabotage:2005} a network with multiple connections between nodes (representing travel destinations) is given, a player called Runner attempts to travel between two given nodes in the network and a player called Blocker attempts to sabotage the first player by removing nodes in the network after every move of Runner. If Runner cannot reach his destination, Blocker wins. Otherwise, Runner wins. The focus of \cite{jfak.sabotage:2005} is on the complexities of solving such games (determine the winner). The modal semantics proposed in \cite[p.\ 271]{jfak.sabotage:2005} models removing a connection between destinations as removing a state in a corresponding Kripke model:
\[ M_s \models \dia{-}\phi \quad \text{iff} \quad \text{there is a } t\neq s \text{ such that } (M\!\!-\!t)_s \models \phi \] where $M\!\!-\!t$ is the model restriction of $M$ to $\Domain(M)\setminus\{t\}$.  
The inequality requirement indicates a relation to the difference operator proposed in \cite{derijke:1992}. We can also see this as `poisoning' or colouring that state, i.e., forbidding passage through it, as in \cite{GrossiR19}. Issues with axiomatizing sabotage modal logic are presented in \cite{AucherBG18}.

Another semantics for the sabotage modality, propagated by Carlos Areces and collaborators in \cite{arecesetal:2012,fervari:2014}, interprets the sabotage modality as removing a pair from the accessibility relation in a model. We then get:
\[ M_s \models \dia{-}\phi \quad \text{iff} \quad \text{there is a } (t,u) \in R \text{ such that } M^{-tu}_s \models \phi \] where $M^{-tu} = M|R^{-tu}$, the restriction of $M$ to the relation $R^{-tu}$ defined as $R^{-tu} = R \setminus \{(t,u)\}$. 

Instead of removing a pair from the relation, these authors also considered other relation-changing updates (and thus moving away from the motivating sabotage game to other applications). They also propose a modality adding a pair to the relation, also known as making a {\em bridge} between states:
\[ M_s \models \dia{+}\phi \quad \text{iff} \quad \text{there is a } (t,u) \notin R \text{ such that } M^{+tu}_s \models \phi \] where $M^{+tu}$ is as $M$ except with relation $R^+ = R \union \{(t,u)\}$. Or we can change the direction of the arrow, known as a {\em swap} of the states involved in the arrow: 
\[ M_s \models \dia{{\pm}}\phi \quad \text{iff} \quad \text{there is a } (t,u) \in R \text{ such that } M^{{\pm}tu}_s \models \phi \] where $M^{{\pm}tu}$ is as $M$ except with relation $R^{{\pm}tu} = R \setminus \{(t,u)\} \union \{(u,t)\}$. 

\begin{example}
Consider the (anonymous) single-agent model $M$ consisting of states $\{s,t,u\}$ with $V(p) = \{u\}$ and $R = \{(s,t),(s,u)\}$, as visualized below. We also depict a number of updated models according to the sabotage, bridge, and swap semantics. 
\begin{center}
\scalebox{0.78}{
\begin{tikzpicture}
\node (0) at (0,0) {$0(s)$};
\node (0b) at (1,-.5) {$M$};
\node (1) at (2,0) {$0(t)$};
\node (2) at (0,2) {$1(u)$};
\draw[->] (0) -- (1);
\draw[->] (0) -- (2);
\end{tikzpicture}
\quad
\begin{tikzpicture}
\node (0) at (0,0) {$0(s)$};
\node (0b) at (1,-.5) {$M^{-st}$};
\node (1) at (2,0) {$0(t)$};
\node (2) at (0,2) {$1(u)$};
\draw[->] (0) -- (2);
\end{tikzpicture}
\quad
\begin{tikzpicture}
\node (0) at (0,0) {$0(s)$};
\node (0b) at (1,-.5) {$M^{-su}$};
\node (1) at (2,0) {$0(t)$};
\node (2) at (0,2) {$1(u)$};
\draw[->] (0) -- (1);
\end{tikzpicture}
\quad
\begin{tikzpicture}
\node (0) at (0,0) {$0(s)$};
\node (0b) at (1,-.5) {$M^{+ut}$};
\node (1) at (2,0) {$0(t)$};
\node (2) at (0,2) {$1(u)$};
\draw[->] (0) -- (1);
\draw[->] (0) -- (2);
\draw[->] (2) -- (1);
\end{tikzpicture}
\quad
\begin{tikzpicture}
\node (0) at (0,0) {$0(s)$};
\node (0b) at (1,-.5) {$M^{+ts}$};
\node (1) at (2,0) {$0(t)$};
\node (2) at (0,2) {$1(u)$};
\draw[->] (0) -- (1);
\draw[->] (0) -- (2);
\draw[->,bend right=20] (1) to (0);
\end{tikzpicture}
\quad
\begin{tikzpicture}
\node (0) at (0,0) {$0(s)$};
\node (0b) at (1,-.5) {$M^{\pm{st}}$};
\node (1) at (2,0) {$0(t)$};
\node (2) at (0,2) {$1(u)$};
\draw[<-] (0) -- (1);
\draw[->] (0) -- (2);
\end{tikzpicture}
}
\end{center}
For example, $M_s \not\models \Box p \vel \Box \neg p$, whereas $M_s \models \dia{-}\Box p$ because  $M^{-st}_s \models \Box p$, but also $M_s \models \dia{-}\Box \neg p$ because  $M^{-su}_s \models \Box \neg p$. Therefore $M_s \not\models [-]\Box p$. Also, $M_s \not \models \Dia \Dia \neg p$ but $M_s \models \dia{+}\Dia\Dia \neg p$ because $M^{+ut}_s \models \Dia\Dia \neg p$, or if you wish because $M^{+ts}_s \models \Dia\Dia \neg p$. Finally $M_s \models [\pm](\Box p \vel \Box\neg p)$ because $M^{{\pm}st}_s \models \Box p$ and (not depicted) $M^{{\pm}su}_s \models \Box \neg p$. 
\end{example}

Sabotage, bridge and swap are all examples of relational change and such modal update logics go by the name of {\em relation-changing modal logics}. Note that the relational pair that is removed, added, or reversed is not selected according to a logical property, the update is semantic and not syntactic. Also note that it is one pair only, it is particular and not universal. Finally, the sabotage, swap, and bridge updates by Carlos Areces et al.\ are global updates (there is no relation between the removed pair and the actual state). However, these authors also proposed local versions of sabotage, swap, and bridge. Johan van Benthem's original sabotage semantics \cite{jfak.sabotage:2005} is also a local update: we remove a state {\em different from the actual state}. All updates in dynamic epistemic logic that we have seen in prior sections are global. 

In such sabotage logics and related logics we can {\em count}. If you remove a state from the domain, the resulting domain contains one state less, so if you do this three times it contains three states less. We can therefore count how many states there are in the (finite) domain, just as in for example the graded modal logics of \cite{fine:1972}. For example, in a three-state model, $[-]^3\bot$ is true in the \cite{jfak.sabotage:2005} semantics: you can twice remove a state different from the actual state, but not thrice. These logics with counting require more refined notions of bisimulation, because the size of a model is not an invariant for the usual notion of bisimulation. For example, a model is standardly bisimilar to two copies of it. But now no longer. A different notion of bisimulation for sabotage logic is proposed in \cite{AucherBG18}, and for relation-changing logics in general in \cite{ArecesFH15,FervariTZ21}, including for a variation of the $\dia{-}$ and $\dia{+}$ semantics above, namely where the removal of a pair $(t,u)$ is also defined if the pair was already absent, and the addition of a pair $(t,u)$ is also defined if it was already present.

In such logics we can also give {\em names} to the states in domain. If you remove a state from the domain different from the actual state, this allows you to refer to the actual state in the logical language, as in hybrid logics.  In \cite{ArecesFHM16}, relation-changing logics are modelled as hybrid logics.

When we can count worlds and give them names, the logics may lack (finitary) Hilbert-style axiomatizations and may be undecidable, which is indeed the case \cite{AucherBG18,ArecesFHM18,ArecesFHM17}. 

\subsection{Dynamic epistemic logics and relation-changing logics}

Let us now relate the domain-changing or relation-changing modal update logics to the dynamic epistemic logics presented in this survey. In relation-changing modal logics we can consider arbitrary iterations or fixpoints of the structural operations wherein a mere pair of the accessibility relation is changed  \cite{Lavalette04,BenthemI08}. In this section we investigate to what extent such iterated transformations correspond to dynamic epistemic updates.

First consider refinement modal logic RML. Removing a pair from the accessibility relation of a model is a refinement of that model. Removing any finite number of pairs is also a refinement of that model. As is removing any infinite number of pairs of the relation, or the entire relation, which may be a limit or fixpoint (models need not have countable domains). It may be interesting to investigate on which class of models (all finite models?) we have that \begin{quote} $\rmldia\phi$ is equivalent to $\dia{-}^*\phi$.\end{quote}
Similarly, the quantifier in simulation modal logic can be seen as arbitrary iteration of the bridge modality, such that it is envisageable for some model class that \begin{quote}$\dia{\preceq}\phi$ is equivalent to $\dia{+}^*\phi$. \end{quote} 

Now consider arrow update logics AUL and AUML, wherein pairs of the accessibility relation come with logical conditions: arrows with source or target formulas. In sabotage logic we delete pairs, whereas in AUL we retain pairs, but this is mere perspective. Retaining all pairs $(s,t)$ with source condition $\phi$ and target condition $\psi$ in a relation $R_a$ (arrow $\phi \imp_a \psi$) is the same as \emph{deleting} all pairs that satisfy $\neg\phi$ at the source or $\neg\psi$ at the target (arrows $\neg\phi \imp_a \top$ and $\top \imp_a \neg\psi$). 

Assuming a single agent $a$, arbitrarily iterated removal of arrows $\top \imp_a \neg p$ appears to correspond to the `arrow eliminating' semantics for public announcement of \cite{gerbrandyetal:1997,kooi.jancl:2007}. Whereas removing all pairs with different values for $p$ at source and target, so preserving all pairs satisfying $p \imp_a p$ or $\neg p \imp_a \neg p$, corresponds to the standard public announcement semantics, in the model refinement sense of \cite{jfaketal.jancl:2007}. What works for propositional variables need not work for arbitrary formulas. Public announcement of $\phi$ is not merely iterated sabotage of arrows $\top \imp_a \neg\phi$, because of Moorean phenomena.

That public announcement can be viewed as iterated removal of states or arrows was suggested by Johan van Benthem. This seems not explicitly addressed in any of the relevant publications  \cite{BenthemI08,AucherBG18,jfaketal:2020}. A `stepwise public announcement' (one pair in the accessibility relation only) was proposed in \cite{AucherBG18} as a variation on the (standard) truthful public announcement semantics (Section 6.2, {\em Stepwise versions of DEL}, in \cite{AucherBG18}). In  \cite{jfaketal:2020}, public announcement and `stepwise removal' (sabotage) are combined in a single logic, but there is no suggestion that one can be defined in terms of the other.

\subsection{Open problems and further research}

These are all very open research projects and not specific open problems. We list:
\begin{itemize}
\item A decidable and axiomatizable relation-changing modal logic
\item A refinement modal logic with refinements as iterated sabotage
\item A public announcement logic with announcements as iterated sabotage
\end{itemize}

\section{Outlook}

We have presented many different dynamic epistemic logics with quantification over information change: arbitrary public announcement logic APAL with quantification over announcements, arbitrary action model logic AAML with quantification over action models, arbitrary arrow update logic AAUL with quantification over arrow updates, and refinement modal logic RML with quantification over refinements. And then the many variations on each of those, with all different logical properties. Some corners of this spectrum are not yet well understood, in particular the various ways to obtain decidable APAL versions, the various simulation modal logics and how they relate to refinement modal logics, and the different complexities of very similar logics quantifying over epistemic actions, whether as action models, arrow updates, or refinements. Promising current developments are the combination of group epistemic modalities with quantifiers, and the investigation of iterated epistemic actions or quantifiers in relation to epistemic planning. Further developments could involve incorporating features of relation-changing modal logics (sabotage, bridge, swap) with dynamic epistemic logics, and there are subjects beyond the scope of this survey such as succinctness, where we can expect logics with quantifiers to be more succinct than the same logic without. Applications to gossip protocols, security protocols, and epistemic planning are to be developed much further. Let a hundred flowers bloom.

\section*{Acknowledgements}

This investigation is a follow-up of a prior survey of logics with quantifiers over information change, entitled `Quantifying Notes' \cite{hvd.wollic:2012}, that also focussed on open problems. Some open problems in \cite{hvd.wollic:2012} have since then been resolved and are now presented as results, for example on GAL and CAL expressivity. Other open problems have come to the fore.

I acknowledge helpful comments and encouragement from: Thomas {\AA}gotnes, Philippe Balbiani, Johan van Benthem, Thomas Bolander, St\'ephane Demri, Raul Fervari, Sa\'ul Fern\'andez Gonz\'alez, Tim French, Rustam Galimullin, Rob van Glabbeek, Davide Grossi, Wiebe van der Hoek, Sophie Pinchinat, and Igor Sedl\'ar. This survey got under way while I was under lockdown in France in the same apartment complex as Hao Wu and Jinsheng Chen. Their support was invaluable and unforgettable. 

Together with Thomas Bolander, Jan van Eijck and Ramanujam, in 2015 I organized a Lorentz Center workshop entitled {\em To Be Announced}, with, on the request of the organization, a subtitle {\em Synthesis of Epistemic Protocols} to avoid confusion. See:
\begin{center}{\footnotesize \url{https://www.lorentzcenter.nl/to-be-announced-synthesis-of-epistemic-protocols.html}}.\end{center} On various occasions I gave invited seminars entitled {\em To Be Announced}. ``Can we at least make this  into `\emph{To Be Announced}', with quotes?'' ``Please don't.'' The predictable confusion with publishers, supporting staff and events servers has always been hilarious  and I hope it will be never-ending.

\bibliographystyle{plainurl}
\bibliography{biblio2022}

\end{document}